\documentclass[fleqn,usenatbib]{mnras}

\usepackage{newtxtext,newtxmath}
\usepackage{hyperref}

\usepackage[T1]{fontenc}
\usepackage{fontawesome}

\DeclareRobustCommand{\VAN}[3]{#2}
\let\VANthebibliography\thebibliography
\def\thebibliography{\DeclareRobustCommand{\VAN}[3]{##3}\VANthebibliography}

\DeclareRobustCommand{\DA}[3]{#2}
\let\DAthebibliography\thebibliography
\def\thebibliography{\DeclareRobustCommand{\DA}[3]{##3}\DAthebibliography}

\usepackage{graphicx}	
\usepackage{amsmath}	
\usepackage{wasysym}
\usepackage{newtxtext,newtxmath}
\usepackage{bm}

\newcommand{\pyaneti}{\texttt{pyaneti}}
\newcommand{\logr}{$\log R'_{\rm HK}$}

\newcommand{\lbe}{$\lambda_{\rm e}$}
\newcommand{\lbp}{$\lambda_{\rm p}$}
\newcommand{\pgp}{$P_{\rm GP}$}
\newcommand{\gcm}{${\rm g\,cm^{-3}}$}
\newcommand{\ms}{${\rm m\,s^{-1}}$}
\newcommand{\kms}{${\rm km\,s^{-1}}$}

\newcommand{\citlalicue}{\texttt{citlalicue}}
\newcommand{\citlalatonac}{\texttt{citlalatonac}}
\newcommand{\vsini}{$v \sin i$}
\newcommand{\logg}{$\log g$}

\newcommand{\halpha}{$\mathrm{H}_{\alpha}$}

\newcommand{\mearth}{$M_{\oplus}$}
\newcommand{\rearth}{$R_{\oplus}$}

\newcommand{\feh}{\ensuremath{[\mbox{Fe}/\mbox{H}]}}
\newcommand{\teff}{\ensuremath{T_{\mathrm{eff}}}}

\newcommand{\tess}{\emph{TESS}}
\newcommand{\ngts}{NGTS}

\newcommand{\jwst}{\emph{JWST}}
\newcommand{\plato}{\emph{PLATO}}
\newcommand{\serval}{\texttt{SERVAL}}
\newcommand{\yarara}{\texttt{YARARA}}
\newcommand{\monthyear}{\@monthname~\number\year}

\newcommand{\target}{TOI-451}
\newcommand{\targetb}{\target\,b}
\newcommand{\targetc}{\target\,c}
\newcommand{\targetd}{\target\,d}
\newcommand{\targetbcd}{\target\,b, c and d}

\newcommand{\tzerob}[1][$\mathrm{days}$]{ $ 10312.4427_{-0.0090}^{+0.0067} $~#1 } 
\newcommand{\pb}[1][$\mathrm{days}$]{ $ 1.8587033_{-0.0000114}^{+0.0000093} $~#1 }

\newcommand{\bb}[1][]{ $ 0.16_{-0.11}^{+0.16} $~#1 } 
\newcommand{\rhostarb}[1][$\mathrm{g\,cm^{-3}}$]{ $ 1.84_{-0.22}^{+0.20} $~#1 } 
\newcommand{\kb}[1][$\mathrm{m\,s^{-1}}$]{ $2.6_{-1.2}^{+1.1} $~#1 } 
\newcommand{\tzeroc}[1][$\mathrm{days}$]{ $ 10314.6376 \pm 0.0025 $~#1 } 
\newcommand{\pc}[1][$\mathrm{days}$]{ $ 9.192463 \pm 0.000017 $~#1 }

\newcommand{\bc}[1][]{ $ 0.26_{-0.16}^{+0.14} $~#1 } 
 
\newcommand{\kc}[1][$\mathrm{m\,s^{-1}}$]{ $ 1.2_{-0.8}^{+1.0} $~#1 } 
\newcommand{\tzerod}[1][$\mathrm{days}$]{ $ 10314.9694 \pm 0.0014 $~#1 } 
\newcommand{\pd}[1][$\mathrm{days}$]{ $ 16.364962 \pm 0.000015 $~#1 }

\newcommand{\bd}[1][]{ $ 0.382_{-0.095}^{+0.085} $~#1 } 
 
\newcommand{\kd}[1][$\mathrm{m\,s^{-1}}$]{ $ 2.7 \pm 1.2 $~#1 } 
\newcommand{\rprstarb}[1][]{ $ 0.01983 \pm 0.00072 $~#1 } 
\newcommand{\rprstarc}[1][]{ $ 0.03166 \pm 0.00078 $~#1 } 
\newcommand{\rprstard}[1][]{ $ 0.04243_{-0.00073}^{+0.00078} $~#1 } 
\newcommand{\qonetess}[1][]{ $ 0.35_{-0.15}^{+0.26} $~#1 } 
\newcommand{\qtwotess}[1][]{ $ 0.40_{-0.24}^{+0.34} $~#1 } 
\newcommand{\qonengts}[1][]{ $ 0.21_{-0.14}^{+0.24} $~#1 } 
\newcommand{\qtwongts}[1][]{ $ 0.35_{-0.25}^{+0.38} $~#1 } 
\newcommand{\rvserval}[1][$\mathrm{km\,s^{-1}}$]{ $ 0.0157_{-0.0031}^{+0.0029} $~#1 } 
\newcommand{\ngtsoffset}[1][$\mathrm{km\,s^{-1}}$]{ $ 1.0012 \pm 0.0018 $~#1 } 
\newcommand{\rvjitterrvserval}[1][$\mathrm{km\,s^{-1}}$]{ $ 2.9_{-0.9}^{+1.1} $~#1 } 
\newcommand{\rvjitterngts}[1][$\mathrm{km\,s^{-1}}$]{ $ 1.39_{-0.20}^{+0.21} $~#1 }

\newcommand{\Azero}[1][]{ $ 6.1_{-3.1}^{+3.0} $~#1 } 
\newcommand{\Aone}[1][]{ $ 28.0_{-2.8}^{+3.5} $~#1 } 
\newcommand{\Atwo}[1][]{ $ -8.6_{-1.0}^{+0.8} $~#1 } 
 
\newcommand{\lambdae}[1][]{ $ 11.5 \pm 1.2 $~#1 } 
\newcommand{\lambdap}[1][]{ $ 0.363_{-0.022}^{+0.025} $~#1 } 
\newcommand{\PGP}[1][]{ $ 5.078_{-0.020}^{+0.021} $~#1 } 
\newcommand{\arstarb}[1][]{ $ 6.95_{-0.29}^{+0.24} $~#1 } 
 
\newcommand{\idegb}[1][$\mathrm{deg}$]{ $ 88.7_{-1.4}^{+0.9} $~#1 } 
 
\newcommand{\Teqb}[1][$\mathrm{K}$]{ $ 1474_{-40}^{+45} $~#1 } 
 
\newcommand{\ab}[1][$\mathrm{AU}$]{ $ 0.0274 \pm 0.0017 $~#1 } 
\newcommand{\rpb}[1][$\mathrm{R_{\oplus}}$]{ $ 1.84 \pm 0.11 $~#1 } 
\newcommand{\trtb}[1][$\mathrm{hours}$]{ $ 2.059_{-0.078}^{+0.066} $~#1 } 
 
\newcommand{\Fpb}[1][$\mathrm{F_{\oplus}}$]{ $ 786_{-82}^{+101} $~#1 } 
 
\newcommand{\mpb}[1][$\mathrm{M_{\oplus}}$]{ $ 4.7_{-2.2}^{+2.1} $~#1 } 
\newcommand{\pdenb}[1][$\mathrm{g\,cm^{-3}}$]{ $ 4.1_{-1.9}^{+2.2} $~#1 } 
\newcommand{\pgrab}[1][$\mathrm{cm\,s^{-2}}$]{ $ 1210_{-573}^{+576} $~#1 } 
\newcommand{\pgratwob}[1][$\mathrm{cm\,s^{-2}}$]{ $ 1352_{-635}^{+658} $~#1 } 
\newcommand{\TSMb}[1][]{ $ 41_{-13}^{+36} $~#1 } 
\newcommand{\arstarc}[1][]{ $ 20.18_{-0.85}^{+0.70} $~#1 } 
 
\newcommand{\idegc}[1][$\mathrm{deg}$]{ $ 89.27_{-0.43}^{+0.47} $~#1 } 
 
\newcommand{\Teqc}[1][$\mathrm{K}$]{ $ 865_{-24}^{+27} $~#1 } 
 
\newcommand{\ac}[1][$\mathrm{AU}$]{ $ 0.0795 \pm 0.0050 $~#1 } 
\newcommand{\rpc}[1][$\mathrm{R_{\oplus}}$]{ $ 2.93 \pm 0.16 $~#1 } 
\newcommand{\trtc}[1][$\mathrm{hours}$]{ $ 3.461_{-0.078}^{+0.086} $~#1 } 
 
\newcommand{\Fpc}[1][$\mathrm{F_{\oplus}}$]{ $ 93_{-10}^{+12} $~#1 } 
 
\newcommand{\mpc}[1][$\mathrm{M_{\oplus}}$]{ $ 3.7_{-2.4}^{+3.2} $~#1 } 
\newcommand{\pdenc}[1][$\mathrm{g\,cm^{-3}}$]{ $ 0.81_{-0.53}^{+0.72} $~#1 } 
\newcommand{\pgrac}[1][$\mathrm{cm\,s^{-2}}$]{ $ 380_{-248}^{+328} $~#1 } 
\newcommand{\pgratwoc}[1][$\mathrm{cm\,s^{-2}}$]{ $ 426_{-278}^{+368} $~#1 } 
\newcommand{\TSMc}[1][]{ $ 123_{-57}^{+231} $~#1 } 
\newcommand{\arstard}[1][]{ $ 29.6_{-1.3}^{+1.0} $~#1 } 
 
\newcommand{\idegd}[1][$\mathrm{deg}$]{ $ 89.26 \pm 0.20 $~#1 } 
 
\newcommand{\Teqd}[1][$\mathrm{K}$]{ $ 714_{-19}^{+22} $~#1 } 
 
\newcommand{\ad}[1][$\mathrm{AU}$]{ $ 0.1168_{-0.0074}^{+0.0072} $~#1 } 
\newcommand{\rpd}[1][$\mathrm{R_{\oplus}}$]{ $ 3.93 \pm 0.20 $~#1 } 
\newcommand{\trtd}[1][$\mathrm{hours}$]{ $ 4.094_{-0.045}^{+0.050} $~#1 } 
 
\newcommand{\Fpd}[1][$\mathrm{F_{\oplus}}$]{ $ 43.2_{-4.5}^{+5.6} $~#1 } 
 
\newcommand{\mpd}[1][$\mathrm{M_{\oplus}}$]{ $ 10.2_{-4.5}^{+4.6} $~#1 } 
\newcommand{\pdend}[1][$\mathrm{g\,cm^{-3}}$]{ $ 0.92_{-0.41}^{+0.46} $~#1 } 
\newcommand{\pgrad}[1][$\mathrm{cm\,s^{-2}}$]{ $ 576_{-255}^{+276} $~#1 } 
\newcommand{\pgratwod}[1][$\mathrm{cm\,s^{-2}}$]{ $ 645_{-286}^{+307} $~#1 } 
\newcommand{\TSMd}[1][]{ $ 90_{-29}^{+71} $~#1 }

\defcitealias{Newton2021}{N21}
\defcitealias{Gaia2020}{G20}

\title[Mass estimates of the young TOI-451 planets]
{Mass estimates of the young TOI-451 transiting planets: Multidimensional Gaussian Process on stellar spectroscopic and photometric signals}

\author[Barragán et al.]{
Oscar~Barragán$^{1,2}$\thanks{\href{mailto:oscar.barragan@physics.ox.ac.uk}{oscar.barragan@physics.ox.ac.uk}, \href{mailto:oscar.barragan@warwick.ac.uk}{oscar.barragan@warwick.ac.uk}},
Manuel~Mallorqu\'in$^{3,4}$, 
Jorge~Fernández-Fernández$^{2,5}$, 
Faith~Hawthorn$^{2,5}$,
\and
Alix~V.~Freckelton$^{6}$,
Marina~Lafarga$^{2,5}$,
Michael~Cretignier$^{1}$,
Yoshi~N.~E.~Eschen$^{2,5}$,
Samuel~Gill$^{2,5}$,
\and
Víctor~J.~S.~Béjar$^{3,4}$,
Nicolas~Lodieu$^{3,4}$,
Haochuan~Yu$^{1}$,
Thomas~G.~Wilson$^{2,5}$,
David~Anderson$^{7}$,
\and
Ioannis~Apergis$^{2,5}$,
Matthew~Battley$^{8,9}$,
Edward~M.~Bryant$^{2}$,
Pía~Cortés-Zuleta$^{10}$,
Edward~Gillen$^{8}$,
\and
James~S.~Jenkins$^{11,12}$,
Baptiste~Klein$^{1}$,
James~McCormac$^{2,5}$,
Annabella~Meech$^{13}$,
Erik~Meier-Valdés$^{1}$,
\and
Maximiliano~Moyano$^{7}$,
Annelies~Mortier$^{6}$,
Felipe~Murgas$^{3,4}$,
Louise~D.~Nielsen$^{14}$,
Suman~Saha$^{13,14}$,
\and
José~I.~Vines$^{7}$,
Richard~West$^{2,5}$,
Peter~J.~Wheatley$^{2,5}$,
and~Suzanne~Aigrain$^{1}$
\\
$^{1}$ Sub-department of Astrophysics, Department of Physics, University of Oxford, Oxford, OX1 3RH, UK \\
$^{2}$ Department of Physics, University of Warwick, Coventry CV4 7AL, UK \\
$^{3}$ Instituto de Astrof\'isica de Canarias (IAC), Calle V\'ia L\'actea s/n, 38205 La Laguna, Tenerife, Spain \\
$^{4}$ Departamento de Astrof\'isica, Universidad de La Laguna (ULL), 38206 La Laguna, Tenerife, Spain \\
$^{5}$ Centre for Exoplanets and Habitability, University of Warwick, Gibbet Hill Road, Coventry CV4 7AL, UK \\
$^{6}$ School of Physics and Astronomy, University of Birmingham, Edgbaston, Birmingham B15 2TT, UK \\
$^{7}$ Instituto de Astronom\'ia, Universidad Cat\'olica del Norte, Angamos 0610, 1270709, Antofagasta, Chile \\
$^{8}$ Astronomy Unit, Queen Mary University of London, G.O. Jones Building, Bethnal Green, London E1 4NS, UK \\
$^{9}$ Observatoire de Gen\`eve, Universit\'e de Gen\`eve, Chemin Pegasi, 51, 1290 Versoix, Switzerland \\
$^{10}$ SUPA School of Physics and Astronomy, University of St Andrews, North Haugh, St Andrews KY16 9SS, UK \\
$^{11}$ Instituto de Estudios Astrof\'isicos, Facultad de Ingenier\'ia y Ciencias, Universidad Diego Portales, Av. Ej\'ercito Libertador 441, Santiago, Chile \\
$^{12}$ Centro de Excelencia en Astrof\'isica y Tecnolog\'ias Afines (CATA), Camino El Observatorio 1515, Las Condes, Santiago, Chile \\
$^{13}$ Center for Astrophysics, Harvard \& Smithsonian, 60 Garden St, Cambridge, MA 02138, US \\
$^{14}$ University Observatory, Faculty of Physics, Ludwig-Maximilians-Universit\"at M\"unchen, Scheinerstr. 1, 81679 Munich, Germany \\
}

\date{Accepted XXX. Received YYY; in original form ZZZ}

\pubyear{2024}

\begin{document}
\label{firstpage}
\pagerange{\pageref{firstpage}--\pageref{lastpage}}
\maketitle

\begin{abstract}
The young TOI-451 planetary system, aged 125 Myr, provides a unique opportunity to test theories of planetary internal structures and atmospheric mass loss through examination of its three transiting planets. We present an exhaustive photometric and spectroscopic follow-up to determine the orbital and physical properties of the system.
We perform multidimensional Gaussian Process regression with the code \texttt{pyaneti} on spectroscopic time-series and NGTS/LCO light curves to disentangle the stellar and planetary signal in ESPRESSO radial velocities.
{We show how contemporaneous photometry serves as an activity indicator to inform RV modelling within a multidimensional Gaussian Processes framework. We argue that this can be exploited when spectroscopic observations are adversely affected by low signal-to-noise and/or poor sampling.}
We estimate the Doppler semi-amplitudes of $k_{\rm b}=$\kb, $k_{\rm c}=$\kc\, and $k_{\rm d}=$\kd. This translates in 2-$\sigma$ mass estimates for TOI-451\,b and d of $M_{\rm b}=$\mpb\ and $M_{\rm d}=$\mpd; as well as a mass upper limit for TOI-451\,c of $M_{\rm c} <11.5\,M_\oplus$.
The derived planetary properties suggest that planets c and d contain significant hydrogen-rich envelopes. 
The inferred parameters of TOI-451\,b are consistent with either a rocky world that still retains a small hydrogen envelope or a water world.
These insights make the TOI-451 system an ideal laboratory for future follow-up studies aimed at measuring atmospheric compositions, detecting atmospheric mass-loss signatures, and further exploring planetary formation and evolution processes.
\end{abstract}




\section{Introduction}

One of the current challenges of exoplanetary science is the characterisation of young exoplanets ($<1$\,Gyr) due to their host star's high activity levels. 
Exoplanets in nearby young stellar associations are especially valuable for studying early planetary evolution, as they are subject to rapid processes such as orbital migration, tidal circularisation, atmospheric escape, and thermal cooling—all of which significantly impact their physical and orbital characteristics. 
The well-constrained ages of these systems allow for robust estimates of the timescales over which these mechanisms operate \citep[e.g.,][]{Owen2013}.
Planets in these environments also provide critical insights to understand the origin of the bimodal radius distribution (the ``radius valley'') observed among close-in super-Earths and sub-Neptunes \citep{Fulton2017}. 
They serve as a crucial test for distinguishing between the photoevaporation-driven mass loss scenario \citep{owenwu2017} and the core-powered mass loss hypothesis \citep{Ginzburg2016}, which predict atmospheric evolution on markedly different timescales $\sim$100 Myr and $\sim$1 Gyr, respectively.

Radial velocity (RV) follow-up of transiting exoplanets is a standard approach that allows us to characterise, at first order, planetary nature. 
However, this is not straightforward for young star systems. 
Young stars are characterised by intense magnetic activity, which manifests in various phenomena on their surfaces \cite[e.g.,][]{Lanza2006}. A prominent feature of this activity is the presence of starspots—regions with distinct temperature and brightness—on the stellar surface \cite[e.g.,][]{Huerta2008}. These starspots differ in size and distribution, and as the star rotates, they cause significant photometric fluctuations, often much larger than the signals produced by transiting exoplanets \cite[e.g.,][]{Morris2020}.
Surface activity also impacts the stellar spectra used in RV measurements \cite[e.g.,][]{Saar1997}. Variations in the spectral line profiles, such as changes in their shape, depth, or symmetry, as well as the appearance of spurious emission or absorption features, can distort the observed spectrum. 
These alterations introduce apparent RV variations that do not correspond to actual motion, complicating the detection of planetary signals. In the case of young stars, such activity-driven RV signals can span from several tens to several hundreds of metres per second \citep[e.g.,][]{Barragan2019,Barragan2022,Suarez2022,Zicher2022}. 

Multidimensional Gaussian Process \citep[multi-GP; ][]{Rajpaul2015} regression is nowadays a standard to detect the RV signals of planets orbiting young and/or active stars \citep[][]{Barragan2019,Barragan2022,Luque2023,Mayo2019,Nardiello2022,Zicher2022}.
\citet{Aigrain2012} proposed the $FF'$ framework to connect the activity-induced photometric flux and RV variations, taking into account its different dependencies on the evolution of the active regions on the stellar surface.
The multi-GP framework of \citet{Rajpaul2015} appeared as a generalisation of the $FF'$ framework to allow the use of spectral activity indicators instead of photometry. 
This multi-GP approach is also a solution to the lack of contemporaneous photometry to RV observations by exploiting the simultaneity of the activity indicators that are extracted from the same spectra that the RVs are computed \citep{Rajpaul2015}\footnote{In this work we refer to events occurring at the exact same time as simultaneous, while contemporaneous denotes events within the same period,  but not necessarily at the same instant.}. 
However, recent studies have shown that the multi-GP framework fails to model the stellar signal if the spectroscopic observations do not sample well the stellar signal time scales \citep[e.g.,][]{Barragan2024,Fridlund2024}.

Recent studies have used contemporaneous photometry to sample the stellar signal with better cadence to help disentangle the stellar and planetary signals from the RVs \citep[][]{Gonzalez2024,Suarez2023,Beard2025}.
These studies use a sequential modelling approach, where a onedimensional (1D) GP regression is performed on the photometry, and the posteriors of the GP kernel hyperparameters are then used as priors when modelling the RV time-series \citep{Haywood2014}.
In this approach, different functions are used to model the stellar signal in photometry and RVs, though they have common covariance properties.
Thus, this fails to exploit the connection between photometry and RVs fully. Furthermore, 1D GP regressions on photometry and RVs yield different hyperparameters  \citep[see ][]{pyaneti2,Nicholson2022}, so training a 1D GP on photometry before applying it to RVs does not necessarily improve the performance.
To our knowledge, a tailored multi-GP regression including contemporaneous photometry and RV observations is still to be tested. 
We explore this scenario in this manuscript by modelling contemporaneous spectroscopic and photometric time-series of the young star \target.

Using observations collected by NASA’s Transiting Exoplanet Survey Satellite \citep[\tess;][]{Ricker2015}, \citet[][hereafter \citetalias{Newton2021}]{Newton2021} identified a planetary system consisting of three small transiting planets orbiting the star \target, a G star belonging to of the young (125 Myr) Pisces–Eridanus stellar stream. The transiting planets, \targetbcd, have periods of 1.86, 9.2, and 16.4\,d and radii of 1.9, 3.1, and 4.1\,\rearth, respectively.
Given that the host star is relatively bright ($V=10.94$), the target was identified as well-suited for additional follow-up observations, such as RV measurements to determine planetary masses, as well as transmission spectroscopy to probe the planets' atmospheres \citep{Batalha2019}. 
Further detailed characterisation of a system like \target\ can provide critical constraints for models of planetary formation and early evolution.
We now present the results of an intense ground-based follow-up campaign of spectroscopic and photometric observations of this system.

This paper is structured as follows: The photometric and spectroscopic data of  \target\ are detailed in Section~\ref{sec:data}. The analytical methods applied to these data are outlined in Section~\ref{sec:datanalaysis}. 
A discussion of the findings is provided in Section~\ref{sec:discussion}, and the paper concludes with a summary of the key outcomes in Section~\ref{sec:conclusions}.
This manuscript is part of a series of papers under the project \emph{GPRV: Overcoming stellar activity in radial velocity planet searches} funded by the European Research Council (ERC, P.I.~S.~Aigrain). 

\begin{table}
\caption{Main identifiers and parameters for \target.  \label{tab:parstellar} 
}
\begin{center}
\begin{tabular}{lcc} 
\hline
\noalign{\smallskip}
Parameter & Value &  Source \\
\noalign{\smallskip}
\hline
\noalign{\smallskip}
\multicolumn{3}{l}{\emph{Main identifiers}} \\
\noalign{\smallskip}
Gaia DR3  & 4844691297067063424 &  \citetalias{Gaia2020}  \\
TYC & 7577-172-1 & \citet[][]{Hog2000} \\
2MASS & J04115194-3756232 &  \citet[][]{Cutri2003} \\
TESS Input Catalog & 257605131 & \citet{Stassun2019} \\
Spectral type & G8V  & This work \\ 
\noalign{\smallskip}
\hline
\noalign{\smallskip}
\multicolumn{3}{l}{\emph{Equatorial coordinates, proper motion, and parallax}} \\
\noalign{\smallskip}
$\alpha$(J2000.0) &  $04^{\rm h}\,11^{\rm m}\,51.9469^{\rm s}$  &  \citetalias{Gaia2020} \\
$\delta$(J2000.0) & $-37^{\rm h}\,56^{\rm m}\,23.2192^{\rm s}$  &  \citetalias{Gaia2020} \\
$\mu_\alpha$\,(mas\,${\rm yr^{{-1}}}$) & $ 	-11.061 \pm 0.010$ &   \citetalias{Gaia2020}  \\
$\mu_\delta$\,(mas\,${\rm yr^{{-1}}}$) & $  12.347 \pm 0.014$ &    \citetalias{Gaia2020} \\
$\pi$\,(mas) & $ 	8.0993 \pm 0.0108$ &   \citetalias{Gaia2020} \\
Distance\,(pc) & $123.47 \pm 0.16$ &  \citetalias{Gaia2020} \\
\noalign{\smallskip}
\hline
\noalign{\smallskip}
\multicolumn{3}{l}{\emph{Magnitudes}} \\
B &	$ 11.64 \pm 0.07$ &	 \citet[][]{Hog2000} \\  		
V &	$ 10.94 \pm 0.06$ &	  \citet[][]{Hog2000} \\ 
Gaia G & $10.7498 \pm 0.0008 $ & \citetalias{Gaia2020}  \\ 	  		
J & $9.636 \pm 0.024$ &
	\citet{Cutri2003} 	 \\ 	  		
H &	$ 9.287 \pm 0.022$ &
	\citet{Cutri2003} 	 \\ 	  		
K &	$  9.190 \pm 0.023$ &
	\citet{Cutri2003} 	 \\
W1 & $9.137 \pm 0.024$ & AllWISE \\
W2 & $9.173 \pm 0.020$ & AllWISE \\
W3 & $9.117 \pm 0.027$ & AllWISE \\
W4 & $8.632 \pm 0.292$ & AllWISE \\
\noalign{\smallskip}
\hline
\noalign{\smallskip}
\multicolumn{3}{l}{\emph{Stellar parameters}} \\
$T_{\rm eff}$ (K) & $5490 \pm 115$ & This work \\
\logg (cgs, dex) & $4.53 \pm 0.22 $  &  This work \\
{[Fe/H]} (dex) &  $0.02 \pm 0.06$  & This work \\
$A_{\rm V}$ &  $0.018 \pm 0.016$  & This work \\
\vsini\ (\kms) & $8.70 \pm 0.96$ & This work \\
$\sin i$ & $1.04 \pm 0.13 $ & This work \\ 
$v_{\rm mic}$ (\kms) & $0.97 \pm 0.07$ & This work \\
$v_{\rm mac}$ (\kms) & $3.31 \pm 0.11$ & This work \\
Luminosity, $L_\star$ (L$_\odot$) & $0.59\pm0.05$ & This work \\
Mass, $M_\star$ (M$_\odot$) & $0.93\pm0.04$ & This work \\
Radius, $R_\star$ (R$_\odot$) & $0.85\pm0.03$ & This work \\
Density, $\rho_\star$ (\gcm) & $2.15 \pm 0.25$ & This work \\
$P_{\rm rot}$ (d) & $5.1\pm0.1$ & This work \\
Age (Myr) & $125 \pm 8$ & \citetalias{Newton2021} \\
\hline
\multicolumn{3}{l}{\footnotesize  \citetalias{Gaia2020} corresponds to \citet{Gaia2020}.}
\end{tabular}
\end{center}
\end{table}

\section{\target\ data}
\label{sec:data}

\subsection{\tess\ data}
\label{sec:tessdata}

\tess\ initially observed \target\ (TIC 257605131) during its first cycle, covering sectors 4 and 5 (from 2018-Oct-18 to 2018-Dec-11). These observations, together with supplementary ground and space-based data, enabled \citetalias{Newton2021} to discover and confirm three transiting planets orbiting \target. 
These planets, named \targetbcd, have periods of 1.9, 9.2 and 16.4\,d, with corresponding radii of $1.91 \pm 0.12$, $3.1 \pm 0.13$, and $4.07 \pm 0.15$ \rearth, respectively. For a detailed description of the discovery and validation of this planetary system, see \citetalias{Newton2021}.

Two years later, \tess\ returned to observe \target\ during its extended mission, as part of Cycle 3 in sector 31 (from 2020-Oct-21 to 2020-Nov-19). 
\citet{Barragan2021b} and \citet{Kokori2023} used all available \tess\ data to refine the planetary ephemerides and radii. These two manuscripts are the ones that present the last analyses performed on transits of \target. For this paper, we use the same \tess\ data as described and processed in \citet{Barragan2021b}. 

\target\ will be re-observed by \tess\ in Sectors 106 and 107 (July and August 2026), according to the \tess-point Web Tool\footnote{\url{https://heasarc.gsfc.nasa.gov/wsgi-scripts/TESS/TESS-point_Web_Tool/TESS-point_Web_Tool/wtv_v2.0.py}}.
It is also worth noting that \target\ is planned to be observed by the \textit{PLATO} mission \citep[][]{Rauer2024} in its first two years of observations \citep[see][and further discussion in Sect.~\ref{sec:futureobservations}]{Eschen2024}.

\subsection{NGTS photometry}
\label{sec:ngtsphotometry}

The Next Generation Transit Survey \citep[\ngts;][]{wheatley18ngts} is a photometric facility consisting of twelve independently steerable robotic telescopes situated at ESO's Paranal Observatory in Chile. Each \ngts\ telescope has a 20\,cm diameter aperture and observes a very wide field-of-view ($2.8 \times 2.8\,\mathrm{degrees}$) using a custom filter \citep[520-890\,nm;][]{wheatley18ngts}. Each telescope has Andor iKon L cameras with deep-depleted, red-sensitive, and back-sided illuminated CCDs.
The precision of \ngts\ observations is scintillation limited for bright stars \citep[$G < 11\,\mathrm{mag}$;][]{OBrien2022} and by using multiple \ngts\ telescopes to simultaneously observe the same star, we can achieve photometric precisions on the order of 100\,ppm-per-30\,minutes \citep{bryant20multicam}.

We observed \target\ with \ngts\ with two objectives. First, to obtain contemporaneous photometry to the ESPRESSO data to monitor the stellar variability of the host star. For these observations, TOI-451 was observed for 30\,min on every night that the telescope opened between 2023-08-18 to 2024-02-09, using the same telescope observing at a similar airmass and with precise auto-guiding to ensure consistent photometry night-to-night.
Such monitoring observations have been previously performed by \ngts\ to monitor stellar variability in support of spectroscopic observations \citep[e.g.][]{Ahrer2023wasp39}. 
Because we are interested in the signal evolution on time scales of days, we binned the data to one point per night.
The \ngts\ monitoring photometry is shown and discussed in Sections~\ref{sec:stellarsignal}~and~\ref{sec:rvanalysis}. NGTS time series can be accesed in the online version of Table~\ref{tab:harps}.


The second objective was to obtain transits of planets c and d to refine the ephemerides and planet radii. We observed two transits of TOI-451~c, observing the transit egress on the night of 2024-10-15 using six \ngts\ telescopes and a full transit on the night of 2024-10-24 using five \ngts\ telescopes. We observed an egress of TOI-451~d using five \ngts\ telescopes on 2024-12-13. The three \ngts\ observations were all performed using an exposure time of 10\,seconds, resulting in a total of 10894,  9710, and 10320 images for the three nights of observations, respectively. 
The \ngts\ images were reduced and the photometry was extracted using a custom version of the standard \ngts\ pipeline, which is described in \citet{wheatley18ngts}. This custom version is designed to obtain high-precision photometry for a specific target star. 
We flattened the transits by fitting a second-order polynomial to the out-of-transit data using \citlalicue\ \citep[][]{pyaneti2}.
The \ngts\ transit photometry is shown and discussed in Sections~\ref{sec:transitanalysis} and \ref{sec:final}.
Orange colour will be used in this paper for all the \ngts-related data/models. 

\subsection{LCO photometry}
\label{sec:lcophotometry}

We also monitored \target\ during 112 days between 14 November 2023 and 5 March 2024, using the two 40\,cm telescopes of Las Cumbres Observatory Global Telescope \citep[\textit{LCOGT};][]{Brown2013} at Las Campanas Observatory. We observed the target in the $V$-band and obtained about 78 useful epochs (about 2 epochs per 3 days), each of them with typically 10 individual exposures of 60s per epoch. 
The 40\,cm telescopes have a 3k$\times$2k SBIG CCD camera with a pixel scale of 0.571\,arcsec providing a field of view of 29.2$\times$19.5\,arcmin. Weather conditions were clear, and the average seeing varies from 3 to 6\,arcsec. Raw data was processed using the \texttt{BANZAI} pipeline \citep{McCully2018}, which includes bad pixel, bias, dark and flat field corrections for each night. We performed aperture differential photometry using AstroImageJ \citep{Collins2017}, adopting an aperture of 10 pixels (5.7\,arcsec).
The LCO photometry is shown and analysed in Sections~\ref{sec:stellarsignal}~and~\ref{sec:rvanalysis}. LCO time series can be acceded in the online version of Table~\ref{tab:harps}.
Purple colour will be used in this paper for all the LCO-related data/models. 

\subsection{Spectroscopic observations}
\label{sec:espressodata}

We acquired 134 ESPRESSO spectra of \target\ between October 2023 and March 2024. ESPRESSO \citep{Pepe2021} is a high-resolution spectrograph mounted on the 8.2 m Very Large Telescope (VLT; Paranal, Chile). The observations were taken as part of the ESO P112 programs 112.261J.001 (P.I. Barragán) and  112.25HY.001 (P.I. Mallorquín).
Each observation has an integration time of 900\,s, a median resolving power of 140 000, and a wavelength range of 380–788 nm. Observations were taken in high-resolution mode.
Given the complex stellar activity in \target, we reduced the spectra with different frameworks to cross validate the extracted RVs and activity indicators. Below, we describe the different methods used to reduce the spectra.

We note that 26 archival HARPS observations of \target\ were taken in 2019 (under the program: 0103.C-0759(A) by P.I. Benatti).
These observations show peak-to-peak variations of the order of 100~\ms, indicating activity-induced signals.
However, as these observations were taken several years before the ESPRESSO observations, the stellar signal has evolved, and these data are not helpful to constrain the stellar signal in the ESPRESSO time-series \citep[][]{Barragan2021b}. 
Furthermore, they do not have sufficient time-sampling or signal-to-noise to constrain the local shape of the stellar signal by themselves. 
Therefore, we decided not to include them in our analysis.

\subsubsection{\textit{DRS}}

We reduced the raw ESPRESSO data with the official Data Reduction Software\footnote{\href{www.eso.org/sci/software/pipelines/espresso/espresso-pipe-recipes.html}{www.eso.org/sci/software/pipelines/espresso/ espresso-pipe-recipes.html}} (DRS) version 3.0.0. The DRS performs standard echelle spectrum reduction, including bias and dark subtraction, optimal order extraction, bad pixel correction, flat-fielding, deblazing, wavelength calibration, and order merging \citep[see][for details]{Pepe2021}. 
We also used the DRS to compute cross-correlation functions (CCFs) with the default ESPRESSO G9 mask. For each observation, the CCFs are computed on an order-by-order basis, and a final CCF is obtained by coadding the individual-order CCFs. A Gaussian function is then fitted to each coadded CCF, and the RV of each observation is obtained from the centroid of the best-fit Gaussian.
The DRS provides activity indicators extracted from the CCF profile: the full-width-at-half-maximum (FWHM), also obtained from the Gaussian fit, and the bisector span (BIS), which measures the asymmetry of the CCF by computing the difference in average RV between the top and the bottom parts of the CCF profile, as described in \citet{Queloz2001}.
The DRS also provides standard chromospheric activity indicators such as \logr\ and \halpha.
The DRS RV time-series has a root-mean-square of 54\,\ms\ and a typical uncertainty of 2.0\,\ms. Figure~\ref{fig:rvtimeseries} shows the DRS RV data with teal green hexagons (teal green colour will be used in this paper for all the DRS-related data/models). 
Table~\ref{tab:harps} contains all the time series described in this section.

\subsubsection{\textit{Yarara}}


YARARA is a post-processing methodology that aims to improve the RV extraction and RV precision by applying post-processing correction on the spectra time-series. Although primarily designed to search for \ms\ signals around Sun-like activity stars \citep{Cretignier2023,Stalport2023,Dalal2024,Nari2025}, the pipeline can be applied to any RV datasets and has been employed in Rossiter-McLaughin studies \citep{Yu2025}. Recently, \yarara\ has been upgraded to extract stellar atmospheric parameters \citep{Cretignier2024b}.

During the pre-processing step, the S1D order-merged spectra from the official DRS are linearly interpolated on a common wavelength grid. The continuum of the spectra is then fitted by using \texttt{RASSINE} \citep{Cretignier2020b}. The fit of the continuum is mandatory to perform the colour correction as explained in \citet{CretignierPhD} or \citet{Barragan2024}. We did not use the telluric correction of the pipeline since its effect is expected to be very small in comparison to the stellar activity signal. Furthermore, the CCF masks used later were designed to exclude spectral regions contaminated by telluric bands. Ultimately, only the cosmic ray correction was kept in the spectra cleaning process. Because the star is a moderate rotator, we did not try to perform a tailored line selection as that process requires narrow, well-separated lines \citep[see][]{Cretignier2020}. Instead, we used the standard G9 CCF mask of the ESPRESSO DRS to extract RVs and CCF momenta from the \yarara\ processed spectra. 
The main difference between the \yarara- and DRS-extracted RVs is that the former is done on the S1D spectra while the latter is done order by order.
The \yarara\ RV time-series has a root-mean-square of 59\,\ms\ and a typical uncertainty of 2.0\,\ms. Figure~\ref{fig:rvtimeseries} shows the \yarara\ RV data as vermilion circles (vermilion colour will be used in this paper for all the \yarara-related data/models). 
Table~\ref{tab:harps} contains all the time series described in this section.

\subsubsection{SERVAL}

\begin{figure*}
    \centering
    \includegraphics[width=0.98\textwidth]{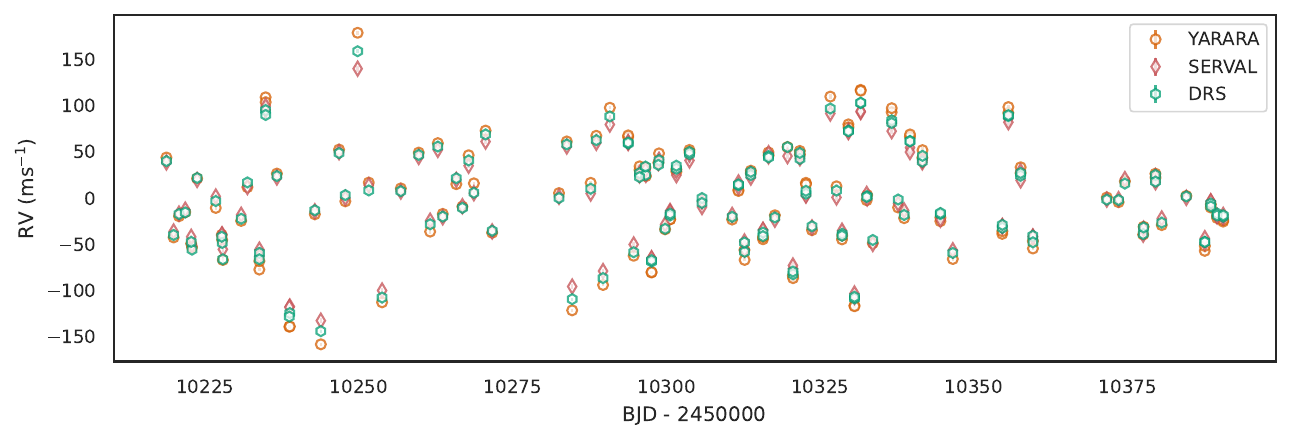}
    \caption{
    Mean-subtracted RV time-series as obtained with \yarara\ (vermilion circles), \serval\ (red diamonds), and DRS (teal green hexagons).
    }
    \label{fig:rvtimeseries}
\end{figure*}

We also extracted RVs from the ESPRESSO spectra with \serval\ \citep{SERVAL}. \serval\ computes a template spectrum from the observations themselves, rather than using a pre-existing digitised mask to compute the CCF as in the DRS. The template is constructed by co-adding the observed spectra in the barycentric rest-frame. For the reduction, we only use the orders from 10 to 160, since the rest have a relatively low signal-to-noise.
The \serval\ analysis also produces time series of activity indicators, in this case we extracted the differential line width (DLW) and the chromatic index (CI) \citep[see][for more details]{SERVAL}. 
The \serval\ RV time-series has a root-mean-square of 51\,\ms\ and a typical uncertainty of 2.8\,\ms. Figure~\ref{fig:rvtimeseries} shows the \serval\ RV data as red diamonds (red colour will be used in this paper for all the \serval-related data/models).
Table~\ref{tab:harps} contains all the time series described in this section.

\begin{table}
\begin{center}
\caption{Stellar time series. Columns list, in order, the observation time, the measured value, the associated uncertainty, the physical units, and a label. The label (fifth column) specifies both the measured quantity (e.g., NGTS photometry, RV serval, FWHM, etc).  The full table is provided in machine-readable format in the supplementary material.\label{tab:harps}}
\begin{tabular}{ccccc}
\hline\hline
Time & Value  &  $\sigma_{\rm Value}$ & Units & time-series \\
(${\rm BJD_{TDB}}$ - 2\,450\,000) &  &   &  &    \\
\hline
10218.692220 & 0.05559 & 0.00200 &  \kms\ & RV serval \\
10219.863082 & -0.01920 & 0.00236 &  \kms\ & RV serval \\
10220.738287 & 0.00016 & 0.0040 &  \kms\ & RV serval \\
10221.758381 & 0.00447 & 0.00178 &  \kms\ & RV serval \\
10222.856319 & -0.03405 & 0.00199  &  \kms\ & RV serval \\
$\cdots$  & $\cdots$  & $\cdots$  & $\cdots$  & $\cdots$   \\
\hline
\end{tabular}
\end{center}
\end{table}

\section{Data analysis}
\label{sec:datanalaysis}

\subsection{Stellar parameters}
\label{sec:stellarparameters}

Because of the high number of ESPRESSO spectra obtained in this work, we proceeded to estimate new stellar atmospheric and physical parameters. We also decided to adopt the stellar age of $125 \pm 8$\,Myr found by \citetalias{Newton2021} and stellar rotation of $P_{\rm rot} = 5.1 \pm 0.1$\,d \citep{Newton2021,Barragan2021b}.

\subsubsection{Atmospheric parameters}

To obtain stellar atmospheric parameters for TOI-451, we used the combined spectrum from the \yarara-processed ESPRESSO observations, achieving an SNR of 568. We used the PAWS pipeline \citep{Freckelton2024}, which employs the functionality of the \texttt{iSpec} package \citep{Blanco-Cuaresma2019}. The pipeline first determined initial estimates of the stellar parameters using the curve-of-growth equivalent widths method. 
These were used as input to the spectral synthesis method to subsequently determine the final atmospheric stellar parameters presented in Table~\ref{tab:parstellar}. The atmospheric parameter set determined by the PAWS pipeline consists of $T_{\rm eff}$, \logg, {[Fe/H]}, $v_{\rm mic}$, $v_{\rm mac}$, and \vsini. 
Both methods in the PAWS pipeline were conducted using the \texttt{SPECTRUM} line list \citep{gray1994} and the ATLAS set of model atmospheres \citep{kurucz2005}.

\subsubsection{\target's mass and radius}

To estimate \target's mass and radius, we performed Spectral Energy Distribution (SED) modelling using the software \texttt{ARIADNE}
\citep[][]{ariadne}.  
We used four stellar atmospheric models: {\tt {Phoenix~v2}} \citep{2013A&A...553A...6H}, {\tt {BtSettl}} \citep{2012RSPTA.370.2765A}, \citet{Castelli2004}, and \citet{Kurucz-93} to model \target's SED.
We used broad-band photometry from  2MASS J, H, and K,  {\it WISE} W1 and W2, the  {\it Johnson} B and V magnitudes, and {\it Gaia} G magnitude and parallax from Gaia DR3 (see Table~\ref{tab:parstellar}).
The parameter space was explored used the nested sampling algorithm implemented in \texttt{dynesty} \citep{dynesty}.
We set Gaussian priors for \teff, \logg,  \feh, and distance from our atmospheric analysis (see Table~\ref{tab:parstellar}).
Stellar radius ($R_\star$) and extinction ($A_V$) were treated as free parameters. 
We obtained $R_\star = 0.85 \pm 0.03\,R_\odot$ and $A_v = 0.018 \pm 0.016$\,mag.
\texttt{ARIADNE} also estimated a stellar mass of $M_\star = 0.93 \pm 0.04\,M_\odot$ using the \texttt{isochrones} package with the MIST stellar evolution tracks \citep{isochrones,choi2016}.
The corresponding results are summarised in Table~\ref{tab:parstellar}.

To cross-check the \texttt{ARIADNE} results, we also used the \texttt{PARAM\,1.3}\footnote{\url{http://stev.oapd.inaf.it/cgi-bin/param_1.3}} online calculator. 
We used the \teff\ and \feh\ values from our spectroscopic analysis together with the visual magnitude and parallax given in Table~\ref{tab:parstellar} as input for \texttt{PARAM\,1.3}.
Given the well-constrained age of this star given by \citetalias{Newton2021}, we set stellar age priors between 100 and 150 Myr and the \texttt{PARSEC} isochrones \citep{Bressan2012}.
We recovered a mass and radius of $M_\star = 0.96 \pm 0.02 M_\odot $ and $R_\star = 0.83 \pm 0.02 R_\odot$.
These results are in agreement with our \texttt{ARIADNE} estimates and \citetalias{Newton2021}.

\subsection{Transit analysis}
\label{sec:transitanalysis}

The transit modelling includes the photometric data from \tess\ and \ngts\ described in Sect.~\ref{sec:data}. To enhance computational efficiency, we constrain the analysis to temporal windows encompassing up to three hours on either side of each transit mid-point. 
Given that the \tess\ and \ngts\ time-series were acquired with short cadences (less than 2 minutes), the light curves can be accurately represented by instantaneous evaluations of the transit models \citep[cf., e.g.,][]{Gandolfi2018}.

To model the transits of \targetbcd, we need to sample and set priors for the following parameters for each planet: time of mid-transit, $T_0$; orbital period, $P_{\rm orb}$; orbital eccentricity, $e$, angle of periastron, $\omega_\star$; and the scaled planetary radius $R_{\rm p}/R_\star$.
We also sample for the stellar density, $\rho_\star$; and the limb darkening parameters $q_1$ and $q_2$ for each band \citep[following][models and parametrisations]{Mandel2002,Kipping2013}. The scaled semi-major axis for each planet, $a/R_\star$, is recovered from $\rho_\star$ and Kepler's third law \citep[see e.g.,][]{Winn2010}.
The model also includes a photometric jitter term per data set to penalise the likelihood.
We assume circular orbits for all planets. 

We set wide uniform priors for all the parameters based on previous analyses and physical boundaries. 
For all the subsequent analyses, we used \pyaneti\ \citep{pyaneti,pyaneti2}. In all our runs, we sample the parameter space with 250 walkers using the Markov chain Monte Carlo (MCMC) ensemble sampler algorithm implemented in \pyaneti\ which is based on \citet{emcee}. 
We ran the MCMC and check for convergence each 5000 iterations. We define convergence when the \citet{Gelman1992} statistics is less than 1.01. 
Once convergence is reached, posterior distributions are created with the last 5000 iterations of converged chains. Chains are thinned by a factor of 10, giving a distribution of 125\,000  points for each sampled parameter. 

\begin{figure}
\centering
\includegraphics[width=0.45\textwidth]{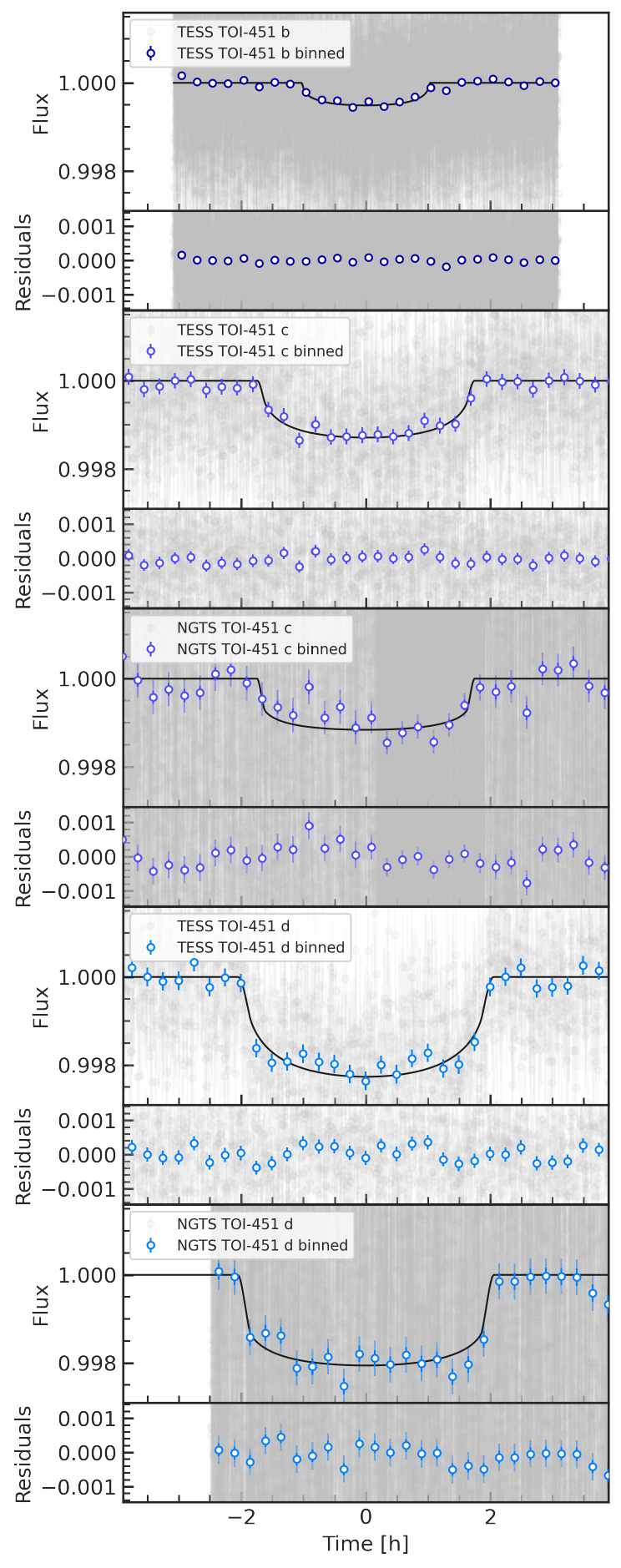}
\caption{
Detrended light curves phase-folded at the period of each planet (individual data points in transparent grey, phase-binned data in solid colour) together with the best-ﬁt transit model (black line). Residuals are shown in the lower inset. The $x$- and $y$-axes in each panel are shown with the same range to facilitate signal comparisons.}
\label{fig:transits}    
\end{figure}

Figure~\ref{fig:transits} shows the phase-folded transit data and inferred models for the three planets.
The major advantage of the inclusion of the \ngts\ data is that it improves the ephemerides for \targetc\ and d in comparison with the latest estimates of the system (2 to 3 times more precise). 
These new observations also show that there are no significant long-term transit time variations in the system.
This is of high importance for future transit observations. 
We use the results from the transit modelling as priors for our analyses of the RVs, which are presented in Sect.~\ref{sec:rvanalysis}.
The final inferred and derived transit parameters are given in Sect.~\ref{sec:final}.

\subsection{Stellar time-series characterisation}
\label{sec:stellarsignal}

\subsubsection{Periodograms}
\label{sec:periodograms}

\begin{figure}
\centering
\includegraphics[width=0.47\textwidth]{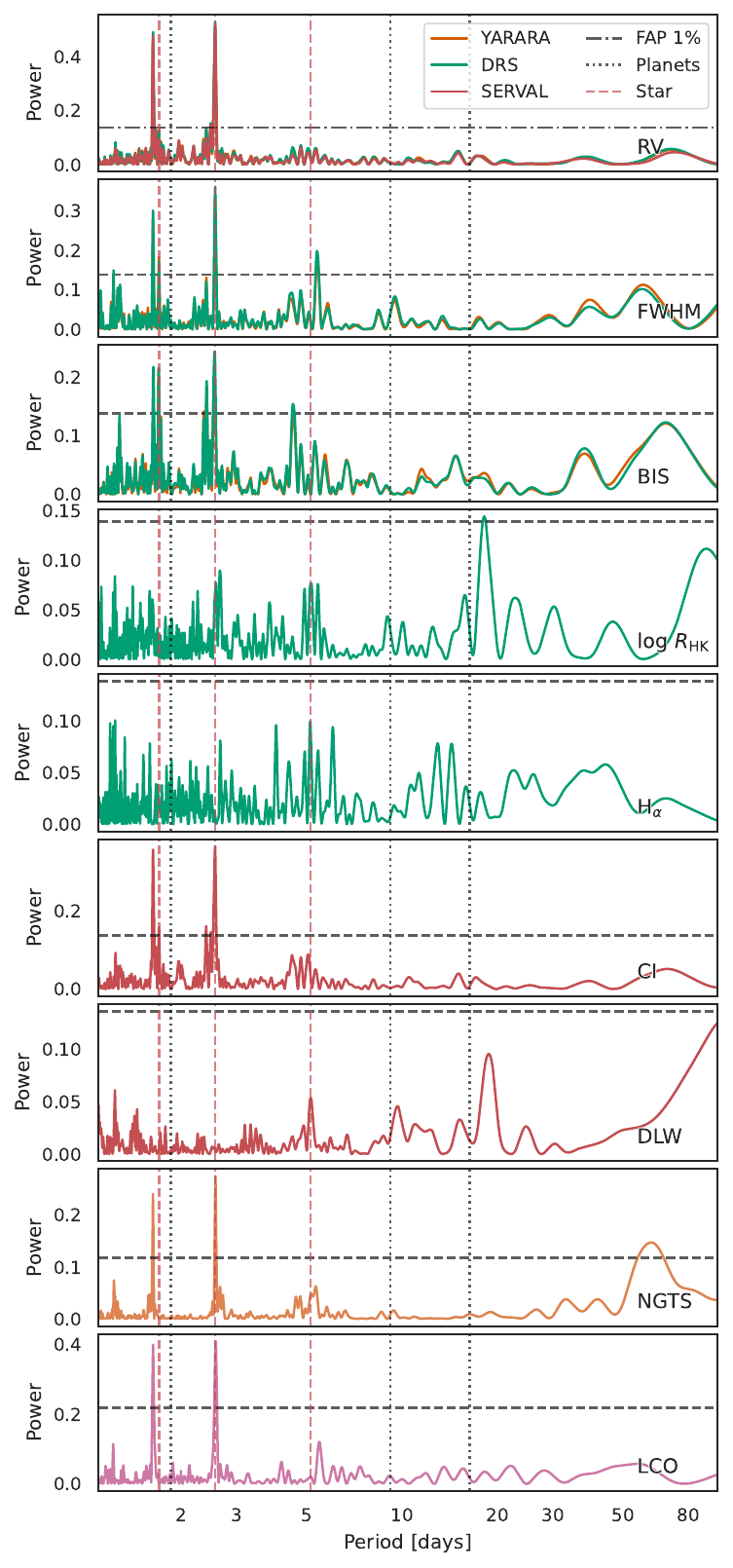}
\caption{
GLS periodograms of the stellar time-series.  The horizontal dashed line indicates the 1\% False Alarm Probability (FAP). The vertical red dashed lines represent the fundamental, first and second harmonics of the rotation period of the star. stellar-related signals, while the vertical black dotted lines mark the orbital periods of \targetbcd.}
\label{fig:periodiograms}
\end{figure}

As a first check to test the information contained in our stellar time-series, we ran a General Lomb-Scargle \citep[GLS;][]{GLS} periodogram on them. 
Figure~\ref{fig:periodiograms} shows the GLS periodogram of all the contemporaneous stellar time-series described previously. 
We can see that most of them peak around 2.5 and 1.7\,d, which correspond to the first and second harmonics of the stellar rotation period \citep[$P_{\rm rot}\sim 5.1$\,d, see][]{Newton2021,Barragan2021b}.
This suggests that the stellar signal contained in the time series has relatively high harmonic complexity.
We note that the chromospheric activity indicators (\logr\ and \halpha) and the \texttt{SERVAL} DLW do not have significant peaks.
This suggests that these time series are not dominated by rotational signals or other periodic phenomena.

We also searched for evidence of the planetary signals in the raw RV time series.
We found no substantial peaks corresponding to the orbital period of \targetbcd\ in any of the RV time series.
This is expected given the relatively large activity-induced variations in comparison with the expected small Doppler semi-amplitudes of the planets.

\subsubsection{Gaussian Processes}
\label{sec:gps1d}

We performed several analyses of the light curve and spectroscopic time-series to analyse and characterise the time scales over which the stellar signal evolves. 
Hereafter, we will only show the results obtained from the \serval\ time-series, DRS CCF activity indicators, two chromospheric indicators (\halpha\ and \logr) and LCO and NGTS light curves. 
We note that we performed similar analyses with the DRS and \yarara\ RV time-series, obtaining similar results. We choose to present the results of the \serval\ RVs because they have the smaller peak-to-peak variations (see Sect~\ref{sec:espressodata}).

We first perform 1D GP regressions on the different contemporaneous activity indicators and light curves. Given the quasi-periodic nature of the stellar signal, we describe the covariance between two data points at times $t_i$ and $t_j$ for each time-series using
\begin{equation}
    \gamma_{\rm 1D} = A^2 \gamma_{i,j},
\end{equation}
where $A$ represents the amplitude factor, and $\gamma_{i,j}$ is given by the Quasi-Periodic (QP) kernel
\begin{equation}
   \gamma_{{\rm QP},i,j} = \exp 
    \left[
    - \frac{\sin^2[\pi(t_i - t_j)/P_{\rm GP}]}{2 \lambda_{\rm P}^2}
    - \frac{(t_i - t_j)^2}{2 \lambda_{\rm e}^2}
    \right],
    \label{eq:gamma}
\end{equation}
where the hyperparameters are: $P_{\rm GP}$, which is the characteristic GP period; $\lambda_{\rm P}$, controlling harmonic complexity; and $\lambda_{\rm e}$, which is the evolution timescale.

For each run, we sample six parameters: the four kernel hyperparameters ($A$, $P_{\rm GP}$, $\lambda_{\rm e}$, $\lambda_{\rm P}$), one offset as the mean function, and an additional jitter term added to the diagonal of the covariance matrix. We use broad uniform priors for these parameters with the following ranges:
For $\lambda_{\rm e}$ between 0.1 and 100\,d (half the observational window); for $\lambda_{\rm P}$ between 0.1 and 5 (0.1 corresponds to high harmonic complexity and 5 to a quasi-coherent signal), and for $P_{\rm GP}$ between 4.8 and 5.3\,d (to include the expected stellar period from the \tess\ photometry).
The parameter space is explored using the MCMC sampler built into \pyaneti, following the same MCMC setup outlined in Sect.~\ref{sec:transitanalysis}.

\begin{figure*}
    \centering
    \includegraphics[width=\linewidth]{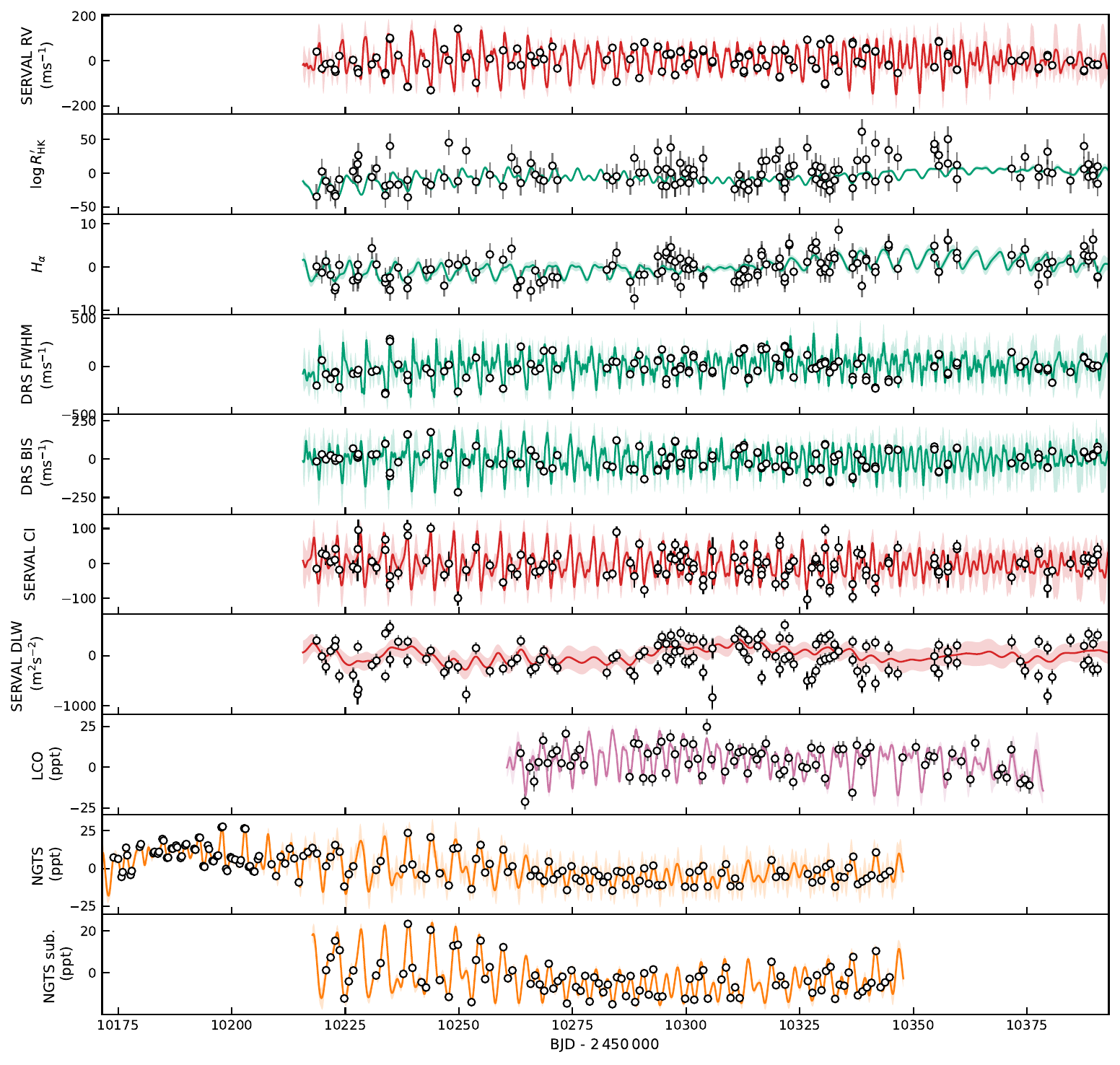}
    \caption{
    1D GP regression on each of the contemporaneous \target\ time-series.
    Measurements are shown as black circles
    with error bars, with semitransparent extension accounting for the inferred jitter. 
    Solid coloured lines show the 
    posterior predictive mean and 3-$\sigma$ credible interval of the inferred GP model.}
    \label{fig:timeseries1dgp}
\end{figure*}

\begin{figure*}
    \centering
    \includegraphics[width=\linewidth]{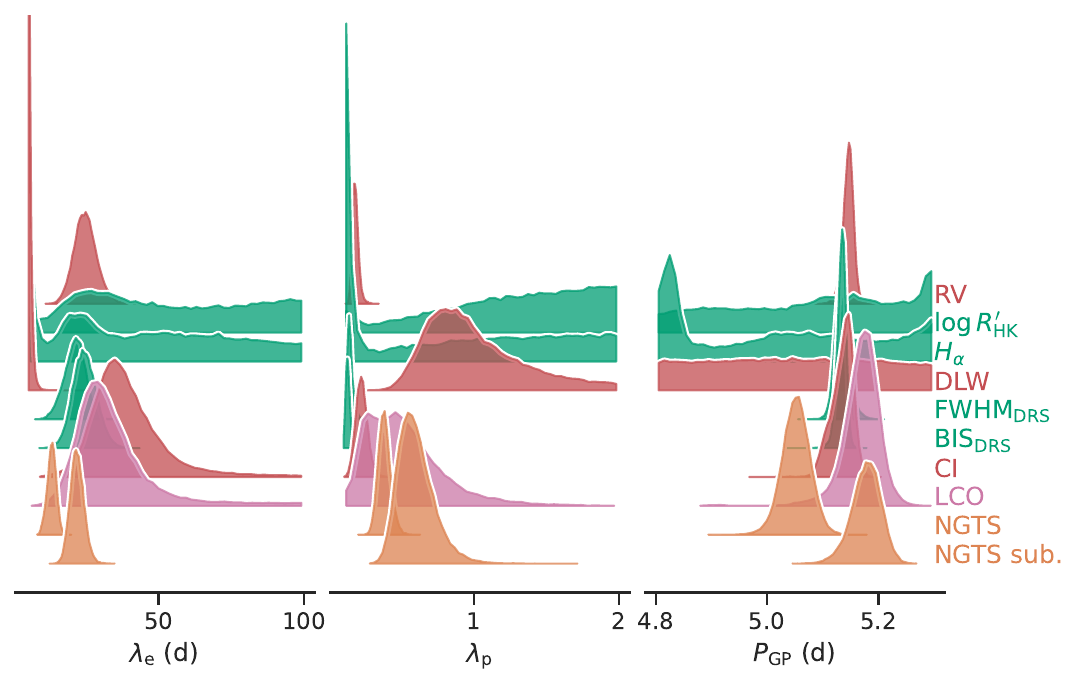}
    \caption{Posterior distributions for \pgp (left), \lbe\ (centre), and \lbp\ (right). Different time-series results are differentiated with a label and offset.}
    \label{fig:pos1dgp}
\end{figure*}

Table~\ref{tab:1gphp} shows the inferred \pgp, \lbe, and \lbp\ hyperparameters for all the stellar time-series. Figure~\ref{fig:timeseries1dgp} shows the \target\ stellar time-series with the inferred GP model. Figure~\ref{fig:pos1dgp} and Table~\ref{tab:1gphp} summarise the inferred posterior distributions for the \pgp, \lbe, and \lbp\ parameters for all the runs.

\begin{table}
\begin{center}
\caption{Recovered hyperparameters for 1D GP regression for different stellar time-series.  \label{tab:1gphp}} 
\begin{tabular}{lccccc}
\hline\hline
Time-series  & \pgp\ [d] & \lbe\ [d] & \lbp\   \\
\hline
RV       & $ 5.15_{-0.01}^{+0.01} $ & $ 24.6_{-3.7}^{+3.9} $ & $ 0.18_{-0.02}^{+0.02} $ \\
DRS FWHM & $ 5.13 \pm 0.01 $ & $ 21_{-3.8}^{+3.4} $ & $ 0.14_{-0.02}^{+0.02} $ \\
DRS BIS  & $ 5.13 \pm 0.01 $ & $ 24_{-3.9}^{+4.2} $ & $0.13_{-0.01}^{+0.02} $ \\
Serval contrast & $ 5.14_{-0.02}^{+0.01} $ & $ 37_{-8.3}^{+12} $ &  $ 0.22_{-0.04}^{+0.04} $ \\
LCO   &  $ 5.17 \pm 0.03 $ & $ 31_{-7.9}^{+12} $ & $ 0.48_{-0.21}^{+0.28} $ \\
NGTS  & $ 5.05 \pm 0.02 $ & $ 13_{-1.6}^{+1.5} $ &  $ 0.38_{-0.04}^{+0.04} $ \\
NGTS sub. & $ 5.10\pm 0.02 $ & $ 16 \pm 1.6 $ &  $ 0.38_{-0.04}^{+0.04} $ \\
\hline
\end{tabular}
\end{center}
\end{table}

We first discuss the results on the parameter \pgp, which is related to the rotation period of the star. 
The first thing to note is that the chromospheric activity indicators (\logr\ and \halpha) and the \serval\ DLW do not have a constrained solution for the period (as also indicated in the periodogram analysis). Their posterior distribution, as seen in Fig.~\ref{fig:pos1dgp}, is constrained purely from the uniform prior, indicating that the inferred values are not constrained by the data. Therefore we do not include their inferred parameters in Table~\ref{tab:1gphp}.
A possible reason for this is the relatively low signal-to-noise of each spectrum, resulting in fairly noisy estimates of these quantities measurements with relatively high noise. 
Such relatively high noise degrades our ability to recover patterns in the data \citep[see also discussion in][]{Barragan2023}.
The RV, CCF activity indicators, the light curves, and \serval\ CI provide a constrained period that peaks around 5.1\,d. All the posteriors are consistent with each other within 1-$\sigma$ except for the \ngts\ light curve (see Fig.~\ref{fig:pos1dgp}). The period recovered from the latter is 3-$\sigma$ away from that obtained from the other stellar time-series.
We note that, as shown in Fig.~\ref{fig:timeseries1dgp}, the \ngts\ data starts approximately 50\,d before the ESPRESSO observations, corresponding to $\sim 10$ rotations of the star, during which the active regions evolved significantly.
This could explain the lower \pgp\ and \lbe\ measured in the \ngts\ light curve.
To test this hypothesis, we modelled a subset of the \ngts\ data that starts at the same time as the ESPRESSO observations, which we will call \ngts\ sub. We ran the same 1D GP setup discussed in this section. Figures~\ref{fig:timeseries1dgp} and \ref{fig:pos1dgp} show that for \ngts\ sub., the recovered \pgp\ is fully consistent with the rest of the contemporaneous stellar time-series. 
From these results, we conclude that the RVs, the CCF activity indicators, the light curves, and the \serval\ CI all trace rotationally modulated features of the stellar surface.

We continue by analysing the parameter \lbe, which quantifies the coherence of the periodic signal and is related to the lifetime of the active regions \citep{Nicholson2022}.
The first thing to notice in Table~\ref{tab:1gphp} is that the value of \lbe\ is at least twice the value of \pgp\ for all the time-series with detected \pgp. This implies that the local coherence of the periodic signal is maintained for at least two stellar rotations.
On the other hand, Table~\ref{tab:1gphp} shows a wide range of values for \lbe\ for the different time-series. 
A possible explanation for this is that the different time series are sensitive to different types of surface features with different lifetimes \citep[as seen also for Sun-like stars, see][]{Klein2024}.
\citet{Barragan2023} also noticed that relatively low signal-to-noise time-series can give larger \lbe\ values. 
This is because low signal to noise time-series are less sensitive to small evolutionary changes in the signals, and larger \lbe\ values are favoured by the GP's built-in Occam's razor\footnote{Larger values of \lbe\ reduce the determinant of the covariance matrix $|\mathbf{K}|$, and the GP likelihood includes $|\mathbf{K}|^{-1}$.}.

Finally, we examine the complexity of the inferred stellar signal, as quantified by the parameter \lbp, which can be understood as the length scale of the periodic signal.
It is expected that \emph{RV-like} time-series (i.e., those that depends on the change of location of the active regions from the red- to the blue-shifted stellar hemisphere, and vice-versa, like the RV or BIS) should have a higher harmonic complexity than the \emph{photometry-like}  (i.e., those which are good tracers of the areas covered by active regions on the stellar surface) in 1D GP regressions \citep[as expected from the $FF'$ framework of \citet{Aigrain2012}, see also][]{pyaneti2,Barragan2023}.
We would thus expect that the RV and BIS would yield a relatively small \lbp\ value compared to the \emph{photometry-like} time-series.
This relation is not observed for all the time-series in Table~\ref{tab:1gphp} and Figure~\ref{fig:pos1dgp}.
As a comparison point, we can take the \lbp\ value obtained from the RV time series.
Most notoriously, FWHM gives a significantly smaller \lbp than the RVs.
\citet{Barragan2024} presented a similar case of unexpected \lbp\ values for FWHM, and argued that this is due to overfitting.
These kinds of results indicate an overfit of the data where the white noise is fitted as fast evolving red noise (behaviour caused by small values of \lbp). 
The risk of overfitting is compounded in the present case by the high harmonic complexity of the signal (already noted from the GLS periodogram analysis in Sect.~\ref{sec:periodograms}), by the relatively sparse separated data compared to \target's short rotation period and by the rapid evolution of the signal \citep[][]{Barragan2021b} compared to the gaps in the observations.
These results hint that the activity indicators extracted from the ESPRESSO observations may be of limited use to characterise the stellar signal and separate it from the planetary signals, because all these time series suffer from the same, relatively poor time-sampling.
From Figure~\ref{fig:timeseries1dgp} we can see how the FWHM, BIS, and \serval\ CI predictive distributions evolve relatively fast compared to the gaps between consecutive points. 
This behaviour is characteristic of overfitting \citep[see][]{Barragan2023,Blunt2023}.

The \ngts\ and LCO light curves give \lbp\ values that are consistently larger than those obtained from the RVs, but this is what we expect from the $FF'$ framework. 
This is because they have a better sampling than the spectroscopic time series. 
The  \lbp\ values obtained from these light curves are more reliable, as they have better time-sampling than the spectroscopic time-series.
From Figure~\ref{fig:timeseries1dgp} we can see that the light curves evolve more smoothly, similarly to the \tess\ light curves from previous seasons \citep{Barragan2021b}. 

From this detailed 1D GP analysis, we can conclude that the stellar signal seems better characterised by the light curves than the ESPRESSO time-series. 
The most important variables that determine the ability to characterise the signal are the signal-to-noise ratio and the time sampling. 
The latter must be tight compared to the evolutionary time scales of the signal, both periodic and long-term \citep[see discussion in][]{Barragan2024}. These two criteria are better kept in our light curves.
In the next section, we explore how different combinations of these time series affect our ability to recover the planetary signals.

\subsection{RV planetary analysis}
\label{sec:rvanalysis}

We now proceed to run diverse multi-GP models where we include the signal of the three transiting planets in the RV time-series and model the stellar signal of each time series as a linear combination of a single latent GP and its time derivative \citep[see][for more details]{Rajpaul2015,pyaneti2}.
For all our runs, we use the QP kernel shown in Sect.~\ref{sec:stellarsignal} and its corresponding derivatives \citep[][]{Rajpaul2015,pyaneti2}.
We modified the \pyaneti\ version presented in \citet{pyaneti} to allow for modelling of contemporaneous time-series with different time stamps and number of observations.

We create $N-$dimensional GP models, including $N$ time-series $\mathcal{A}_i$,  as
\begin{equation}
\begin{matrix}
 \mathcal{A}_1 =  A_{1} G(t) + B_{1} \dot{G}(t) \\
\vdots \\
\mathcal{A}_N =  A_{N} G(t) + B_{N} \dot{G}(t), \\
\end{matrix}
\label{eq:gps}
\end{equation}
\noindent
where the parameters $A_{1}$, $B_{1}$, $\dots$, $A_{N}$, $B_{N}$ serve as free coefficients that connect each time-series to both $G(t)$ and its time derivative, $\dot{G}(t)$. In this framework, $G(t)$ represents an unobserved (latent) variable, which can be conceptually understood as the time-dependent coverage of active regions across the projected visible stellar disc.

We model the RV and BIS as \emph{RV-like} time series that are described by $G(t)$ and $\dot{G}(t)$, while the rest of the time series are assumed \emph{photometry-like}, i.e., modelled only by $G(t)$.

The mean function for the RV time-series consists of the planetary signals. 
We use 3 Keplerian signals, where every Keplerian signal $i$ is parametrised by a time of minimum conjunction (equivalent to the time of mid-transit for transiting planets), $T_{0,i}$; orbital period, $P_{{\rm orb},i}$; orbital eccentricity, $e_i$; angle of periastron, $\omega_{\star,i}$; and Doppler semi-amplitude, $K_i$. 
We also include one offset and a jitter term to penalise the imperfections of our model. 

We first ran 1D GP models on the RV time-series only, following the same setup as described in Sect.~\ref{sec:stellarsignal}. We ran two different flavours of RV-only model: one with uniform priors for all the GP hyper-parameters, and one with a GP trained with the \ngts\ light curve (i.e., using the values of Table~\ref{tab:1gphp} for NGTS sub.).
We also performed multi-GP regressions using different combinations of RV and activity indicators and/or light curves.
We used the same MCMC setup as the one described in Sect.~\ref{sec:transitanalysis}.
We set Gaussian priors on the planet ephemerides based on the results obtained in Sect.~\ref{sec:transitanalysis}. We set uniform priors for the reminder parameters. We also assumed circular orbits for the three planets. 
For all the GP hyperparameters, we set uniform priors with the same ranges described in Sect.~\ref{sec:stellarsignal} (except for the GP-trained case in which we set Gaussian priors).

\begin{figure*}
    \centering
    \includegraphics[width=\linewidth]{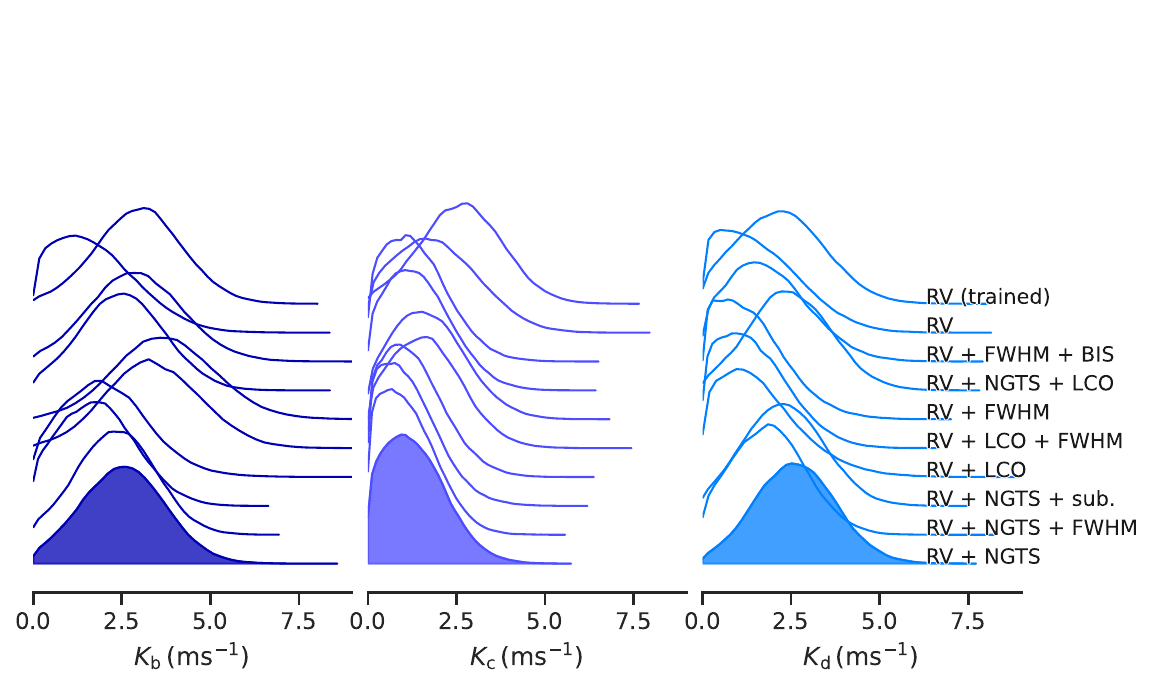}
    \caption{Posterior distributions for the Doppler semi-amplitudes for \targetbcd. Each posterior distribution corresponds to a different model labelled to the right. We highlight the adopted distributions (RV + NGTS) in this work with a coloured filled distribution.  }
    \label{fig:poskas}
\end{figure*}

Figure~\ref{fig:poskas} shows the posterior distributions of the Doppler semi-amplitudes of \targetbcd\ for all the multi-GP runs.
Interestingly, the approach employing the 1D GP training delivers the most precise outcomes alongside 3-$\sigma$ detections. 
Although this result might appear preferable, we need to cautious, as such detections may be caused by overfitting, by forcing covariances that are not representative of the intrinsic covariance in the RV. 
By visually inspecting the inferred posterior distributions, we cannot distinguish which one gives the better value or the most precise. All inferred posteriors give distributions that agree well with each other.
Following the suggestions of \citet{Rajpaul2024}, the present analysis shows that diverse activity models produce consistent results, suggesting that the true Doppler semi-amplitudes reside within the inferred credible intervals.

We conclude that it is not possible to obviously distinguish which method provides the optimal model for adoption in this work. 
To differentiate between methods and identify the best model would require a quantitative comparison of the various multi-GP models, which lies beyond the scope of this manuscript.
To select the final model to use in this manuscript, we refer to our results from the one-dimensional GP regressions of the stellar time series (Sect.~\ref{sec:gps1d}). We concluded that the ESPRESSO spectroscopic activity indicators fail to constrain the stellar activity signal robustly by themselves. 
Meanwhile, the NGTS/LCO light curves successfully constrain the timescales of the stellar signal by themselves. 
Therefore, we adopt the two-dimensional GP regression of RV + NGTS as our final model. We select the NGTS over the LCO observations because they cover a larger time span.

\subsection{Final joint model}
\label{sec:final}

For completeness, we ran a final joint model of transit photometry and RV to characterise \target's planetary signals.
Based on the analyses presented in this section, our final model for \target\ is the transit model described in Section~\ref{sec:transitanalysis}, together with the RV-\ngts\ multi-GP model described in Sect.~\ref{sec:rvanalysis}.
The whole set of sampled parameters and priors are shown in Table~\ref{tab:pars}.

We also re-run the same model, but this time allowing for eccentric orbits for the three planets. We obtained Doppler semi-amplitudes consistent with the circular orbit case with $k_{\rm b} = 2.7 \pm 1.2$\ms,  $k_{\rm c} = 1.4_{-0.9}^{+1.2}$\ms, and $k_{\rm d} = 2.9 \pm 1.4$\ms, with corresponding eccentricities of $e_{\rm b} = 0.15_{-0.10}^{+0.16}$, $e_{\rm b} = 0.21_{-0.16}^{+0.28}$, and $e_{\rm d} = 0.09_{-0.07}^{+0.13}$. These eccentricities cannot be distinguished from circular orbits. This result is supported by a Akaike Information Criterion difference of 13 on favour of the circular orbits model. We therefore conclude that our dataset does not allow us to set constrains on the eccentric orbits of the planets and we adopt the model with circular orbits.

\begin{figure*}
    \centering
    \includegraphics[width=\textwidth]{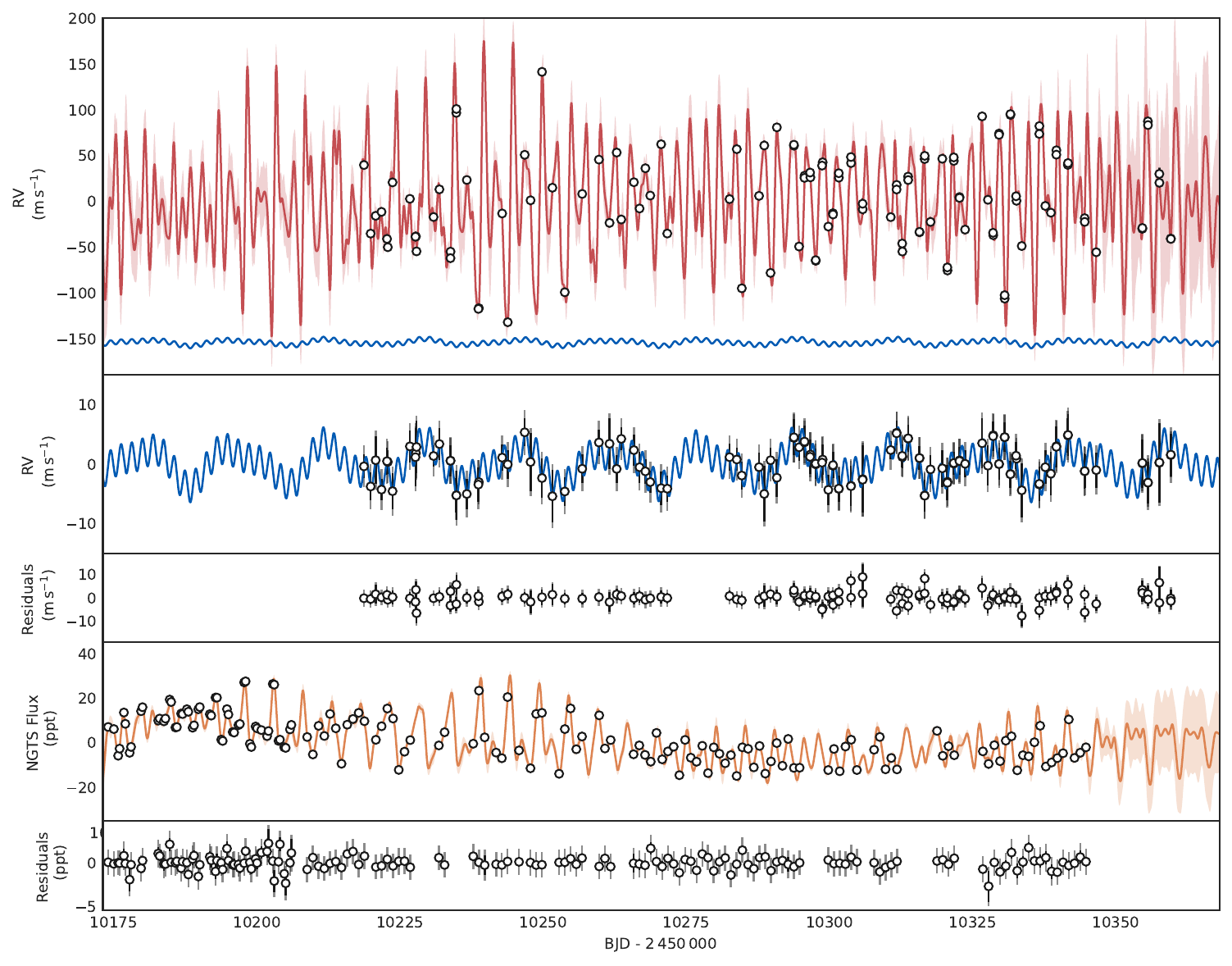}\\
    \includegraphics[width=\textwidth]{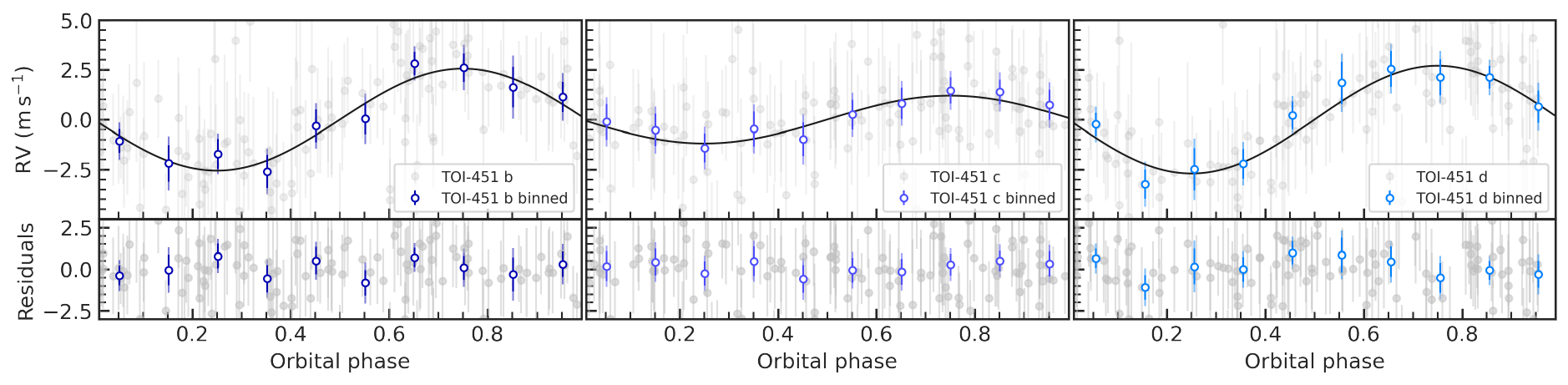}\\
    \caption{
    \emph{Top:}
    \target's RV and \ngts\ time-series after being corrected by inferred offsets. 
    The plot shows (from top to bottom) RV data together with full and planetary signal inferred models; RV data with stellar signal model subtracted; RV residuals; \ngts\ data together with inferred stellar model, and
    \ngts\ residuals. The solid lines show the inferred full model coming from our multi-GP, light-shaded areas showing the corresponding GP model’s 3-$\sigma$ credible intervals. 
    \emph{Bottom:}
    Phase-folded RV signals for \targetbcd\ following the subtraction of the systemic velocities, stellar signal, and other planets. Nominal RV observations are shown as light grey points. Solid colourful points show binned data to 1/10 of the orbital phase.
    }
    \label{fig:timeseries}
\end{figure*}

Figure~\ref{fig:timeseries} shows the RV and \ngts\ time-series, together with the phase-folded planetary Doppler signal.
Table~\ref{tab:pars} shows the inferred sampled parameters, defined as the median and 68.3\% credible interval of the posterior distribution.
Table~\ref{tab:derived} also shows the derived planetary and orbital parameters.

\begin{table*}
\begin{center}
  \caption{Sampled parameters and priors for final modelling.\label{tab:pars}}  
  \begin{tabular}{lccc}
  \hline
  \hline
  \noalign{\smallskip}
  Parameter & Prior$^{(a)}$ & Final value$^{(b)}$ & Maximum value$^{(c)}$  \\
  \noalign{\smallskip}
  \hline
  \noalign{\smallskip}
  \multicolumn{4}{l}{\emph{\bf \targetb's parameters }} \\
  \noalign{\smallskip}
    Orbital period $P_{\mathrm{orb}}$ (days)  & $\mathcal{U}[1.8586,1.8589]$ &\pb[] & $\cdots$ \\
    Transit epoch $T_0$ (BJD$_\mathrm{TDB}-$2\,450\,000)  & $\mathcal{U}[10312.3110,10312.5110]$ & \tzerob[] & $\cdots$ \\  
    Scaled planet radius $R_\mathrm{p}/R_{\star}$  &$\mathcal{U}[0.0,0.05]$ & \rprstarb[] & $\cdots$ \\
    Impact parameter, $b$ & $\mathcal{U}[0,1]$ & \bb[] & $\cdots$ \\
    Orbital eccentricity, $e$ & $\mathcal{F}[0]$ & 0 & $\cdots$ \\
    Angle of periastron, $\omega_\star$ (deg) & $\mathcal{F}[90]$ & $90$ & $\cdots$ \\
    Doppler semi-amplitude variation $K$ (m s$^{-1}$) & $\mathcal{U}[0,50]$ & \kb[] & 5.3  \\
  \multicolumn{4}{l}{\emph{\bf \targetc's parameters }} \\
    Orbital period $P_{\mathrm{orb}}$ (days)  & $\mathcal{U}[9.1920,9.1930]$ &\pc[] & $\cdots$ \\
    Transit epoch $T_0$ (BJD$_\mathrm{TDB}-$2\,450\,000)  & $\mathcal{U}[10314.5279,10314.7279]$ & \tzeroc[] & $\cdots$ \\  
    Scaled planet radius $R_\mathrm{p}/R_{\star}$  &$\mathcal{U}[0.0,0.05]$ & \rprstarc[] & $\cdots$ \\
    Impact parameter, $b$ & $\mathcal{U}[0,1]$ & \bc[] & $\cdots$ \\
    Orbital eccentricity, $e$ & $\mathcal{F}[0]$ & 0 & $\cdots$ \\
    Angle of periastron, $\omega_\star$ (deg) & $\mathcal{F}[90]$ & $90$ & $\cdots$ \\
    Doppler semi-amplitude variation $K$ (m s$^{-1}$) & $\mathcal{U}[0,50]$ & \kc[] & 3.7 \\
  \multicolumn{4}{l}{\emph{\bf \targetd's parameters }} \\
    Orbital period $P_{\mathrm{orb}}$ (days)  & $\mathcal{U}[16.3648,16.3651]$ &\pd[] & $\cdots$ \\
    Transit epoch $T_0$ (BJD$_\mathrm{TDB}-$2\,450\,000)  & $\mathcal{U}[10314.8635,10315.0635]$ & \tzerod[] & $\cdots$ \\  
    Scaled planet radius $R_\mathrm{p}/R_{\star}$  &$\mathcal{U}[0.0,0.05]$ & \rprstard[] & $\cdots$ \\
    Impact parameter, $b$ & $\mathcal{U}[0,1]$ & \bd[] & $\cdots$ \\
    Orbital eccentricity, $e$ & $\mathcal{F}[0]$ & 0 & $\cdots$ \\
    Angle of periastron, $\omega_\star$ (deg) & $\mathcal{F}[90]$ & $90$ & $\cdots$ \\
    Doppler semi-amplitude variation $K$ (m s$^{-1}$) & $\mathcal{U}[0,50]$ & \kd[] & 5.5 \\
    \multicolumn{4}{l}{\emph{ \bf GP hyperparameters}} \\
   GP Period $P_{\rm GP}$ (days) &  $\mathcal{U}[4.9,5.3]$ & \PGP[] & $\cdots$ \\
    $\lambda_{\rm p}$ &  $\mathcal{U}[0.1,5]$ &  \lambdap[] & $\cdots$ \\
    $\lambda_{\rm e}$ (days) &  $\mathcal{U}[5,100]$ &  \lambdae[] & $\cdots$ \\
    $A_{\rm RV}$ (\ms)  &  $\mathcal{U}[0,100]$ & \Azero & $\cdots$ \\
    $B_{\rm RV}$ (\ms\,d$^{-1}$) &  $\mathcal{U}[-100,100]$ & \Aone & $\cdots$ \\
    $A_{\rm NGTS}$ (ppm) &  $\mathcal{U}[-100,100]$ & \Atwo & $\cdots$ \\
    $B_{\rm NGTS}$ (ppm\,d$^{-1}$) &  $\mathcal{F}[0]$ & 0 & $\cdots$ \\
    \multicolumn{4}{l}{\emph{ \bf Other parameters}} \\
    Stellar density $\rho_\star$ (\gcm) & $\mathcal{U}[0.1,5]$ & \rhostarb[] & $\cdots$ \\ 
    \tess\ limb-darkening coefficient $q_1$  &$\mathcal{U}[0,1]$ & \qonetess & $\cdots$ \\ 
    \tess\ limb-darkening coefficient $q_2$  &$\mathcal{U}[0,1]$ & \qtwotess & $\cdots$ \\ 
    \ngts\  limb-darkening coefficient $q_1$  &$\mathcal{U}[0,1]$ & \qonengts & $\cdots$ \\ 
    \ngts\ limb-darkening coefficient $q_2$  &$\mathcal{U}[0,1]$ & \qtwongts & $\cdots$ \\ 
    Offset RV (\kms) & $\mathcal{U}[ -1, 1]$ & \rvserval[] & $\cdots$ \\
    Offset \ngts\ & $\mathcal{U}[ 0.9, 1.1]$ & \ngtsoffset[] & $\cdots$ \\
    Jitter term $\sigma_{\rm RV}$ (\ms) & $\mathcal{J}[1,100]$ & \rvjitterrvserval[] & $\cdots$ \\
    Jitter term $\sigma_{\rm NGTS}$ (ppm) & $\mathcal{J}[1,100]$ & \rvjitterngts[] & $\cdots$ \\
    \noalign{\smallskip}
    \hline
\multicolumn{4}{l}{\footnotesize $^a$ $\mathcal{F}[a]$ refers to a fixed value $a$, $\mathcal{U}[a,b]$ to an uniform prior between $a$ and $b$, and $\mathcal{J}[a,b]$ to the modified Jeffrey's prior as  defined by}\\
\multicolumn{4}{l}{\footnotesize  \citet[eq.~16]{Gregory2005}.$^b$ Inferred parameters and errors are defined as the median and 68.3\% credible interval of the posterior distribution.} \\
\multicolumn{4}{l}{$^c$ Maximum values reported as the 99\% upper credible limit.}\\
  \end{tabular}
\end{center}
\end{table*}

\begin{table*}
\begin{center}
  \caption{Derived parameters for the \target\ planets. \label{tab:derived}}
  \begin{tabular}{lccc}
  \hline
  \hline
  Parameter & \targetb's  & \targetc's & \targetd \\
   & values & values & values \\
  \hline
  \noalign{\smallskip}
    Planet mass $M_\mathrm{p}$ ($M_{\rm \oplus}$) &  \mpb[]   &  \mpc[]  &  \mpd[] \\
    Maximum$^{(d)}$ planet mass ($M_{\rm \oplus}$) &  9.7  &  11.5  &  20.7 \\
    Planet radius $R_\mathrm{p}$ ($R_{\rm \oplus}$) &  \rpb[] &  \rpc[] & \rpd[] \\
    Planet density $\rho_{\rm p}$ (g\,cm$^{-3}$) &  \pdenb[] &  \pdenc[] & \pdend[] \\
    Scaled semi-major axis  $a/R_\star$ &  \arstarb[]  &  \arstarc[]  & \arstard[] \\
    Semi-major axis  $a$ (AU) &  \ab[] &  \ac[] & \ad[] \\
    Orbit inclination $i_\mathrm{p}$ ($^{\circ}$) &  \idegb[] &  \idegc[] & \idegd[] \\
    Transit duration $\tau_{14}$ (hours) & \trtb[]  & \trtc[] & \trtd[] \\
    Planet surface gravity $g_{\rm p}$ (${\rm cm\,s^{-2}}$)$^{(b)}$ & \pgrab[] & \pgrac[] & \pgrad[] \\
    Planet surface gravity $g_{\rm p}$ (${\rm cm\,s^{-2}}$)$^{(c)}$ & \pgratwob[] & \pgratwoc[] & \pgratwod[] \\
    Equilibrium temperature  $T_\mathrm{eq}$ (K)$^{(d)}$  &   \Teqb[] &   \Teqc[] & \Teqd[] \\
    Received irradiance ($F_\oplus$) & \Fpb[] & \Fpc[] & \Fpd[] \\
    TSM$^{(e)}$ & \TSMb[] & \TSMc[] & \TSMd[] \\
   \noalign{\smallskip}
  \hline
  \multicolumn{4}{l}{$^a$ Maximum values reported as the 99\% upper credible limit. $^b$ Derived using $g_{\rm p} = G M_{\rm p} R_{\rm p}^{-2}$.}\\
  \multicolumn{4}{l}{$^c$ Derived using sampled parameters following \citet{Sotuhworth2007}. $^d$ Assuming a zero albedo.}\\
  \multicolumn{4}{l}{$^e$ Transmission spectroscopy metric (TSM) by \citet{Kempton2018}.}\\
  \end{tabular}
\end{center}
\end{table*}


\section{Discussion}
\label{sec:discussion}

\subsection{Injection tests}
\label{sec:injectiontests}

It has been shown that the combination of complex models and the window function of the observations can create spurious planet-like signals in RV time-series, specially for active stars \citep[e.g.,][]{Rajpaul2016}. 
In order to check the reliability of our RV detection, we performed numerical simulations similar to \citet{Barragan2019} and \citet{Zicher2022}, and suggested by \citet{Rajpaul2024}. 

We use \citlalatonac\ \citep{pyaneti2} to simulate synthetic RV and NGTS time-series.
We  utilise the median predictive distribution obtained from the multi-GP model for each of this time series as the base of our stellar signal.
We then added correlated noise, to mimic instrumental systematics, as a GP generated by a squared exponential kernel with a length-scale of one day, and the same amplitude as the jitter term obtained from the real data for each time series.
We then use the same time stamps from the real observations to create synthetic data, and added white noise for each synthetic datum according to the nominal measurement uncertainty .
We did this 100 times to obtain 100 simulated activity-only RV and NGTS time-series, with similar noise properties and same time-sampling as the real data.

\begin{figure*}
    \centering
    \includegraphics[width=0.98\linewidth]{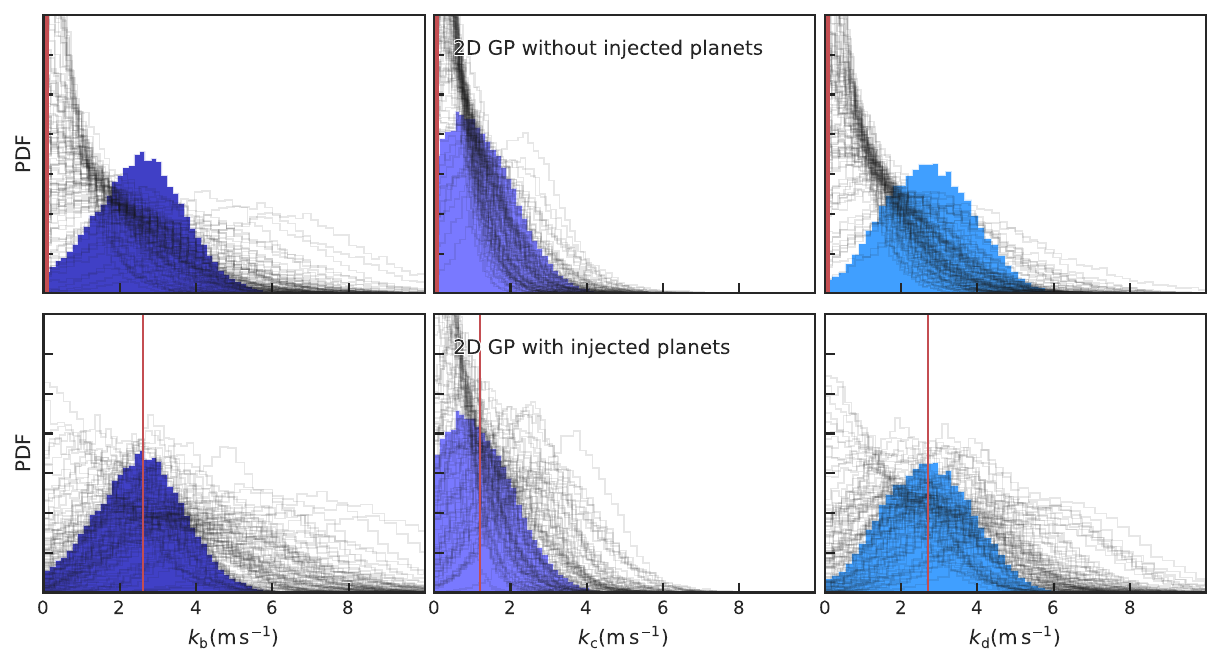}
    \caption{Results of the injection-recovery tests. 
The top panel shows the recovered posteriors for \targetbcd\ obtained from the activity-only time series. 
The bottom panel displays the posteriors derived from the time series that include both stellar activity and injected planetary signals, modelled with a 2D GP. 
In each panel, the filled histograms correspond to the posteriors from the original dataset, while the thin grey lines represent the posteriors obtained from the simulated datasets. Vertical lines show the injected Doppler semi-amplitude for each case. }
    \label{fig:injectiontests}
\end{figure*}

We first explore the question: could we have detected the planetary signals if there were no signals?  We modelled each activity-only RV synthetic dataset (i.e., with no planetary signals injected) using a three-planet and 2-dimensional GP configuration as described in Sect.~\ref{sec:rvanalysis} with RV and NGTS-like model.
For each simulation, we plot the posterior over the semi-amplitudes for the three `planets' in the top column of Figure~\ref{fig:injectiontests}, compared to the posterior obtained from the real dataset. We then count the fraction of the simulations where we would have claimed 2-$\sigma$ detection for the three planets.
For planet b, this occurred in 5\% of the times, for planet c in 1\% of the cases, and for planet d in 1\% of the cases. 
This suggests that 2-$\sigma$ estimates are unlikely to appear if the Doppler semi-amplitudes were zero. This is expected because we know a priori that \targetbcd\ exist, and we expect that there are Doppler signals larger than zero that is consistent with the ephemeris of the transiting signal.

We then repeat the same exercise, using the same 100 synthetic datasets, but this time, injecting three Keplerian signals with the median planet parameters reported in Table~\ref{tab:pars}. 
Again, we modelled these synthetic datasets using \pyaneti\ with the same configuration as described in Sect.~\ref{sec:rvanalysis}. The resulting posteriors are shown in the bottom row of Figure~\ref{fig:injectiontests}. 
First we ask the question, how often are \targetbcd\ detected at 2-$\sigma$ for a given confidence level from these simulations? We find that \targetbcd\ are detected 46\%, 15\%, and 43\% of the time, respectively. 
We then ask the question, what fraction of the time the injected semi-amplitudes are within 2-$\sigma$ of the recovered posterior distributions? We find that this happens on 96\%, 98\%, and 99\% of the times for \targetbcd, respectively.
For Gaussian distributions we would expect this to be close to 95\%, making these  results consistent with expectations.
These injection tests suggest that the detected signals are unlikely to be entirely spurious. Nevertheless, our sensitivity to the planetary signatures appears to be significantly affected by stochastic stellar variability and instrumental white noise. Consequently, these results should be interpreted with caution.

\subsection{Planets properties and mass-loss perspectives}

In this section we explore the implications of the observed properties of the \target\ planets in the context of composition and atmospheric mass loss.
Tables~\ref{tab:pars} and \ref{tab:derived} present the inferred planetary and orbital parameters that we will use to discuss the nature of the \target\ worlds. 
It is important to note that we did not detect any of the three transiting planets in the RVs with a significance $>$ 3-$\sigma$. 
We report the maximum Doppler semi-amplitudes and masses for the three planets in Tables~\ref{tab:pars} and \ref{tab:derived}.
The masses of planets \targetb\ and d are $M_{\rm b}=$\mpb\ and $M_{\rm d}=$\mpd\ (2-$\sigma$ confidence). \targetc\  posterior distribution is less constrained and has a maximum mass of $11.5$\,\mearth. 
For this reason, we will only explore the possible properties for \targetb\ and d. 
We will base our conclusions on the median values, but we are aware that the parameters credible intervals allow wider possibilities.

\begin{figure*}
    \includegraphics[width=1\textwidth]{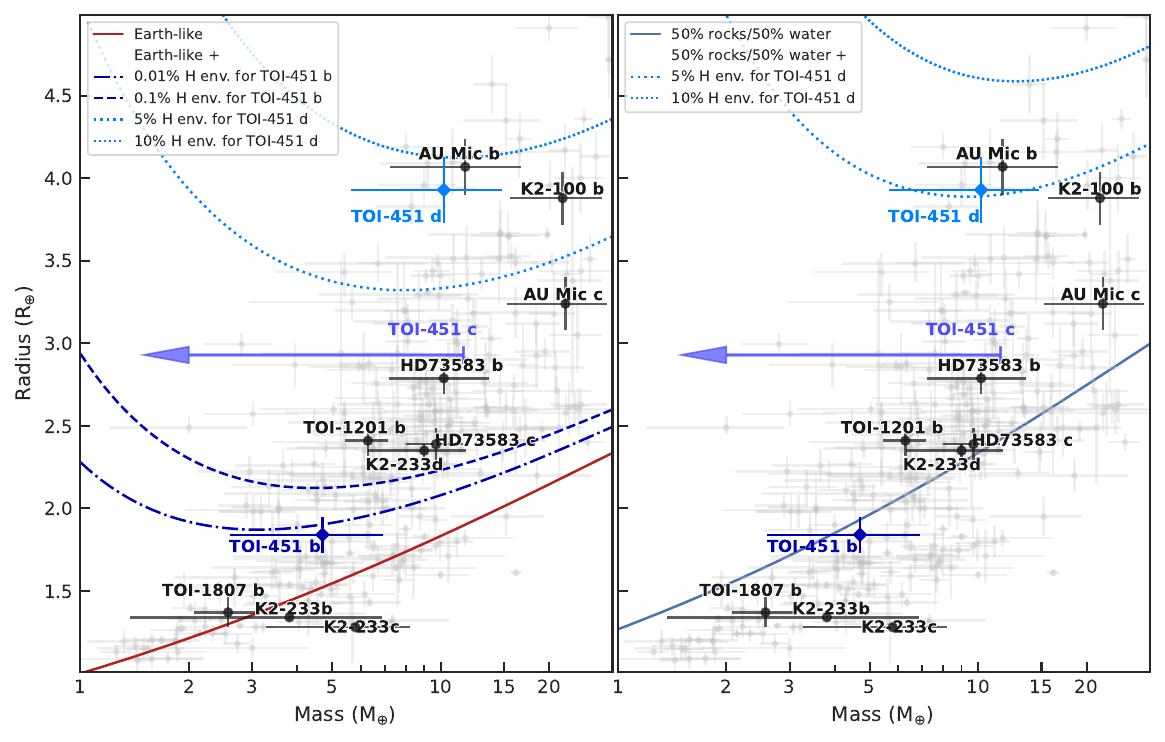}
    \caption{Mass \emph{vs} radius diagram for small exoplanets ($1 < R_{\rm p} < 5\, R_\oplus$ and $1 < M_{\rm p} < 32\, M_\oplus$). Grey points with error bars show planets with mass and radius measurements better than 50\% \citep[As in the NASA Exoplanet Archive on Dec 17, 2025 \url{https://exoplanetarchive.ipac.caltech.edu/}, ][]{NASAexoplanet}. 
    Black points and labels refer to young exoplanets ($< 1$\,Gyr) with mass and radius measurements. 
    \targetb\ and d are shown with colourful diamonds and properly labelled.
    An arrow denotes the 99\% upper credible limit for \targetc, illustrating the extent of possible values.
    Solid lines represent two-layer core models as given by \citet{Zeng2016}.  
    Non-solid lines correspond to cores surrounded by a hydrogen envelope properly labelled in each inset.
    This plot was created using the same code used to create the mass-radius diagram in \citet[][]{Barragan2018b}.
    }
    \label{fig:mr}
\end{figure*}

Figure~\ref{fig:mr} shows mass-radius diagrams for small exoplanets ($1 < R_{\rm p} < 5\, R_\oplus$ and $1 < M_{\rm p} < 32 M_\oplus$).
Planets \targetb\ and d are shown with relative composition models. 
\targetc\ is shown with an arrow to indicate the 99\% upper credible limit.
The plot also highlights other young exoplanets with reported masses and radii.
The models assume planets with a Earth-like (1/3 iron and 2/3 silicates) and water-rich (50\% rocky and 50\% water) cores.  We also show the hydrogen envelope mass fraction models for both core composition cases.
To compute the solid core composition curves for rocky and water-rich planets, we adopted the theoretical mass-radius relations by \citet{Zeng2016} for an Earth-like rocky core and for a water-rich core.
To compute the hydrogen composition curves for each planet, we assumed a two layer model with a gaseous envelope on top of a solid core. 
For the rocky core, we again adopted the models by \citet{Zeng2016} described above. For the gaseous envelope, we adopted the envelope structure model by \citet{Chen2016EvolutionaryMESA}, based on \texttt{MESA} simulations and for which they provide a polynomial fit valid for planet ages over 100\,Myr and envelope mass fractions under 20\%.

The first thing to note from Fig.~\ref{fig:mr} is that there is an apparent dichotomy of young exoplanets that coincides with the radius gap \citep{Fulton2017,VanEylen2018}.
\targetb\ is the only young exoplanet that lies within it, making it an excellent target to test mass-loss theories.
If we assume that \targetb\ has an Earth-like core surrounded by a hydrogen envelope, the planet would need an hydrogen envelope of 0.01\% of mass to explain its radius. Given the age of \target, one would expect that the star has just past the peak of its magnetic activity and the most intense phase of photo-evaporation could have already passed \citep[][]{owenwu2017}. However, atmospheric mass-loss could still be occurring by photo-evaporation until roughly 1\,Gyr \citep[][]{Rogers2021,King2021} and/or core-powered mass loss \citep[][]{Ginzburg2016} at longer time scales. This would make of \targetb\ a planet that is on the verge of losing what remains of its atmosphere and move to the rocky side of the radius valley.
The inferred characteristics of \targetb\ suggest that the planet could still host a small hydrogen envelope.
Therefore this planet is an excellent test to search for signatures of ongoing evaporation, for example by searching for escaping Helium in transmission spectroscopy \citep[e.g.,][]{Zhang2022}.
However, \targetb's nature could be different if we assume a different core composition. If \targetb\ instead has a water-rich core, then the planet is consistent with being a solid world with no need of an envelope (see Fig.~\ref{fig:mr}).
This scenario would imply that photo-evaporation has fully removed any initial envelope within the first 125\,Myr.

Because of their radii, \targetc\ and d are expected to be planets with a volatile-rich envelope rather than being solid \citep[as suggested by previous works e.g.,][]{Fulton2017,VanEylen2018}.
Depending on the assumed core composition for \targetd, its radius can be explained by different fractions of hydrogen envelopes. Ranging from 8\% by mass for an Earth-like core, to 5\% by mass for a water rich core.
Because of its youth, extended atmosphere, and host star brightness, \targetd\ is an excellent candidate to perform transmission spectroscopy. 
We note that \targetd\ has a Transmission spectroscopic metric (TSM) of \TSMd, that is consistent with the threshold of 90 suggested by \citet{Kempton2018}. This makes \targetd\ a highly valuable target for the \jwst. Furthermore, the mass provided in this work will be useful to interpret atmospheric observations \citep{Batalha2019}.
We note that observations of \targetc\ and d are scheduled to be observed with the \jwst\ \citep[GO 5959;][]{Feinstein2024}.

\begin{figure*}
    \includegraphics[width=1\textwidth]{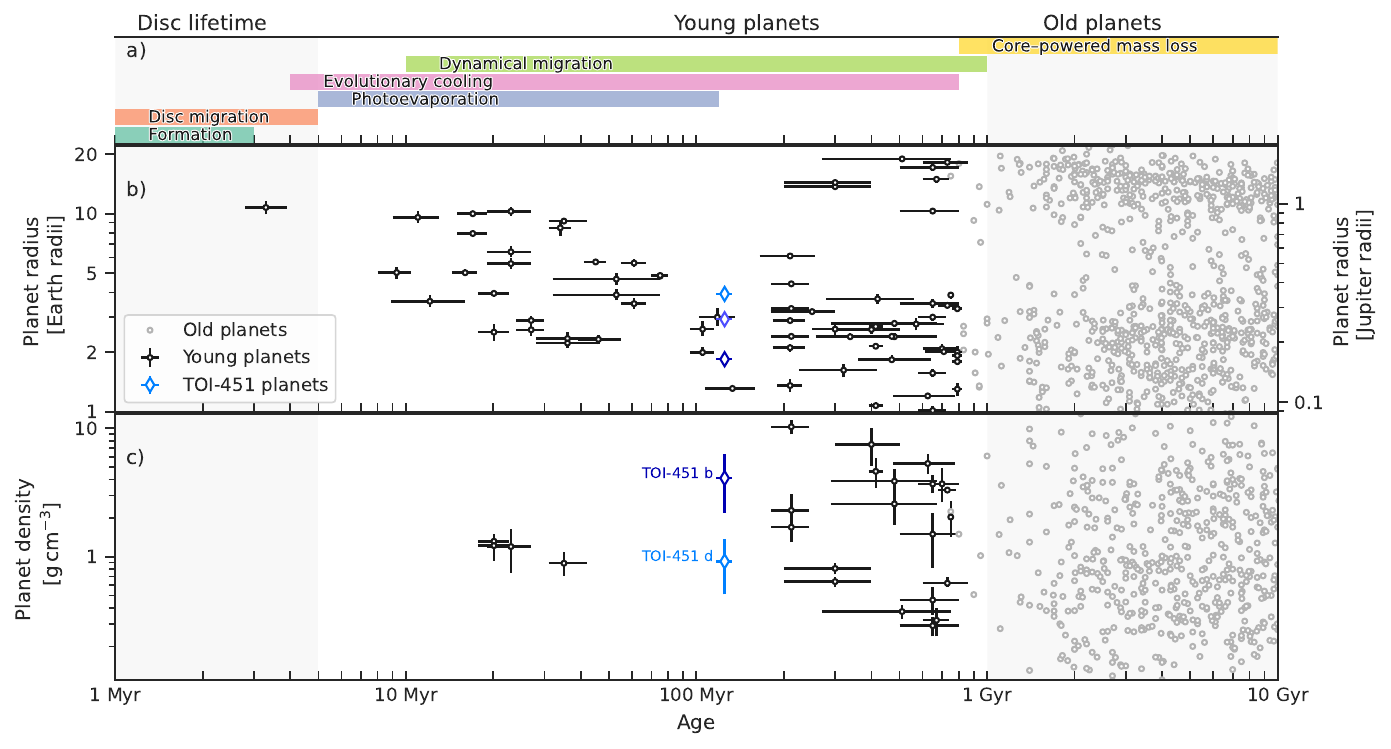}
    \caption{ a) Timescales of key planetary evolutionary processes.
b) Planetary radius and c) planetary density as functions of system age. Old exoplanets are shown as grey dots, and known young exoplanets are indicated by dark dots \citep[As in the NASA Exoplanet Archive on Dec 17, 2025, \url{https://exoplanetarchive.ipac.caltech.edu/},][]{NASAexoplanet}. 
Our radius and density estimates for \targetbcd\ are shown with diamonds.
    }
    \label{fig:ageden}
\end{figure*}

Figure~\ref{fig:ageden} shows the key-role of the \target\ system within the broader context of planetary evolution and atmospheric mass-loss processes. 
The current population of transiting planets displays a trend of decreasing radii with age that aligns well with the expected outcomes of the mass-loss mechanisms discussed in the previous paragraphs (see Panel b, Fig.~\ref{fig:ageden}).
Within the photo-evaporation framework, one would expect an evolutionary path in which planets become progressively denser as they lose a substantial fraction of their primordial envelopes during the first $\sim$100\,Myr.
Notably, the subset of well-characterised young exoplanets with ages younger than 100\,Myr tends to exhibit lower bulk densities compared to those in the 100\,Myr–1\,Gyr range (see Panel c, Fig.~\ref{fig:ageden}).
Although this pattern is still supported by a small number of systems, it provides valuable clues into atmospheric evolution. Additional well-characterised young planets are required to confirm whether this constitutes a robust evolutionary trend.
The \target\ system, with its precisely determined age and multiple transiting planets, emerges as a cornerstone laboratory for testing photo-evaporation efficiency. 
In particular, \targetb\ may represent a planet that has just passed through an intense stage of high-energy irradiation. The inferred density values derived here will therefore serve as a key reference for future follow-up observations and modelling efforts. Nonetheless, improved mass constraints will be crucial to refine the present interpretation.

\subsection{On follow-up observations}
\label{sec:futureobservations}

\target\ is a keystone target that can help to understand the evolution of multi-planetary systems, but additional follow-up observations are needed to improve the precision of the planetary and orbital parameters. 
In particular, additional RV monitoring to refine the planet masses is needed. 
The results presented in this paper suggest that \targetb\ could still retain a small fraction of primordial atmosphere, or being a water world. Both scenarios are interesting for testing planetary evolution and mass-loss theories.
Improvement of masses for \targetc\ and d are also essential. Being young worlds with significant atmospheric envelopes, they are excellent candidates to perform transmission spectroscopy studies. But the interpretations of such are only useful if  masses are known \citep{Batalha2019}.

How many more RV measurements would we need to improve the mass measurements of \targetbcd? To answer this question, we used \texttt{citlalatonac} \citep[][]{pyaneti2} to simulate multiple realisations of additional ESPRESSO observations and analysed them in the same way as we analysed the existing data in this paper.  
We used the GP hyper-parameters obtained in Table~\ref{tab:pars} to create the stellar signal as samples of a multi-GP. 
We simulated synthetic time-series of activity-induced RV and NGTS-like photometry and added the expected Keplerian signals for the three planets and sampled the data assuming \target\ is observed from Paranal with a maximum airmass of 1.7 in an intensive 5 months campaign, and added realistic white and instrumental red noise (based on our existing ESPRESSO data). 
We then attempted to recover the stellar and planetary signals, using the same multi-GP framework and combining the existing and simulated data, as we have done with the available real data. By varying the number of synthetic observations, we established that we need around 125 more epochs spread over 5 months to model the stellar signal robustly and to detect the signal of planets to 3-sigma confidence or better.

By the end of 2026 ESA's upcoming M-Class mission, \plato\ \citep[][]{Rauer2024} is planned to be launched. \plato\ is observing a fixed field in the Southern hemisphere \citep{nascimbeni2025} for at least the first two years of its mission. The spacecraft has 26 cameras that partially overlap resulting in stars within the field being observed by up to 24 cameras in the centre and down to 6 at the edges. \target\ lies within the Southern \plato\ field \citep{Eschen2024} and is a target of the \plato\ Input Catalogue \citep[PIC;][]{montalto2021}. Currently it is estimated that \plato\ shall observe this target with 6 cameras for at least the first two years of its mission. Due to \plato\ observing the target with more cameras than \tess\ and for a longer duration,  it improves the sensitivity to detecting smaller and long-orbital period planets. For \target\ the estimated noise recorded in the PIC, results in sensitivities allowing to detect planets as small as 0.5\,\rearth and planets at orbital periods of up to 200\,d if they transit this star \citep{Eschen2024}.

\section{Conclusions}
\label{sec:conclusions}

In this study, we performed an intensive photometric and spectroscopic ground-based follow-up of \target\ to further characterise the properties of the \target\ planetary system.
Our analysis provides 2-$\sigma$ masses estimates for \targetb\ and d that allowed us to infer, at first order, their possible nature, as well as a upper mass limit for \targetc.

Our analysis highlights \targetb\ as uniquely positioned within the radius gap, making it a crucial target for testing atmospheric mass-loss theories. 
If composed primarily of an Earth-like rocky core, \targetb\ would retain a minimal hydrogen envelope, potentially indicating ongoing atmospheric-loss processes. 
Conversely, if \targetb\ has a rocky-ice composition, its radius can be explained without the need of an atmospheric envelope, suggesting efficient mass loss within the first 125\,Myr.
Planet \targetd\ is likely enveloped by a substantial volatile-rich atmosphere. Depending on its core composition, hydrogen envelopes between 5\% and 8\% by mass could explain its observed radius. Its youth, substantial atmosphere, and brightness of its host star position \targetd\ as an optimal candidate for transmission spectroscopy. 
Planned \jwst\ observations of \targetc\ and d will further refine our understanding, leveraging the mass constraints provided by this study to better interpret atmospheric characterization results.

This manuscript was also a proof-of-concept on the use of contemporaneous spectroscopic-photometry observations within the context of multi-GP regression. 
We showed the potential of using contemporaneous photometry as an activity indicator.
However, we note that our photometric sampling strategy was similar to the spectroscopic one, of taking one point per night.
It is still to be tested the usefulness of having continuous and high precision contemporaneous photometry, e.g., space photometry, when modelling stellar signals in RVs. 
We foresee that contemporaneous and high-cadence photometry will be useful in cases where the stellar activity evolves faster or similar to the time-scales in which the spectroscopic observations are taken (e.g., young stars).
This scenario has to be explore further given the upcoming \plato\ mission that will provide high-quality continuos photometry that could be paired with contemporaneous RV follow-ups. 
The \plato\ observing strategy combined with its 24-camera strategy will deliver high-quality light curves with high signal-to-noise astrophysical signals \citep[e.g.,][]{republic}.

\section*{Acknowledgements}
Based on observations collected at the European Organisation for Astronomical Research in the Southern Hemisphere under ESO programmes 0112.C-0347(A) and 0112.C-0131(A).
This work is based in part on data collected under the NGTS project at the ESO La Silla Paranal Observatory. The NGTS facility is operated by a consortium of institutes with support from the UK Science and Technology Facilities Council (STFC) under projects ST/M001962/1, ST/S002642/1 and ST/W003163/1.
This work makes use of observations from the Las Cumbres Observatory global telescope network.
This publication made use of data products from the Wide-field
Infrared Survey Explorer, which is a joint project of the University of California, Los Angeles, and the Jet Propulsion Laboratory/California Institute of Technology, funded by the National Aeronautics and Space Administration.
This research has made use of the NASA Exoplanet Archive, which is operated by the California Institute of Technology, under contract with the National Aeronautics and Space Administration under the Exoplanet Exploration Program.
We thank the anonymous referees for their careful reading of the manuscript and comments that helped to improve the quality of the work.
O.B., H.Y., B.K., and S.A. acknowledge that this publication is part of a project that has received funding from the European Research Council (ERC) under the European Union’s Horizon 2020 research and innovation programme (Grant agreement No. 865624).
M.M., V.B., and N.L. acknowledge financial support from the Agencia Estatal de Investigaci\'on (AEI/10.13039/501100011033) of the Ministerio de Ciencia e Innovaci\'on  through project PID2022-137241NB-C41,
A.V.F. acknowledges the support of the IOP through the Bell Burnell Graduate Scholarship Fund.
A.M. acknowledges funding from a UKRI Future Leader Fellowship, grant number MR/X033244/1 and a UK Science and Technology Facilities Council (STFC) small grant ST/Y002334/1.
S.S. acknowledges Fondo Comité Mixto-ESO Chile ORP 025/2022.
Y.N.E.E. acknowledges support from a Science and Technology Facilities Council (STFC) studentship, grant number ST/Y509693/1.
E.M.V. acknowledges financial support from the Swiss National Science Foundation (SNSF) Postdoc. Mobility Fellowship under grant no. P500PT\_225456/1.
M.L. acknowledged that this research was in part funded by the UKRI (Grant EP/X027562/1).
F. M. acknowledges the financial support from the Agencia Estatal de Investigaci\'{o}n del Ministerio de Ciencia, Innovaci\'{o}n y Universidades (MCIU/AEI) through grant PID2023-152906NA-I00.
JSJ gratefully acknowledges support by FONDECYT grant 1240738 and from the ANID BASAL project FB210003.
This work made use of \texttt{numpy} \citep[][]{numpy}, \texttt{matplotlib} \citep[][]{matplotlib}, and \texttt{pandas} \citep{pandas} libraries.
This work made use of Astropy:\footnote{\url{http://www.astropy.org}} a community-developed core Python package and an ecosystem of tools and resources for astronomy \citep{astropy1, astropy2,astropy3}. 
O.B., H.Y., B.K., E.M.V., M.C., and S.A. acknowledge that they miss that fusion restaurant that provided the calories to write this paper.

\section*{Data Availability}

The codes used in this manuscript are freely available at \url{https://github.com/oscaribv}.
The spectroscopic measurements that appear in Table~\ref{tab:harps} are available as supplementary material in the online version of this manuscript.
All \tess\ data are available via the MAST archive.

%
\bibliographystyle{mnras} 
\bibliography{refs} 

@ARTICLE{Haywood2014,
       author = {{Haywood}, R.~D. and {Collier Cameron}, A. and {Queloz}, D. and
         {Barros}, S.~C.~C. and {Deleuil}, M. and {Fares}, R. and {Gillon}, M. and
         {Lanza}, A.~F. and {Lovis}, C. and {Moutou}, C.},
        title = "{Planets and stellar activity: hide and seek in the CoRoT-7 system}",
      journal = {\mnras},
         year = "2014",
        month = "Sep",
       volume = {443},
       number = {3},
        pages = {2517-2531},
          doi = {10.1093/mnras/stu1320},
archivePrefix = {arXiv},
       eprint = {1407.1044},
 primaryClass = {astro-ph.EP},
       adsurl = {https://ui.adsabs.harvard.edu/abs/2014MNRAS.443.2517H}
}

@ARTICLE{nascimbeni2025,
       author = {{Nascimbeni}, V. and {Piotto}, G. and {Cabrera}, J. and {Montalto}, M. and {Marinoni}, S. and {Marrese}, P.~M. and {Aerts}, C. and {Altavilla}, G. and {Benatti}, S. and {B{\"o}rner}, A. and {Deleuil}, M. and {Desidera}, S. and {Gizon}, L. and {Goupil}, M.~J. and {Granata}, V. and {Heras}, A.~M. and {Magrin}, D. and {Malavolta}, L. and {Mas-Hesse}, J.~M. and {Osborn}, H.~P. and {Pagano}, I. and {Paproth}, C. and {Pollacco}, D. and {Prisinzano}, L. and {Ragazzoni}, R. and {Ramsay}, G. and {Rauer}, H. and {Tkachenko}, A. and {Udry}, S.},
        title = "{The PLATO field selection process: II. Characterization of LOPS2, the first long-pointing field}",
      journal = {\aap},
         year = 2025,
        month = feb,
       volume = {694},
          eid = {A313},
        pages = {A313},
          doi = {10.1051/0004-6361/202452325},
archivePrefix = {arXiv},
       eprint = {2501.07687},
 primaryClass = {astro-ph.EP},
       adsurl = {https://ui.adsabs.harvard.edu/abs/2025A&A...694A.313N}
}

@ARTICLE{montalto2021,
       author = {{Montalto}, M. and {Piotto}, G. and {Marrese}, P.~M. and {Nascimbeni}, V. and {Prisinzano}, L. and {Granata}, V. and {Marinoni}, S. and {Desidera}, S. and {Ortolani}, S. and {Aerts}, C. and {Alei}, E. and {Altavilla}, G. and {Benatti}, S. and {B{\"o}rner}, A. and {Cabrera}, J. and {Claudi}, R. and {Deleuil}, M. and {Fabrizio}, M. and {Gizon}, L. and {Goupil}, M.~J. and {Heras}, A.~M. and {Magrin}, D. and {Malavolta}, L. and {Mas-Hesse}, J.~M. and {Pagano}, I. and {Paproth}, C. and {Pertenais}, M. and {Pollacco}, D. and {Ragazzoni}, R. and {Ramsay}, G. and {Rauer}, H. and {Udry}, S.},
        title = "{The all-sky PLATO input catalogue}",
      journal = {\aap},
         year = 2021,
        month = sep,
       volume = {653},
          eid = {A98},
        pages = {A98},
          doi = {10.1051/0004-6361/202140717},
archivePrefix = {arXiv},
       eprint = {2108.13712},
 primaryClass = {astro-ph.EP},
       adsurl = {https://ui.adsabs.harvard.edu/abs/2021A&A...653A..98M}
}

@ARTICLE{Lanza2006,
       author = {{Lanza}, Antonino F.},
        title = "{On the time dependence of differential rotation in young late-type stars}",
      journal = {\mnras},
         year = 2006,
        month = dec,
       volume = {373},
       number = {2},
        pages = {819-826},
          doi = {10.1111/j.1365-2966.2006.11085.x},
       adsurl = {https://ui.adsabs.harvard.edu/abs/2006MNRAS.373..819L}
}

@ARTICLE{Saar1997,
       author = {{Saar}, Steven H. and {Donahue}, Robert A.},
        title = "{Activity-Related Radial Velocity Variation in Cool Stars}",
      journal = {\apj},
         year = 1997,
        month = aug,
       volume = {485},
       number = {1},
        pages = {319-327},
          doi = {10.1086/304392},
       adsurl = {https://ui.adsabs.harvard.edu/abs/1997ApJ...485..319S}
}

@ARTICLE{Huerta2008,
       author = {{Huerta}, Marcos and {Johns-Krull}, Christopher M. and {Prato}, L. and {Hartigan}, Patrick and {Jaffe}, D.~T.},
        title = "{Starspot-Induced Radial Velocity Variability in LkCa 19}",
      journal = {\apj},
         year = 2008,
        month = may,
       volume = {678},
       number = {1},
        pages = {472-482},
          doi = {10.1086/526415},
archivePrefix = {arXiv},
       eprint = {0711.2505},
 primaryClass = {astro-ph},
       adsurl = {https://ui.adsabs.harvard.edu/abs/2008ApJ...678..472H}
}

@ARTICLE{Morris2020,
       author = {{Morris}, Brett M.},
        title = "{A Relationship between Stellar Age and Spot Coverage}",
      journal = {\apj},
         year = 2020,
        month = apr,
       volume = {893},
       number = {1},
          eid = {67},
        pages = {67},
          doi = {10.3847/1538-4357/ab79a0},
archivePrefix = {arXiv},
       eprint = {2002.09135},
 primaryClass = {astro-ph.SR},
       adsurl = {https://ui.adsabs.harvard.edu/abs/2020ApJ...893...67M}
}

@ARTICLE{Klein2024,
       author = {{Klein}, Baptiste and {Aigrain}, Suzanne and {Cretignier}, Michael and {Al Moulla}, Khaled and {Dumusque}, Xavier and {Barrag{\'a}n}, Oscar and {Yu}, Haochuan and {Mortier}, Annelies and {Rescigno}, Federica and {Cameron}, Andrew Collier and {L{\'o}pez-Morales}, Mercedes and {Meunier}, Nad{\`e}ge and {Sozzetti}, Alessandro and {O'Sullivan}, Niamh K.},
        title = "{Investigating stellar activity through eight years of Sun-as-a-star observations}",
      journal = {\mnras},
         year = 2024,
        month = jul,
       volume = {531},
       number = {4},
        pages = {4238-4262},
          doi = {10.1093/mnras/stae1313},
archivePrefix = {arXiv},
       eprint = {2405.12065},
 primaryClass = {astro-ph.EP},
       adsurl = {https://ui.adsabs.harvard.edu/abs/2024MNRAS.531.4238K}
}

@ARTICLE{GLS,
       author = {{Zechmeister}, M. and {K{\"u}rster}, M.},
        title = "{The generalised Lomb-Scargle periodogram. A new formalism for the floating-mean and Keplerian periodograms}",
      journal = {\aap},
         year = 2009,
        month = mar,
       volume = {496},
       number = {2},
        pages = {577-584},
          doi = {10.1051/0004-6361:200811296},
archivePrefix = {arXiv},
       eprint = {0901.2573},
 primaryClass = {astro-ph.IM},
       adsurl = {https://ui.adsabs.harvard.edu/abs/2009A&A...496..577Z}
}

@ARTICLE{Zicher2022,
       author = {{Zicher}, Norbert and {Barrag{\'a}n}, Oscar and {Klein}, Baptiste and {Aigrain}, Suzanne and {Owen}, James E. and {Gandolfi}, Davide and {Lagrange}, Anne-Marie and {Serrano}, Luisa Maria and {Kaye}, Laurel and {Nielsen}, Louise Dyregaard and {Rajpaul}, Vinesh Maguire and {Grandjean}, Antoine and {Goffo}, Elisa and {Nicholson}, Belinda},
        title = "{One year of AU Mic with HARPS - I. Measuring the masses of the two transiting planets}",
      journal = {\mnras},
         year = 2022,
        month = may,
       volume = {512},
       number = {2},
        pages = {3060-3078},
          doi = {10.1093/mnras/stac614},
archivePrefix = {arXiv},
       eprint = {2203.01750},
 primaryClass = {astro-ph.EP},
       adsurl = {https://ui.adsabs.harvard.edu/abs/2022MNRAS.512.3060Z}
}

@ARTICLE{pyaneti,
       author = {Barrag\'an, O. and Gandolfi, D. and Antoniciello, G.},
        title = "{PYANETI: a fast and powerful software suite for multiplanet radial
        velocity and transit fitting}",
      journal = {\mnras},
         year = 2019,
        month = Jan,
       volume = {482},
        pages = {1017-1030},
          doi = {10.1093/mnras/sty2472},
 primaryClass = {astro-ph.EP},
       adsurl = {https://ui.adsabs.harvard.edu/#abs/2019MNRAS.482.1017B}
}

@ARTICLE{Barragan2023,
       author = {{Barrag{\'a}n}, Oscar and {Gillen}, Edward and {Aigrain}, Suzanne and {Meech}, Annabella and {Klein}, Baptiste and {Nielsen}, Louise Dyregaard and {Yu}, Haochuan and {O'Sullivan}, Niamh K. and {Nicholson}, Belinda A. and {Lillo-Box}, Jorge},
        title = "{Revisiting K2-233 spectroscopic time-series with multidimensional Gaussian processes}",
      journal = {\mnras},
         year = 2023,
        month = jul,
       volume = {522},
       number = {3},
        pages = {3458-3471},
          doi = {10.1093/mnras/stad1139},
archivePrefix = {arXiv},
       eprint = {2304.06406},
 primaryClass = {astro-ph.EP},
       adsurl = {https://ui.adsabs.harvard.edu/abs/2023MNRAS.522.3458B}
}

@ARTICLE{Blunt2023,
       author = {{Blunt}, Sarah and {Carvalho}, Adolfo and {David}, Trevor J. and {Beichman}, Charles and {Zink}, Jon K. and {Gaidos}, Eric and {Behmard}, Aida and {Bouma}, Luke G. and {Cody}, Devin and {Dai}, Fei and {Foreman-Mackey}, Daniel and {Grunblatt}, Sam and {Howard}, Andrew W. and {Kosiarek}, Molly and {Knutson}, Heather A. and {Rubenzahl}, Ryan A. and {Beard}, Corey and {Chontos}, Ashley and {Giacalone}, Steven and {Hirano}, Teruyuki and {Johnson}, Marshall C. and {Lubin}, Jack and {Akana Murphy}, Joseph M. and {Petigura}, Erik A. and {Van Zandt}, Judah and {Weiss}, Lauren},
        title = "{Overfitting Affects the Reliability of Radial Velocity Mass Estimates of the V1298 Tau Planets}",
      journal = {\aj},
         year = 2023,
        month = aug,
       volume = {166},
       number = {2},
          eid = {62},
        pages = {62},
          doi = {10.3847/1538-3881/acde78},
archivePrefix = {arXiv},
       eprint = {2306.08145},
 primaryClass = {astro-ph.EP},
       adsurl = {https://ui.adsabs.harvard.edu/abs/2023AJ....166...62B}
}

@ARTICLE{choi2016,
       author = {{Choi}, Jieun and {Dotter}, Aaron and {Conroy}, Charlie and
         {Cantiello}, Matteo and {Paxton}, Bill and {Johnson}, Benjamin D.},
        title = "{Mesa Isochrones and Stellar Tracks (MIST). I. Solar-scaled Models}",
      journal = {\apj},
         year = "2016",
        month = "Jun",
       volume = {823},
       number = {2},
          eid = {102},
        pages = {102},
          doi = {10.3847/0004-637X/823/2/102},
archivePrefix = {arXiv},
       eprint = {1604.08592},
 primaryClass = {astro-ph.SR},
       adsurl = {https://ui.adsabs.harvard.edu/abs/2016ApJ...823..102C}
}

@ARTICLE{Rajpaul2015,
       author = {{Rajpaul}, V. and {Aigrain}, S. and {Osborne}, M.~A. and {Reece}, S. and
         {Roberts}, S.},
        title = "{A Gaussian process framework for modelling stellar activity signals in radial velocity data}",
      journal = {\mnras},
         year = "2015",
        month = "Sep",
       volume = {452},
       number = {3},
        pages = {2269-2291},
          doi = {10.1093/mnras/stv1428},
archivePrefix = {arXiv},
       eprint = {1506.07304},
 primaryClass = {astro-ph.EP},
       adsurl = {https://ui.adsabs.harvard.edu/abs/2015MNRAS.452.2269R}
}

@ARTICLE{Gelman1992,
       author = {{Gelman}, Andrew and {Rubin}, Donald B.},
        title = "{Inference from Iterative Simulation Using Multiple Sequences}",
      journal = {Statistical Science},
         year = 1992,
        month = jan,
       volume = {7},
        pages = {457-472},
          doi = {10.1214/ss/1177011136},
       adsurl = {https://ui.adsabs.harvard.edu/abs/1992StaSc...7..457G}
}

@ARTICLE{Barragan2018b,
       author = {{Barrag{\'a}n}, O. and {Gandolfi}, D. and {Dai}, F. and
         {Livingston}, J. and {Persson}, C.~M. and {Hirano}, T. and
         {Narita}, N. and {Csizmadia}, Sz. and {Winn}, J.~N. and {Nespral}, D. and
         {Prieto-Arranz}, J. and {Smith}, A.~M.~S. and {Nowak}, G. and
         {Albrecht}, S. and {Antoniciello}, G. and {Bo Justesen}, A. and
         {Cabrera}, J. and {Cochran}, W.~D. and {Deeg}, H. and {Eigmuller}, Ph. and
         {Endl}, M. and {Erikson}, A. and {Fridlund}, M. and {Fukui}, A. and
         {Grziwa}, S. and {Guenther}, E. and {Hatzes}, A.~P. and {Hidalgo}, D. and
         {Johnson}, M.~C. and {Korth}, J. and {Palle}, E. and {Patzold}, M. and
         {Rauer}, H. and {Tanaka}, Y. and {Van Eylen}, V.},
        title = "{K2-141 b. A 5-M$_{{\ensuremath{\oplus}}}$ super-Earth transiting a K7 V star every 6.7 h}",
      journal = {\aap},
         year = "2018",
        month = "May",
       volume = {612},
          eid = {A95},
        pages = {A95},
          doi = {10.1051/0004-6361/201732217},
archivePrefix = {arXiv},
       eprint = {1711.02097},
 primaryClass = {astro-ph.EP},
       adsurl = {https://ui.adsabs.harvard.edu/abs/2018A&A...612A..95B}
}

@ARTICLE{Winn2010,
       author = {{Winn}, Joshua N.},
        title = "{Transits and Occultations}",
      journal = {arXiv e-prints},
         year = "2010",
        month = "Jan",
          eid = {arXiv:1001.2010},
        pages = {arXiv:1001.2010},
archivePrefix = {arXiv},
       eprint = {1001.2010},
 primaryClass = {astro-ph.EP},
       adsurl = {https://ui.adsabs.harvard.edu/abs/2010arXiv1001.2010W}
}

@ARTICLE{Zeng2016,
       author = {{Zeng}, Li and {Sasselov}, Dimitar D. and {Jacobsen}, Stein B.},
        title = "{Mass-Radius Relation for Rocky Planets Based on PREM}",
      journal = {\apj},
         year = "2016",
        month = "Mar",
       volume = {819},
       number = {2},
          eid = {127},
        pages = {127},
          doi = {10.3847/0004-637X/819/2/127},
archivePrefix = {arXiv},
       eprint = {1512.08827},
 primaryClass = {astro-ph.EP},
       adsurl = {https://ui.adsabs.harvard.edu/abs/2016ApJ...819..127Z}
}

@ARTICLE{Owen2013,
       author = {{Owen}, James E. and {Wu}, Yanqin},
        title = "{Kepler Planets: A Tale of Evaporation}",
      journal = {\apj},
         year = "2013",
        month = "Oct",
       volume = {775},
       number = {2},
          eid = {105},
        pages = {105},
          doi = {10.1088/0004-637X/775/2/105},
archivePrefix = {arXiv},
       eprint = {1303.3899},
 primaryClass = {astro-ph.EP},
       adsurl = {https://ui.adsabs.harvard.edu/abs/2013ApJ...775..105O}
}

@ARTICLE{Sotuhworth2007,
       author = {{Southworth}, John and {Wheatley}, Peter J. and {Sams}, Giles},
        title = "{A method for the direct determination of the surface gravities of transiting extrasolar planets}",
      journal = {\mnras},
         year = "2007",
        month = "Jul",
       volume = {379},
       number = {1},
        pages = {L11-L15},
          doi = {10.1111/j.1745-3933.2007.00324.x},
archivePrefix = {arXiv},
       eprint = {0704.1570},
 primaryClass = {astro-ph},
       adsurl = {https://ui.adsabs.harvard.edu/abs/2007MNRAS.379L..11S}
}

@ARTICLE{emcee,
       author = {{Foreman-Mackey}, Daniel and {Hogg}, David W. and {Lang}, Dustin and
         {Goodman}, Jonathan},
        title = "{emcee: The MCMC Hammer}",
      journal = {\pasp},
         year = "2013",
        month = "Mar",
       volume = {125},
       number = {925},
        pages = {306},
          doi = {10.1086/670067},
archivePrefix = {arXiv},
       eprint = {1202.3665},
 primaryClass = {astro-ph.IM},
       adsurl = {https://ui.adsabs.harvard.edu/abs/2013PASP..125..306F}
}

@ARTICLE{Kipping2013,
   author = {{Kipping}, D.~M.},
    title = "{Efficient, uninformative sampling of limb darkening coefficients for two-parameter laws}",
  journal = {\mnras},
archivePrefix = "arXiv",
   eprint = {1308.0009},
 primaryClass = "astro-ph.SR",
     year = 2013,
    month = nov,
   volume = 435,
    pages = {2152-2160},
      doi = {10.1093/mnras/stt1435},
   adsurl = {https://ui.adsabs.harvard.edu/abs/2013MNRAS.435.2152K}
}

@ARTICLE{Aigrain2012,
       author = {{Aigrain}, S. and {Pont}, F. and {Zucker}, S.},
        title = "{A simple method to estimate radial velocity variations due to stellar activity using photometry}",
      journal = {\mnras},
         year = "2012",
        month = "Feb",
       volume = {419},
       number = {4},
        pages = {3147-3158},
          doi = {10.1111/j.1365-2966.2011.19960.x},
archivePrefix = {arXiv},
       eprint = {1110.1034},
 primaryClass = {astro-ph.SR},
       adsurl = {https://ui.adsabs.harvard.edu/abs/2012MNRAS.419.3147A}
}

@ARTICLE{Rajpaul2016,
       author = {{Rajpaul}, V. and {Aigrain}, S. and {Roberts}, S.},
        title = "{Ghost in the time series: no planet for Alpha Cen B}",
      journal = {\mnras},
         year = "2016",
        month = "Feb",
       volume = {456},
       number = {1},
        pages = {L6-L10},
          doi = {10.1093/mnrasl/slv164},
archivePrefix = {arXiv},
       eprint = {1510.05598},
 primaryClass = {astro-ph.EP},
       adsurl = {https://ui.adsabs.harvard.edu/abs/2016MNRAS.456L...6R}
}

@ARTICLE{Bressan2012,
       author = {{Bressan}, Alessandro and {Marigo}, Paola and {Girardi}, L{\'e}o. and
         {Salasnich}, Bernardo and {Dal Cero}, Claudia and {Rubele}, Stefano and
         {Nanni}, Ambra},
        title = "{PARSEC: stellar tracks and isochrones with the PAdova and TRieste Stellar Evolution Code}",
      journal = {\mnras},
         year = "2012",
        month = "Nov",
       volume = {427},
       number = {1},
        pages = {127-145},
          doi = {10.1111/j.1365-2966.2012.21948.x},
archivePrefix = {arXiv},
       eprint = {1208.4498},
 primaryClass = {astro-ph.SR},
       adsurl = {https://ui.adsabs.harvard.edu/abs/2012MNRAS.427..127B}
}

@ARTICLE{Mandel2002,
   author = {{Mandel}, K. and {Agol}, E.},
    title = "{Analytic Light Curves for Planetary Transit Searches}",
  journal = {\apjl},
   eprint = {astro-ph/0210099},
     year = 2002,
    month = dec,
   volume = 580,
    pages = {L171-L175},
      doi = {10.1086/345520},
   adsurl = {http://adsabs.harvard.edu/abs/2002ApJ...580L.171M}
}

@ARTICLE{Barragan2019,
       author = {{Barrag{\'a}n}, O. and {Aigrain}, S. and {Kubyshkina}, D. and {Gand
        olfi}, D. and {Livingston}, J. and {Fridlund}, M.~C.~V. and
         {Fossati}, L. and {Korth}, J. and {Parviainen}, H. and {Malavolta}, L. and
         {Palle}, E. and {Deeg}, H.~J. and {Nowak}, G. and {Rajpaul}, V.~M. and
         {Zicher}, N. and {Antoniciello}, G. and {Narita}, N. and
         {Albrecht}, S. and {Bedin}, L.~R. and {Cabrera}, J. and
         {Cochran}, W.~D. and {de Leon}, J. and {Eigm{\"u}ller}, Ph and
         {Fukui}, A. and {Granata}, V. and {Grziwa}, S. and {Guenther}, E. and
         {Hatzes}, A.~P. and {Kusakabe}, N. and {Latham}, D.~W. and
         {Libralato}, M. and {Luque}, R. and
         {Monta{\~n}{\'e}s-Rodr{\'\i}guez}, P. and {Murgas}, F. and
         {Nardiello}, D. and {Pagano}, I. and {Piotto}, G. and {Persson}, C.~M. and
         {Redfield}, S. and {Tamura}, M.},
        title = "{Radial velocity confirmation of K2-100b: a young, highly irradiated, and low-density transiting hot Neptune}",
      journal = {\mnras},
         year = 2019,
        month = nov,
       volume = {490},
       number = {1},
        pages = {698-708},
          doi = {10.1093/mnras/stz2569},
archivePrefix = {arXiv},
       eprint = {1909.05252},
 primaryClass = {astro-ph.EP},
       adsurl = {https://ui.adsabs.harvard.edu/abs/2019MNRAS.490..698B}
}

@ARTICLE{Ricker2015,
       author = {{Ricker}, George R. and {Winn}, Joshua N. and {Vanderspek}, Roland and
        {Latham}, David W. and {Bakos}, G{\'a}sp{\'a}r {\'A}. and
        {Bean}, Jacob L. and {Berta-Thompson}, Zachory K. and {Brown},
        Timothy M. and {Buchhave}, Lars and {Butler}, Nathaniel R. and
        {Butler}, R. Paul and {Chaplin}, William J. and {Charbonneau},
        David and {Christensen-Dalsgaard}, J{\o}rgen and {Clampin}, Mark
        and {Deming}, Drake and {Doty}, John and {De Lee}, Nathan and
        {Dressing}, Courtney and {Dunham}, Edward W. and {Endl}, Michael
        and {Fressin}, Francois and {Ge}, Jian and {Henning}, Thomas and
        {Holman}, Matthew J. and {Howard}, Andrew W. and {Ida}, Shigeru
        and {Jenkins}, Jon M. and {Jernigan}, Garrett and {Johnson},
        John Asher and {Kaltenegger}, Lisa and {Kawai}, Nobuyuki and
        {Kjeldsen}, Hans and {Laughlin}, Gregory and {Levine}, Alan M.
        and {Lin}, Douglas and {Lissauer}, Jack J. and {MacQueen},
        Phillip and {Marcy}, Geoffrey and {McCullough}, Peter R. and
        {Morton}, Timothy D. and {Narita}, Norio and {Paegert}, Martin
        and {Palle}, Enric and {Pepe}, Francesco and {Pepper}, Joshua
        and {Quirrenbach}, Andreas and {Rinehart}, Stephen A. and
        {Sasselov}, Dimitar and {Sato}, Bun'ei and {Seager}, Sara and
        {Sozzetti}, Alessandro and {Stassun}, Keivan G. and {Sullivan},
        Peter and {Szentgyorgyi}, Andrew and {Torres}, Guillermo and
        {Udry}, Stephane and {Villasenor}, Joel},
        title = "{Transiting Exoplanet Survey Satellite (TESS)}",
      journal = {Journal of Astronomical Telescopes, Instruments, and Systems},
         year = 2015,
        month = Jan,
       volume = {1},
          eid = {014003},
        pages = {014003},
          doi = {10.1117/1.JATIS.1.1.014003},
       adsurl = {https://ui.adsabs.harvard.edu/#abs/2015JATIS...1a4003R}
}

@PHDTHESIS{CretignierPhD,
       author = {{Cretignier}, Michael},
        title = "{Improving the RV precision to detect the Earth’s twins around the Sun’s siblings}",
       school = {University of Geneva, Switzerland},
         year = 2022,
        month = sep,
        page = 47,
       adsurl = {https://archive-ouverte.unige.ch/unige:164874}
}

@ARTICLE{Dalal2024,
       author = {{Dalal}, S. and {Rescigno}, F. and {Cretignier}, M. and {Anna John}, A. and {Majidi}, F.~Z. and {Malavolta}, L. and {Mortier}, A. and {Pinamonti}, M. and {Buchhave}, L.~A. and {Haywood}, R.~D. and {Sozzetti}, A. and {Dumusque}, X. and {Lienhard}, F. and {Rice}, K. and {Vanderburg}, A. and {Lakeland}, B. and {Bonomo}, A.~S. and {Cameron}, A. Collier and {Damasso}, M. and {Affer}, L. and {Boschin}, W. and {Cooke}, B. and {Cosentino}, R. and {Fabrizio}, L. Di and {Ghedina}, A. and {Harutyunyan}, A. and {Latham}, D.~W. and {L{\'o}pez-Morales}, M. and {Lovis}, C. and {Fiorenzano}, A.~F. Mart{\'\i}nez and {Mayor}, M. and {Nicholson}, B. and {Pepe}, F. and {Stalport}, M. and {Udry}, S. and {Watson}, C.~A. and {Wilson}, T.~G.},
        title = "{Trio of super-Earth candidates orbiting K-dwarf HD 48948: a new habitable zone candidate}",
      journal = {\mnras},
         year = 2024,
        month = jul,
       volume = {531},
       number = {4},
        pages = {4464-4481},
          doi = {10.1093/mnras/stae1367},
       adsurl = {https://ui.adsabs.harvard.edu/abs/2024MNRAS.531.4464D}
}

@ARTICLE{Stalport2023,
       author = {{Stalport}, M. and {Cretignier}, M. and {Udry}, S. and {John}, A. Anna and {Wilson}, T.~G. and {Delisle}, J. -B. and {Bonomo}, A.~S. and {Buchhave}, L.~A. and {Charbonneau}, D. and {Dalal}, S. and {Damasso}, M. and {Di Fabrizio}, L. and {Dumusque}, X. and {Fiorenzano}, A. and {Harutyunyan}, A. and {Haywood}, R.~D. and {Latham}, D.~W. and {L{\'o}pez-Morales}, M. and {Lorenzi}, V. and {Lovis}, C. and {Malavolta}, L. and {Molinari}, E. and {Mortier}, A. and {Pedani}, M. and {Pepe}, F. and {Pinamonti}, M. and {Poretti}, E. and {Rice}, K. and {Sozzetti}, A.},
        title = "{A review of planetary systems around HD 99492, HD 147379, and HD 190007 with HARPS-N}",
      journal = {\aap},
         year = 2023,
        month = oct,
       volume = {678},
          eid = {A90},
        pages = {A90},
          doi = {10.1051/0004-6361/202346887},
archivePrefix = {arXiv},
       eprint = {2308.05669},
 primaryClass = {astro-ph.EP},
       adsurl = {https://ui.adsabs.harvard.edu/abs/2023A&A...678A..90S}
}

@ARTICLE{Nari2025,
       author = {{Nari}, N. and {Dumusque}, X. and {Hara}, N.~C. and {Su{\'a}rez Mascare{\~n}o}, A. and {Cretignier}, M. and {Gonz{\'a}lez Hern{\'a}ndez}, J.~I. and {Stefanov}, A.~K. and {Passegger}, V.~M. and {Rebolo}, R. and {Pepe}, F. and {Santos}, N.~C. and {Cristiani}, S. and {Faria}, J.~P. and {Figueira}, P. and {Sozzetti}, A. and {Zapatero Osorio}, M.~R. and {Adibekyan}, V. and {Alibert}, Y. and {Allende Prieto}, C. and {Bouchy}, F. and {Benatti}, S. and {Castro-Gonz{\'a}lez}, A. and {D'Odorico}, V. and {Damasso}, M. and {Delisle}, J.~B. and {Di Marcantonio}, P. and {Ehrenreich}, D. and {G{\'e}nova-Santos}, R. and {Hobson}, M.~J. and {Lavie}, B. and {Lillo-Box}, J. and {Lo Curto}, G. and {Lovis}, C. and {Martins}, C.~J.~A.~P. and {Mehner}, A. and {Micela}, G. and {Molaro}, P. and {Mordasini}, C. and {Nunes}, N. and {Palle}, E. and {Quanz}, S.~P. and {S{\'e}gransan}, D. and {Silva}, A.~M. and {Sousa}, S.~G. and {Udry}, S. and {Unger}, N. and {Venturini}, J.},
        title = "{Revisiting the multi-planetary system of the nearby star HD 20794: Confirmation of a low-mass planet in the habitable zone of a nearby G-dwarf}",
      journal = {\aap},
         year = 2025,
        month = jan,
       volume = {693},
          eid = {A297},
        pages = {A297},
          doi = {10.1051/0004-6361/202451769},
archivePrefix = {arXiv},
       eprint = {2501.17092},
 primaryClass = {astro-ph.EP},
       adsurl = {https://ui.adsabs.harvard.edu/abs/2025A&A...693A.297N}
}

@ARTICLE{Yu2025,
       author = {{Yu}, H. and {Garai}, Z. and {Cretignier}, M. and {Szab{\'o}}, Gy M. and {Aigrain}, S. and {Gandolfi}, D. and {Bryant}, E.~M. and {Correia}, A.~C.~M. and {Klein}, B. and {Brandeker}, A. and {Owen}, J.~E. and {G{\"u}nther}, M.~N. and {Winn}, J.~N. and {Heitzmann}, A. and {Cegla}, H.~M. and {Wilson}, T.~G. and {Gill}, S. and {Kriskovics}, L. and {Barrag{\'a}n}, O. and {Boldog}, A. and {Nielsen}, L.~D. and {Billot}, N. and {Lafarga}, M. and {Meech}, A. and {Alibert}, Y. and {Alonso}, R. and {B{\'a}rczy}, T. and {Barrado}, D. and {Barros}, S.~C.~C. and {Baumjohann}, W. and {Bayliss}, D. and {Benz}, W. and {Bergomi}, M. and {Borsato}, L. and {Broeg}, C. and {Cameron}, A. Collier and {Csizmadia}, Sz and {Cubillos}, P.~E. and {Davies}, M.~B. and {Deleuil}, M. and {Deline}, A. and {Demangeon}, O.~D.~S. and {Demory}, B. -O. and {Derekas}, A. and {Doyle}, L. and {Edwards}, B. and {Egger}, J.~A. and {Ehrenreich}, D. and {Erikson}, A. and {Fortier}, A. and {Fossati}, L. and {Fridlund}, M. and {Gazeas}, K. and {Gillon}, M. and {G{\"u}del}, M. and {Helling}, Ch and {Isaak}, K.~G. and {Kiss}, L.~L. and {Korth}, J. and {Lam}, K.~W.~F. and {Laskar}, J. and {Lecavelier des Etangs}, A. and {Lendl}, M. and {Magrin}, D. and {Maxted}, P.~F.~L. and {McCormac}, J. and {Mer{\'\i}n}, B. and {Mordasini}, C. and {Nascimbeni}, V. and {O'Brien}, S.~M. and {Olofsson}, G. and {Ottensamer}, R. and {Pagano}, I. and {Pall{\'e}}, E. and {Peter}, G. and {Piazza}, D. and {Piotto}, G. and {Pollacco}, D. and {Queloz}, D. and {Ragazzoni}, R. and {Rando}, N. and {Rauer}, H. and {Ribas}, I. and {Santos}, N.~C. and {Scandariato}, G. and {S{\'e}gransan}, D. and {Simon}, A.~E. and {Smith}, A.~M.~S. and {Sousa}, S.~G. and {Southworth}, R. and {Stalport}, M. and {Steinberger}, M. and {Sulis}, S. and {Udry}, S. and {Ulmer}, B. and {Ulmer-Moll}, S. and {Van Grootel}, V. and {Venturini}, J. and {Villaver}, E. and {Walton}, N.~A. and {Wheatley}, P.~J.},
        title = "{A possible misaligned orbit for the young planet AU Mic c}",
      journal = {\mnras},
         year = 2025,
        month = jan,
       volume = {536},
       number = {3},
        pages = {2046-2063},
          doi = {10.1093/mnras/stae2655},
archivePrefix = {arXiv},
       eprint = {2411.16958},
 primaryClass = {astro-ph.EP},
       adsurl = {https://ui.adsabs.harvard.edu/abs/2025MNRAS.536.2046Y}
}

@ARTICLE{Cretignier2024b,
       author = {{Cretignier}, M. and {Hara}, N.~C. and {Pietrow}, A.~G.~M. and {Zhao}, Y. and {Yu}, H. and {Dumusque}, X. and {Sozzetti}, A. and {Lovis}, C. and {Aigrain}, S.},
        title = "{Stellar surface information from the Ca II H\&K lines - II. Defining better activity proxies}",
      journal = {\mnras},
         year = 2024,
        month = dec,
       volume = {535},
       number = {3},
        pages = {2562-2584},
          doi = {10.1093/mnras/stae2508},
archivePrefix = {arXiv},
       eprint = {2411.00557},
 primaryClass = {astro-ph.SR},
       adsurl = {https://ui.adsabs.harvard.edu/abs/2024MNRAS.535.2562C}
}

@ARTICLE{Cretignier2023,
       author = {{Cretignier}, M. and {Dumusque}, X. and {Aigrain}, S. and {Pepe}, F.},
        title = "{YARARA V2: Reaching sub-m s$^{-1}$ precision over a decade using PCA on line-by-line radial velocities}",
      journal = {\aap},
         year = 2023,
        month = oct,
       volume = {678},
          eid = {A2},
        pages = {A2},
          doi = {10.1051/0004-6361/202347232},
archivePrefix = {arXiv},
       eprint = {2308.11812},
 primaryClass = {astro-ph.EP},
       adsurl = {https://ui.adsabs.harvard.edu/abs/2023A&A...678A...2C}
}

@ARTICLE{Cretignier2020b,
       author = {{Cretignier}, M. and {Francfort}, J. and {Dumusque}, X. and {Allart}, R. and {Pepe}, F.},
        title = "{RASSINE: Interactive tool for normalising stellar spectra. I. Description and performance of the code}",
      journal = {\aap},
         year = 2020,
        month = aug,
       volume = {640},
          eid = {A42},
        pages = {A42},
          doi = {10.1051/0004-6361/202037722},
archivePrefix = {arXiv},
       eprint = {2006.13098},
 primaryClass = {astro-ph.SR},
       adsurl = {https://ui.adsabs.harvard.edu/abs/2020A&A...640A..42C}
}

@ARTICLE{Cretignier2020,
       author = {{Cretignier}, M. and {Dumusque}, X. and {Allart}, R. and {Pepe}, F. and {Lovis}, C.},
        title = "{Measuring precise radial velocities on individual spectral lines. II. Dependance of stellar activity signal on line depth}",
      journal = {\aap},
         year = 2020,
        month = jan,
       volume = {633},
          eid = {A76},
        pages = {A76},
          doi = {10.1051/0004-6361/201936548},
archivePrefix = {arXiv},
       eprint = {1912.05192},
 primaryClass = {astro-ph.EP},
       adsurl = {https://ui.adsabs.harvard.edu/abs/2020A&A...633A..76C}
}

@ARTICLE{Mayo2019,
       author = {{Mayo}, Andrew W. and {Rajpaul}, Vinesh M. and {Buchhave}, Lars A. and {Dressing}, Courtney D. and {Mortier}, Annelies and {Zeng}, Li and {Fortenbach}, Charles D. and {Aigrain}, Suzanne and {Bonomo}, Aldo S. and {Collier Cameron}, Andrew and {Charbonneau}, David and {Coffinet}, Adrien and {Cosentino}, Rosario and {Damasso}, Mario and {Dumusque}, Xavier and {Martinez Fiorenzano}, A.~F. and {Haywood}, Rapha{\"e}lle D. and {Latham}, David W. and {L{\'o}pez-Morales}, Mercedes and {Malavolta}, Luca and {Micela}, Giusi and {Molinari}, Emilio and {Pearce}, Logan and {Pepe}, Francesco and {Phillips}, David and {Piotto}, Giampaolo and {Poretti}, Ennio and {Rice}, Ken and {Sozzetti}, Alessandro and {Udry}, Stephane},
        title = "{An 11 Earth-mass, Long-period Sub-Neptune Orbiting a Sun-like Star}",
      journal = {\aj},
         year = 2019,
        month = oct,
       volume = {158},
       number = {4},
          eid = {165},
        pages = {165},
          doi = {10.3847/1538-3881/ab3e2f},
archivePrefix = {arXiv},
       eprint = {1908.08585},
 primaryClass = {astro-ph.EP},
       adsurl = {https://ui.adsabs.harvard.edu/abs/2019AJ....158..165M}
}

@ARTICLE{astropy1,
       author = {{Astropy Collaboration} and {Robitaille}, Thomas P. and {Tollerud}, Erik J. and {Greenfield}, Perry and {Droettboom}, Michael and {Bray}, Erik and {Aldcroft}, Tom and {Davis}, Matt and {Ginsburg}, Adam and {Price-Whelan}, Adrian M. and {Kerzendorf}, Wolfgang E. and {Conley}, Alexander and {Crighton}, Neil and {Barbary}, Kyle and {Muna}, Demitri and {Ferguson}, Henry and {Grollier}, Fr{\'e}d{\'e}ric and {Parikh}, Madhura M. and {Nair}, Prasanth H. and {Unther}, Hans M. and {Deil}, Christoph and {Woillez}, Julien and {Conseil}, Simon and {Kramer}, Roban and {Turner}, James E.~H. and {Singer}, Leo and {Fox}, Ryan and {Weaver}, Benjamin A. and {Zabalza}, Victor and {Edwards}, Zachary I. and {Azalee Bostroem}, K. and {Burke}, D.~J. and {Casey}, Andrew R. and {Crawford}, Steven M. and {Dencheva}, Nadia and {Ely}, Justin and {Jenness}, Tim and {Labrie}, Kathleen and {Lim}, Pey Lian and {Pierfederici}, Francesco and {Pontzen}, Andrew and {Ptak}, Andy and {Refsdal}, Brian and {Servillat}, Mathieu and {Streicher}, Ole},
        title = "{Astropy: A community Python package for astronomy}",
      journal = {\aap},
         year = 2013,
        month = oct,
       volume = {558},
          eid = {A33},
        pages = {A33},
          doi = {10.1051/0004-6361/201322068},
archivePrefix = {arXiv},
       eprint = {1307.6212},
 primaryClass = {astro-ph.IM},
       adsurl = {https://ui.adsabs.harvard.edu/abs/2013A&A...558A..33A}
}

@ARTICLE{astropy2,
       author = {{Astropy Collaboration} and {Price-Whelan}, A.~M. and {Sip{\H{o}}cz}, B.~M. and {G{\"u}nther}, H.~M. and {Lim}, P.~L. and {Crawford}, S.~M. and {Conseil}, S. and {Shupe}, D.~L. and {Craig}, M.~W. and {Dencheva}, N. and {Ginsburg}, A. and {VanderPlas}, J.~T. and {Bradley}, L.~D. and {P{\'e}rez-Su{\'a}rez}, D. and {de Val-Borro}, M. and {Aldcroft}, T.~L. and {Cruz}, K.~L. and {Robitaille}, T.~P. and {Tollerud}, E.~J. and {Ardelean}, C. and {Babej}, T. and {Bach}, Y.~P. and {Bachetti}, M. and {Bakanov}, A.~V. and {Bamford}, S.~P. and {Barentsen}, G. and {Barmby}, P. and {Baumbach}, A. and {Berry}, K.~L. and {Biscani}, F. and {Boquien}, M. and {Bostroem}, K.~A. and {Bouma}, L.~G. and {Brammer}, G.~B. and {Bray}, E.~M. and {Breytenbach}, H. and {Buddelmeijer}, H. and {Burke}, D.~J. and {Calderone}, G. and {Cano Rodr{\'\i}guez}, J.~L. and {Cara}, M. and {Cardoso}, J.~V.~M. and {Cheedella}, S. and {Copin}, Y. and {Corrales}, L. and {Crichton}, D. and {D'Avella}, D. and {Deil}, C. and {Depagne}, {\'E}. and {Dietrich}, J.~P. and {Donath}, A. and {Droettboom}, M. and {Earl}, N. and {Erben}, T. and {Fabbro}, S. and {Ferreira}, L.~A. and {Finethy}, T. and {Fox}, R.~T. and {Garrison}, L.~H. and {Gibbons}, S.~L.~J. and {Goldstein}, D.~A. and {Gommers}, R. and {Greco}, J.~P. and {Greenfield}, P. and {Groener}, A.~M. and {Grollier}, F. and {Hagen}, A. and {Hirst}, P. and {Homeier}, D. and {Horton}, A.~J. and {Hosseinzadeh}, G. and {Hu}, L. and {Hunkeler}, J.~S. and {Ivezi{\'c}}, {\v{Z}}. and {Jain}, A. and {Jenness}, T. and {Kanarek}, G. and {Kendrew}, S. and {Kern}, N.~S. and {Kerzendorf}, W.~E. and {Khvalko}, A. and {King}, J. and {Kirkby}, D. and {Kulkarni}, A.~M. and {Kumar}, A. and {Lee}, A. and {Lenz}, D. and {Littlefair}, S.~P. and {Ma}, Z. and {Macleod}, D.~M. and {Mastropietro}, M. and {McCully}, C. and {Montagnac}, S. and {Morris}, B.~M. and {Mueller}, M. and {Mumford}, S.~J. and {Muna}, D. and {Murphy}, N.~A. and {Nelson}, S. and {Nguyen}, G.~H. and {Ninan}, J.~P. and {N{\"o}the}, M. and {Ogaz}, S. and {Oh}, S. and {Parejko}, J.~K. and {Parley}, N. and {Pascual}, S. and {Patil}, R. and {Patil}, A.~A. and {Plunkett}, A.~L. and {Prochaska}, J.~X. and {Rastogi}, T. and {Reddy Janga}, V. and {Sabater}, J. and {Sakurikar}, P. and {Seifert}, M. and {Sherbert}, L.~E. and {Sherwood-Taylor}, H. and {Shih}, A.~Y. and {Sick}, J. and {Silbiger}, M.~T. and {Singanamalla}, S. and {Singer}, L.~P. and {Sladen}, P.~H. and {Sooley}, K.~A. and {Sornarajah}, S. and {Streicher}, O. and {Teuben}, P. and {Thomas}, S.~W. and {Tremblay}, G.~R. and {Turner}, J.~E.~H. and {Terr{\'o}n}, V. and {van Kerkwijk}, M.~H. and {de la Vega}, A. and {Watkins}, L.~L. and {Weaver}, B.~A. and {Whitmore}, J.~B. and {Woillez}, J. and {Zabalza}, V. and {Astropy Contributors}},
        title = "{The Astropy Project: Building an Open-science Project and Status of the v2.0 Core Package}",
      journal = {\aj},
         year = 2018,
        month = sep,
       volume = {156},
       number = {3},
          eid = {123},
        pages = {123},
          doi = {10.3847/1538-3881/aabc4f},
archivePrefix = {arXiv},
       eprint = {1801.02634},
 primaryClass = {astro-ph.IM},
       adsurl = {https://ui.adsabs.harvard.edu/abs/2018AJ....156..123A}
}

@Article{numpy,
 title         = {Array programming with {NumPy}},
 author        = {Charles R. Harris and K. Jarrod Millman and St{\'{e}}fan J.
                 van der Walt and Ralf Gommers and Pauli Virtanen and David
                 Cournapeau and Eric Wieser and Julian Taylor and Sebastian
                 Berg and Nathaniel J. Smith and Robert Kern and Matti Picus
                 and Stephan Hoyer and Marten H. van Kerkwijk and Matthew
                 Brett and Allan Haldane and Jaime Fern{\'{a}}ndez del
                 R{\'{i}}o and Mark Wiebe and Pearu Peterson and Pierre
                 G{\'{e}}rard-Marchant and Kevin Sheppard and Tyler Reddy and
                 Warren Weckesser and Hameer Abbasi and Christoph Gohlke and
                 Travis E. Oliphant},
 year          = {2020},
 month         = sep,
 journal       = {Nature},
 volume        = {585},
 number        = {7825},
 pages         = {357--362},
 doi           = {10.1038/s41586-020-2649-2},
 publisher     = {Springer Science and Business Media {LLC}},
 url           = {https://doi.org/10.1038/s41586-020-2649-2}
}

@article{Barragan2021b,
	doi = {10.3847/2515-5172/abef70},
	url = {https://doi.org/10.3847/2515-5172/abef70},
	year = 2021,
	month = {mar},
	publisher = {American Astronomical Society},
	volume = {5},
	number = {3},
	pages = {51},
	author = {Oscar Barrag{\'{a}}n and Suzanne Aigrain and Edward Gillen and Fernando Guti{\'{e}}rrez-Canales},
	title = {{TESS} Re-observes the Young Multi-planet System {TOI}-451: Refined Ephemeris and Activity Evolution},
	journal = {Research Notes of the {AAS}}
}

@ARTICLE{Queloz2001,
       author = {{Queloz}, D. and {Henry}, G.~W. and {Sivan}, J.~P. and {Baliunas}, S.~L. and {Beuzit}, J.~L. and {Donahue}, R.~A. and {Mayor}, M. and {Naef}, D. and {Perrier}, C. and {Udry}, S.},
        title = "{No planet for HD 166435}",
      journal = {\aap},
         year = 2001,
        month = nov,
       volume = {379},
        pages = {279-287},
          doi = {10.1051/0004-6361:20011308},
archivePrefix = {arXiv},
       eprint = {astro-ph/0109491},
 primaryClass = {astro-ph},
       adsurl = {https://ui.adsabs.harvard.edu/abs/2001A&A...379..279Q}
}

@ARTICLE{Zhang2022,
       author = {{Zhang}, Michael and {Knutson}, Heather A. and {Wang}, Lile and {Dai}, Fei and {Barrag{\'a}n}, Oscar},
        title = "{Escaping Helium from TOI 560.01, a Young Mini-Neptune}",
      journal = {\aj},
         year = 2022,
        month = feb,
       volume = {163},
       number = {2},
          eid = {67},
        pages = {67},
          doi = {10.3847/1538-3881/ac3fa7},
archivePrefix = {arXiv},
       eprint = {2110.13150},
 primaryClass = {astro-ph.EP},
       adsurl = {https://ui.adsabs.harvard.edu/abs/2022AJ....163...67Z}
}

@ARTICLE{dynesty,
       author = {{Speagle}, Joshua S.},
        title = "{DYNESTY: a dynamic nested sampling package for estimating Bayesian posteriors and evidences}",
      journal = {\mnras},
         year = 2020,
        month = apr,
       volume = {493},
       number = {3},
        pages = {3132-3158},
          doi = {10.1093/mnras/staa278},
archivePrefix = {arXiv},
       eprint = {1904.02180},
 primaryClass = {astro-ph.IM},
       adsurl = {https://ui.adsabs.harvard.edu/abs/2020MNRAS.493.3132S}
}

@ARTICLE{Gandolfi2018,
       author = {{Gandolfi}, D. and {Barrag{\'a}n}, O. and {Livingston}, J.~H. and {Fridlund}, M. and {Justesen}, A.~B. and {Redfield}, S. and {Fossati}, L. and {Mathur}, S. and {Grziwa}, S. and {Cabrera}, J. and {Garc{\'\i}a}, R.~A. and {Persson}, C.~M. and {Van Eylen}, V. and {Hatzes}, A.~P. and {Hidalgo}, D. and {Albrecht}, S. and {Bugnet}, L. and {Cochran}, W.~D. and {Csizmadia}, Sz. and {Deeg}, H. and {Eigm{\"u}ller}, Ph. and {Endl}, M. and {Erikson}, A. and {Esposito}, M. and {Guenther}, E. and {Korth}, J. and {Luque}, R. and {Monta{\~n}es Rodr{\'\i}guez}, P. and {Nespral}, D. and {Nowak}, G. and {P{\"a}tzold}, M. and {Prieto-Arranz}, J.},
        title = "{TESS's first planet. A super-Earth transiting the naked-eye star {\ensuremath{\pi}} Mensae}",
      journal = {\aap},
         year = 2018,
        month = nov,
       volume = {619},
          eid = {L10},
        pages = {L10},
          doi = {10.1051/0004-6361/201834289},
archivePrefix = {arXiv},
       eprint = {1809.07573},
 primaryClass = {astro-ph.EP},
       adsurl = {https://ui.adsabs.harvard.edu/abs/2018A&A...619L..10G}
}

@ARTICLE{Fulton2017,
       author = {{Fulton}, Benjamin J. and {Petigura}, Erik A. and {Howard}, Andrew W. and {Isaacson}, Howard and {Marcy}, Geoffrey W. and {Cargile}, Phillip A. and {Hebb}, Leslie and {Weiss}, Lauren M. and {Johnson}, John Asher and {Morton}, Timothy D. and {Sinukoff}, Evan and {Crossfield}, Ian J.~M. and {Hirsch}, Lea A.},
        title = "{The California-Kepler Survey. III. A Gap in the Radius Distribution of Small Planets}",
      journal = {\aj},
         year = 2017,
        month = sep,
       volume = {154},
       number = {3},
          eid = {109},
        pages = {109},
          doi = {10.3847/1538-3881/aa80eb},
archivePrefix = {arXiv},
       eprint = {1703.10375},
 primaryClass = {astro-ph.EP},
       adsurl = {https://ui.adsabs.harvard.edu/abs/2017AJ....154..109F}
}

@ARTICLE{VanEylen2018,
       author = {{Van Eylen}, V. and {Agentoft}, Camilla and {Lundkvist}, M.~S. and {Kjeldsen}, H. and {Owen}, J.~E. and {Fulton}, B.~J. and {Petigura}, E. and {Snellen}, I.},
        title = "{An asteroseismic view of the radius valley: stripped cores, not born rocky}",
      journal = {\mnras},
         year = 2018,
        month = oct,
       volume = {479},
       number = {4},
        pages = {4786-4795},
          doi = {10.1093/mnras/sty1783},
archivePrefix = {arXiv},
       eprint = {1710.05398},
 primaryClass = {astro-ph.EP},
       adsurl = {https://ui.adsabs.harvard.edu/abs/2018MNRAS.479.4786V}
}

@ARTICLE{ariadne,
       author = {{Vines}, Jose I. and {Jenkins}, James S.},
        title = "{ARIADNE: measuring accurate and precise stellar parameters through SED fitting}",
      journal = {\mnras},
         year = 2022,
        month = jun,
       volume = {513},
       number = {2},
        pages = {2719-2731},
          doi = {10.1093/mnras/stac956},
archivePrefix = {arXiv},
       eprint = {2204.03769},
 primaryClass = {astro-ph.SR},
       adsurl = {https://ui.adsabs.harvard.edu/abs/2022MNRAS.513.2719V}
}

@ARTICLE{Beard2025,
       author = {{Beard}, Corey and {Robertson}, Paul and {Lubin}, Jack and {Han}, Te and {Holcomb}, Rae and {Premnath}, Pranav and {Butler}, R. Paul and {Dalba}, Paul A. and {Holden}, Brad and {Blake}, Cullen H. and {Diddams}, Scott A. and {Gupta}, Arvind F. and {Halverson}, Samuel and {Krolikowski}, Daniel M. and {Li}, Dan and {Lin}, Andrea S.~J. and {Logsdon}, Sarah E. and {Lubar}, Emily and {Mahadevan}, Suvrath and {McElwain}, Michael W. and {Ninan}, Joe P. and {Paredes}, Leonardo A. and {Roy}, Arpita and {Schwab}, Christian and {Stefansson}, Gudmundur and {Terrien}, Ryan C. and {Wright}, Jason T.},
        title = "{Jitter Across 15 yr: Leveraging Precise Photometry from Kepler and TESS to Extract Exoplanets from Radial Velocity Time Series}",
      journal = {\aj},
         year = 2025,
        month = feb,
       volume = {169},
       number = {2},
          eid = {92},
        pages = {92},
          doi = {10.3847/1538-3881/ad9eb0},
archivePrefix = {arXiv},
       eprint = {2412.11329},
 primaryClass = {astro-ph.EP},
       adsurl = {https://ui.adsabs.harvard.edu/abs/2025AJ....169...92B}
}

@ARTICLE{republic,
       author = {{Barrag{\'a}n}, Oscar and {Aigrain}, Suzanne and {McCormac}, James},
        title = "{REPUBLIC: A variability-preserving systematic-correction algorithm for PLATO's multi-camera light curves}",
      journal = {RAS Techniques and Instruments},
         year = 2024,
        month = jan,
       volume = {3},
       number = {1},
        pages = {198-208},
          doi = {10.1093/rasti/rzae014},
archivePrefix = {arXiv},
       eprint = {2404.06132},
 primaryClass = {astro-ph.IM},
       adsurl = {https://ui.adsabs.harvard.edu/abs/2024RASTI...3..198B}
}

@ARTICLE{Ginzburg2016,
       author = {{Ginzburg}, Sivan and {Schlichting}, Hilke E. and {Sari}, Re'em},
        title = "{Super-Earth Atmospheres: Self-consistent Gas Accretion and Retention}",
      journal = {\apj},
         year = 2016,
        month = jul,
       volume = {825},
       number = {1},
          eid = {29},
        pages = {29},
          doi = {10.3847/0004-637X/825/1/29},
archivePrefix = {arXiv},
       eprint = {1512.07925},
 primaryClass = {astro-ph.EP},
       adsurl = {https://ui.adsabs.harvard.edu/abs/2016ApJ...825...29G}
}

@ARTICLE{Gregory2005,
       author = {{Gregory}, P.~C.},
        title = "{A Bayesian Analysis of Extrasolar Planet Data for HD 73526}",
      journal = {\apj},
         year = 2005,
        month = oct,
       volume = {631},
       number = {2},
        pages = {1198-1214},
          doi = {10.1086/432594},
       adsurl = {https://ui.adsabs.harvard.edu/abs/2005ApJ...631.1198G}
}

@ARTICLE{Kempton2018,
       author = {{Kempton}, Eliza M. -R. and {Bean}, Jacob L. and {Louie}, Dana R. and {Deming}, Drake and {Koll}, Daniel D.~B. and {Mansfield}, Megan and {Christiansen}, Jessie L. and {L{\'o}pez-Morales}, Mercedes and {Swain}, Mark R. and {Zellem}, Robert T. and {Ballard}, Sarah and {Barclay}, Thomas and {Barstow}, Joanna K. and {Batalha}, Natasha E. and {Beatty}, Thomas G. and {Berta-Thompson}, Zach and {Birkby}, Jayne and {Buchhave}, Lars A. and {Charbonneau}, David and {Cowan}, Nicolas B. and {Crossfield}, Ian and {de Val-Borro}, Miguel and {Doyon}, Ren{\'e} and {Dragomir}, Diana and {Gaidos}, Eric and {Heng}, Kevin and {Hu}, Renyu and {Kane}, Stephen R. and {Kreidberg}, Laura and {Mallonn}, Matthias and {Morley}, Caroline V. and {Narita}, Norio and {Nascimbeni}, Valerio and {Pall{\'e}}, Enric and {Quintana}, Elisa V. and {Rauscher}, Emily and {Seager}, Sara and {Shkolnik}, Evgenya L. and {Sing}, David K. and {Sozzetti}, Alessandro and {Stassun}, Keivan G. and {Valenti}, Jeff A. and {von Essen}, Carolina},
        title = "{A Framework for Prioritizing the TESS Planetary Candidates Most Amenable to Atmospheric Characterization}",
      journal = {\pasp},
         year = 2018,
        month = nov,
       volume = {130},
       number = {993},
        pages = {114401},
          doi = {10.1088/1538-3873/aadf6f},
archivePrefix = {arXiv},
       eprint = {1805.03671},
 primaryClass = {astro-ph.EP},
       adsurl = {https://ui.adsabs.harvard.edu/abs/2018PASP..130k4401K}
}

@ARTICLE{OBrien2022,
       author = {{O'Brien}, Sean M. and {Bayliss}, Daniel and {Osborn}, James and {Bryant}, Edward M. and {McCormac}, James and {Wheatley}, Peter J. and {Acton}, Jack S. and {Alves}, Douglas R. and {Anderson}, David R. and {Burleigh}, Matthew R. and {Casewell}, Sarah L. and {Gill}, Samuel and {Goad}, Michael R. and {Henderson}, Beth A. and {Jackman}, James A.~G. and {Lendl}, Monika and {Tilbrook}, Rosanna H. and {Udry}, St{\'e}phane and {Vines}, Jose I. and {West}, Richard G.},
        title = "{Scintillation-limited photometry with the 20-cm NGTS telescopes at Paranal Observatory}",
      journal = {\mnras},
         year = 2022,
        month = feb,
       volume = {509},
       number = {4},
        pages = {6111-6118},
          doi = {10.1093/mnras/stab3399},
archivePrefix = {arXiv},
       eprint = {2111.10321},
 primaryClass = {astro-ph.IM},
       adsurl = {https://ui.adsabs.harvard.edu/abs/2022MNRAS.509.6111O}
}

@ARTICLE{Ahrer2023wasp39,
       author = {{Ahrer}, Eva-Maria and {Stevenson}, Kevin B. and {Mansfield}, Megan and {Moran}, Sarah E. and {Brande}, Jonathan and {Morello}, Giuseppe and {Murray}, Catriona A. and {Nikolov}, Nikolay K. and {Petit dit de la Roche}, Dominique J.~M. and {Schlawin}, Everett and {Wheatley}, Peter J. and {Zieba}, Sebastian and {Batalha}, Natasha E. and {Damiano}, Mario and {Goyal}, Jayesh M. and {Lendl}, Monika and {Lothringer}, Joshua D. and {Mukherjee}, Sagnick and {Ohno}, Kazumasa and {Batalha}, Natalie M. and {Battley}, Matthew P. and {Bean}, Jacob L. and {Beatty}, Thomas G. and {Benneke}, Bj{\"o}rn and {Berta-Thompson}, Zachory K. and {Carter}, Aarynn L. and {Cubillos}, Patricio E. and {Daylan}, Tansu and {Espinoza}, N{\'e}stor and {Gao}, Peter and {Gibson}, Neale P. and {Gill}, Samuel and {Harrington}, Joseph and {Hu}, Renyu and {Kreidberg}, Laura and {Lewis}, Nikole K. and {Line}, Michael R. and {L{\'o}pez-Morales}, Mercedes and {Parmentier}, Vivien and {Powell}, Diana K. and {Sing}, David K. and {Tsai}, Shang-Min and {Wakeford}, Hannah R. and {Welbanks}, Luis and {Alam}, Munazza K. and {Alderson}, Lili and {Allen}, Natalie H. and {Anderson}, David R. and {Barstow}, Joanna K. and {Bayliss}, Daniel and {Bell}, Taylor J. and {Blecic}, Jasmina and {Bryant}, Edward M. and {Burleigh}, Matthew R. and {Carone}, Ludmila and {Casewell}, S.~L. and {Changeat}, Quentin and {Chubb}, Katy L. and {Crossfield}, Ian J.~M. and {Crouzet}, Nicolas and {Decin}, Leen and {D{\'e}sert}, Jean-Michel and {Feinstein}, Adina D. and {Flagg}, Laura and {Fortney}, Jonathan J. and {Gizis}, John E. and {Heng}, Kevin and {Iro}, Nicolas and {Kempton}, Eliza M. -R. and {Kendrew}, Sarah and {Kirk}, James and {Knutson}, Heather A. and {Komacek}, Thaddeus D. and {Lagage}, Pierre-Olivier and {Leconte}, J{\'e}r{\'e}my and {Lustig-Yaeger}, Jacob and {MacDonald}, Ryan J. and {Mancini}, Luigi and {May}, E.~M. and {Mayne}, N.~J. and {Miguel}, Yamila and {Mikal-Evans}, Thomas and {Molaverdikhani}, Karan and {Palle}, Enric and {Piaulet}, Caroline and {Rackham}, Benjamin V. and {Redfield}, Seth and {Rogers}, Laura K. and {Roy}, Pierre-Alexis and {Rustamkulov}, Zafar and {Shkolnik}, Evgenya L. and {Sotzen}, Kristin S. and {Taylor}, Jake and {Tremblin}, P. and {Tucker}, Gregory S. and {Turner}, Jake D. and {de Val-Borro}, Miguel and {Venot}, Olivia and {Zhang}, Xi},
        title = "{Early Release Science of the exoplanet WASP-39b with JWST NIRCam}",
      journal = {\nat},
         year = 2023,
        month = feb,
       volume = {614},
       number = {7949},
        pages = {653-658},
          doi = {10.1038/s41586-022-05590-4},
archivePrefix = {arXiv},
       eprint = {2211.10489},
 primaryClass = {astro-ph.EP},
       adsurl = {https://ui.adsabs.harvard.edu/abs/2023Natur.614..653A}
}

@ARTICLE{wheatley18ngts,
       author = {{Wheatley}, Peter J. and {West}, Richard G. and {Goad}, Michael R. and
         {Jenkins}, James S. and {Pollacco}, Don L. and {Queloz}, Didier and
         {Rauer}, Heike and {Udry}, St{\'e}phane and {Watson}, Christopher A. and
         {Chazelas}, Bruno and {Eigm{\"u}ller}, Philipp and {Lambert}, Gregory and
         {Genolet}, Ludovic and {McCormac}, James and {Walker}, Simon and
         {Armstrong}, David J. and {Bayliss}, Daniel and {Bento}, Joao and
         {Bouchy}, Fran{\c{c}}ois and {Burleigh}, Matthew R. and
         {Cabrera}, Juan and {Casewell}, Sarah L. and {Chaushev}, Alexander and
         {Chote}, Paul and {Csizmadia}, Szil{\'a}rd and {Erikson}, Anders and
         {Faedi}, Francesca and {Foxell}, Emma and {G{\"a}nsicke}, Boris T. and
         {Gillen}, Edward and {Grange}, Andrew and {G{\"u}nther}, Maximilian N. and
         {Hodgkin}, Simon T. and {Jackman}, James and {Jord{\'a}n}, Andr{\'e}s and
         {Louden}, Tom and {Metrailler}, Lionel and {Moyano}, Maximiliano and
         {Nielsen}, Louise D. and {Osborn}, Hugh P. and {Poppenhaeger}, Katja and
         {Raddi}, Roberto and {Raynard}, Liam and {Smith}, Alexis M.~S. and
         {Soto}, Maritza and {Titz-Weider}, Ruth},
        title = "{The Next Generation Transit Survey (NGTS)}",
      journal = {\mnras},
         year = 2018,
        month = apr,
       volume = {475},
       number = {4},
        pages = {4476-4493},
          doi = {10.1093/mnras/stx2836},
archivePrefix = {arXiv},
       eprint = {1710.11100},
 primaryClass = {astro-ph.EP},
       adsurl = {https://ui.adsabs.harvard.edu/abs/2018MNRAS.475.4476W}
}

@ARTICLE{bryant20multicam,
       author = {{Bryant}, Edward M. and {Bayliss}, Daniel and {McCormac}, James and
         {Wheatley}, Peter J. and {Acton}, Jack S. and {Anderson}, David R. and
         {Armstrong}, David J. and {Bouchy}, Fran{\c{c}}ois and
         {Belardi}, Claudia and {Burleigh}, Matthew R. and {Tilbrook}, Rosie H. and
         {Casewell}, Sarah L. and {Cooke}, Benjamin F. and {Gill}, Samuel and
         {Goad}, Michael R. and {Jenkins}, James S. and {Lendl}, Monika and
         {Pollacco}, Don and {Queloz}, Didier and {Raynard}, Liam and
         {Smith}, Alexis M.~S. and {Vines}, Jose I. and {West}, Richard G. and
         {Udry}, Stephane},
        title = "{Simultaneous TESS and NGTS transit observations of WASP-166 b}",
      journal = {\mnras},
         year = 2020,
        month = apr,
       volume = {494},
       number = {4},
        pages = {5872-5881},
          doi = {10.1093/mnras/staa1075},
archivePrefix = {arXiv},
       eprint = {2004.07589},
 primaryClass = {astro-ph.EP},
       adsurl = {https://ui.adsabs.harvard.edu/abs/2020MNRAS.494.5872B}
}

@ARTICLE{gray1994,
       author = {{Gray}, R.~O. and {Corbally}, C.~J.},
        title = "{The Calibration of MK Spectral Classes Using Spectral Synthesis. I. The Effective Temperature Calibration of Dwarf Stars}",
      journal = {\aj},
         year = 1994,
        month = feb,
       volume = {107},
        pages = {742},
          doi = {10.1086/116893},
       adsurl = {https://ui.adsabs.harvard.edu/abs/1994AJ....107..742G}
}

@MISC{isochrones,
       author = {{Morton}, Timothy D.},
        title = "{isochrones: Stellar model grid package}",
         year = 2015,
        month = mar,
          eid = {ascl:1503.010},
        pages = {ascl:1503.010},
archivePrefix = {ascl},
       eprint = {1503.010},
       adsurl = {https://ui.adsabs.harvard.edu/abs/2015ascl.soft03010M}
}

@MISC{Feinstein2024,
       author = {{Feinstein}, Adina and {Welbanks}, Luis and {Ahrer}, Eva-Maria and {Alderson}, Lili and {Barat}, Saugata and {Brande}, Jonathan and {Crossfield}, Ian and {Desert}, Jean-Michel and {Duvvuri}, Girish M. and {Espinoza}, Nestor and {France}, Kevin and {Gao}, Peter and {Guzman Caloca}, Giannina and {Levine}, Garrett and {Livingston}, John and {Lunine}, Jonathan I. and {Luque}, Rafael and {Mann}, Andrew Withycombe and {Mukherjee}, Sagnick and {Murray}, Catriona Anne and {Owen}, James Edward and {Rackham}, Benjamin and {Radica}, Michael and {Rockcliffe}, Keighley Elizabeth and {Rogers}, Leslie and {Seager}, Sara and {Seligman}, Darryl and {Shapiro}, Alexander I. and {Thao}, Pa Chia and {Vissapragada}, Shreyas},
        title = "{KRONOS: Keys to Revealing the Origin and Nature Of sub-neptune Systems}",
 howpublished = {JWST Proposal. Cycle 3, ID. \#5959},
         year = 2024,
        month = feb,
        pages = {5959},
       adsurl = {https://ui.adsabs.harvard.edu/abs/2024jwst.prop.5959F}
}

@ARTICLE{Rajpaul2024,
       author = {{Rajpaul}, Vinesh M. and {Barrag{\'a}n}, Oscar and {Zicher}, Norbert},
        title = "{A non-zero Doppler amplitude is not enough: revisiting the putative radial velocity detection of sub-Venus exoplanet L 98-59b}",
      journal = {\mnras},
         year = 2024,
        month = jun,
       volume = {530},
       number = {4},
        pages = {4665-4675},
          doi = {10.1093/mnras/stae778},
       adsurl = {https://ui.adsabs.harvard.edu/abs/2024MNRAS.530.4665R}
}

@ARTICLE{Rogers2021,
       author = {{Rogers}, James G. and {Owen}, James E.},
        title = "{Unveiling the planet population at birth}",
      journal = {\mnras},
         year = 2021,
        month = may,
       volume = {503},
       number = {1},
        pages = {1526-1542},
          doi = {10.1093/mnras/stab529},
archivePrefix = {arXiv},
       eprint = {2007.11006},
 primaryClass = {astro-ph.EP},
       adsurl = {https://ui.adsabs.harvard.edu/abs/2021MNRAS.503.1526R}
}

@ARTICLE{Newton2021,
       author = {{Newton}, Elisabeth R. and {Mann}, Andrew W. and {Kraus}, Adam L. and {Livingston}, John H. and {Vanderburg}, Andrew and {Curtis}, Jason L. and {Thao}, Pa Chia and {Hawkins}, Keith and {Wood}, Mackenna L. and {Rizzuto}, Aaron C. and {Soubkiou}, Abderahmane and {Tofflemire}, Benjamin M. and {Zhou}, George and {Crossfield}, Ian J.~M. and {Pearce}, Logan A. and {Collins}, Karen A. and {Conti}, Dennis M. and {Tan}, Thiam-Guan and {Villeneuva}, Steven and {Spencer}, Alton and {Dragomir}, Diana and {Quinn}, Samuel N. and {Jensen}, Eric L.~N. and {Collins}, Kevin I. and {Stockdale}, Chris and {Cloutier}, Ryan and {Hellier}, Coel and {Benkhaldoun}, Zouhair and {Ziegler}, Carl and {Brice{\~n}o}, C{\'e}sar and {Law}, Nicholas and {Benneke}, Bj{\"o}rn and {Christiansen}, Jessie L. and {Gorjian}, Varoujan and {Kane}, Stephen R. and {Kreidberg}, Laura and {Morales}, Farisa Y. and {Werner}, Michael W. and {Twicken}, Joseph D. and {Levine}, Alan M. and {Ciardi}, David R. and {Guerrero}, Natalia M. and {Hesse}, Katharine and {Quintana}, Elisa V. and {Shiao}, Bernie and {Smith}, Jeffrey C. and {Torres}, Guillermo and {Ricker}, George R. and {Vanderspek}, Roland and {Seager}, Sara and {Winn}, Joshua N. and {Jenkins}, Jon M. and {Latham}, David W.},
        title = "{TESS Hunt for Young and Maturing Exoplanets (THYME). IV. Three Small Planets Orbiting a 120 Myr Old Star in the Pisces-Eridanus Stream}",
      journal = {\aj},
         year = 2021,
        month = feb,
       volume = {161},
       number = {2},
          eid = {65},
        pages = {65},
          doi = {10.3847/1538-3881/abccc6},
archivePrefix = {arXiv},
       eprint = {2102.06049},
 primaryClass = {astro-ph.EP},
       adsurl = {https://ui.adsabs.harvard.edu/abs/2021AJ....161...65N}
}

@ARTICLE{Pepe2021,
       author = {{Pepe}, F. and {Cristiani}, S. and {Rebolo}, R. and {Santos}, N.~C. and {Dekker}, H. and {Cabral}, A. and {Di Marcantonio}, P. and {Figueira}, P. and {Lo Curto}, G. and {Lovis}, C. and {Mayor}, M. and {M{\'e}gevand}, D. and {Molaro}, P. and {Riva}, M. and {Zapatero Osorio}, M.~R. and {Amate}, M. and {Manescau}, A. and {Pasquini}, L. and {Zerbi}, F.~M. and {Adibekyan}, V. and {Abreu}, M. and {Affolter}, M. and {Alibert}, Y. and {Aliverti}, M. and {Allart}, R. and {Allende Prieto}, C. and {{\'A}lvarez}, D. and {Alves}, D. and {Avila}, G. and {Baldini}, V. and {Bandy}, T. and {Barros}, S.~C.~C. and {Benz}, W. and {Bianco}, A. and {Borsa}, F. and {Bourrier}, V. and {Bouchy}, F. and {Broeg}, C. and {Calderone}, G. and {Cirami}, R. and {Coelho}, J. and {Conconi}, P. and {Coretti}, I. and {Cumani}, C. and {Cupani}, G. and {D'Odorico}, V. and {Damasso}, M. and {Deiries}, S. and {Delabre}, B. and {Demangeon}, O.~D.~S. and {Dumusque}, X. and {Ehrenreich}, D. and {Faria}, J.~P. and {Fragoso}, A. and {Genolet}, L. and {Genoni}, M. and {G{\'e}nova Santos}, R. and {Gonz{\'a}lez Hern{\'a}ndez}, J.~I. and {Hughes}, I. and {Iwert}, O. and {Kerber}, F. and {Knudstrup}, J. and {Landoni}, M. and {Lavie}, B. and {Lillo-Box}, J. and {Lizon}, J. -L. and {Maire}, C. and {Martins}, C.~J.~A.~P. and {Mehner}, A. and {Micela}, G. and {Modigliani}, A. and {Monteiro}, M.~A. and {Monteiro}, M.~J.~P.~F.~G. and {Moschetti}, M. and {Murphy}, M.~T. and {Nunes}, N. and {Oggioni}, L. and {Oliveira}, A. and {Oshagh}, M. and {Pall{\'e}}, E. and {Pariani}, G. and {Poretti}, E. and {Rasilla}, J.~L. and {Rebord{\~a}o}, J. and {Redaelli}, E.~M. and {Santana Tschudi}, S. and {Santin}, P. and {Santos}, P. and {S{\'e}gransan}, D. and {Schmidt}, T.~M. and {Segovia}, A. and {Sosnowska}, D. and {Sozzetti}, A. and {Sousa}, S.~G. and {Span{\`o}}, P. and {Su{\'a}rez Mascare{\~n}o}, A. and {Tabernero}, H. and {Tenegi}, F. and {Udry}, S. and {Zanutta}, A.},
        title = "{ESPRESSO at VLT. On-sky performance and first results}",
      journal = {\aap},
         year = 2021,
        month = jan,
       volume = {645},
          eid = {A96},
        pages = {A96},
          doi = {10.1051/0004-6361/202038306},
archivePrefix = {arXiv},
       eprint = {2010.00316},
 primaryClass = {astro-ph.IM},
       adsurl = {https://ui.adsabs.harvard.edu/abs/2021A&A...645A..96P}
}

@ARTICLE{Eschen2024,
       author = {{Eschen}, Yoshi Nike Emilia and {Bayliss}, Daniel and {Wilson}, Thomas G. and {Kunimoto}, Michelle and {Pelisoli}, Ingrid and {Rodel}, Toby},
        title = "{Viewing the PLATO LOPS2 field through the lenses of TESS}",
      journal = {\mnras},
         year = 2024,
        month = dec,
       volume = {535},
       number = {2},
        pages = {1778-1795},
          doi = {10.1093/mnras/stae2427},
archivePrefix = {arXiv},
       eprint = {2409.13039},
 primaryClass = {astro-ph.EP},
       adsurl = {https://ui.adsabs.harvard.edu/abs/2024MNRAS.535.1778E}
}

@ARTICLE{Gonzalez2024,
       author = {{Gonz{\'a}lez Hern{\'a}ndez}, J.~I. and {Su{\'a}rez Mascare{\~n}o}, A. and {Silva}, A.~M. and {Stefanov}, A.~K. and {Faria}, J.~P. and {Tabernero}, H.~M. and {Sozzetti}, A. and {Rebolo}, R. and {Pepe}, F. and {Santos}, N.~C. and {Cristiani}, S. and {Lovis}, C. and {Dumusque}, X. and {Figueira}, P. and {Lillo-Box}, J. and {Nari}, N. and {Benatti}, S. and {Hobson}, M.~J. and {Castro-Gonz{\'a}lez}, A. and {Allart}, R. and {Passegger}, V.~M. and {Zapatero Osorio}, M. -R. and {Adibekyan}, V. and {Alibert}, Y. and {Allende Prieto}, C. and {Bouchy}, F. and {Damasso}, M. and {D'Odorico}, V. and {Di Marcantonio}, P. and {Ehrenreich}, D. and {Lo Curto}, G. and {Santos}, R. G{\'e}nova and {Martins}, C.~J.~A.~P. and {Mehner}, A. and {Micela}, G. and {Molaro}, P. and {Nunes}, N. and {Palle}, E. and {Sousa}, S.~G. and {Udry}, S.},
        title = "{A sub-Earth-mass planet orbiting Barnard's star}",
      journal = {\aap},
         year = 2024,
        month = oct,
       volume = {690},
          eid = {A79},
        pages = {A79},
          doi = {10.1051/0004-6361/202451311},
archivePrefix = {arXiv},
       eprint = {2410.00569},
 primaryClass = {astro-ph.EP},
       adsurl = {https://ui.adsabs.harvard.edu/abs/2024A&A...690A..79G}
}

@ARTICLE{Barragan2024,
       author = {{Barrag{\'a}n}, Oscar and {Yu}, Haochuan and {Freckelton}, Alix Violet and {Meech}, Annabella and {Cretignier}, Michael and {Mortier}, Annelies and {Aigrain}, Suzanne and {Klein}, Baptiste and {O'Sullivan}, Niamh K. and {Gillen}, Edward and {Nielsen}, Louise Dyregaard and {Mallorqu{\'\i}n}, Manuel and {Zicher}, Norbert},
        title = "{TOI-837 b is a young Saturn-sized exoplanet with a massive 70 M$_{{\ensuremath{\oplus}}}$ core}",
      journal = {\mnras},
         year = 2024,
        month = jul,
       volume = {531},
       number = {4},
        pages = {4275-4292},
          doi = {10.1093/mnras/stae1344},
archivePrefix = {arXiv},
       eprint = {2404.13750},
 primaryClass = {astro-ph.EP},
       adsurl = {https://ui.adsabs.harvard.edu/abs/2024MNRAS.531.4275B}
}

@ARTICLE{Suarez2023,
       author = {{Su{\'a}rez Mascare{\~n}o}, A. and {Gonz{\'a}lez-{\'A}lvarez}, E. and {Zapatero Osorio}, M.~R. and {Lillo-Box}, J. and {Faria}, J.~P. and {Passegger}, V.~M. and {Gonz{\'a}lez Hern{\'a}ndez}, J.~I. and {Figueira}, P. and {Sozzetti}, A. and {Rebolo}, R. and {Pepe}, F. and {Santos}, N.~C. and {Cristiani}, S. and {Lovis}, C. and {Silva}, A.~M. and {Ribas}, I. and {Amado}, P.~J. and {Caballero}, J.~A. and {Quirrenbach}, A. and {Reiners}, A. and {Zechmeister}, M. and {Adibekyan}, V. and {Alibert}, Y. and {B{\'e}jar}, V.~J.~S. and {Benatti}, S. and {D'Odorico}, V. and {Damasso}, M. and {Delisle}, J. -B. and {Di Marcantonio}, P. and {Dreizler}, S. and {Ehrenreich}, D. and {Hatzes}, A.~P. and {Hara}, N.~C. and {Henning}, Th. and {Kaminski}, A. and {L{\'o}pez-Gonz{\'a}lez}, M.~J. and {Martins}, C.~J.~A.~P. and {Micela}, G. and {Montes}, D. and {Pall{\'e}}, E. and {Pedraz}, S. and {Rodr{\'\i}guez}, E. and {Rodr{\'\i}guez-L{\'o}pez}, C. and {Tal-Or}, L. and {Sousa}, S. and {Udry}, S.},
        title = "{Two temperate Earth-mass planets orbiting the nearby star GJ 1002}",
      journal = {\aap},
         year = 2023,
        month = feb,
       volume = {670},
          eid = {A5},
        pages = {A5},
          doi = {10.1051/0004-6361/202244991},
archivePrefix = {arXiv},
       eprint = {2212.07332},
 primaryClass = {astro-ph.EP},
       adsurl = {https://ui.adsabs.harvard.edu/abs/2023A&A...670A...5S}
}

@ARTICLE{Fridlund2024,
       author = {{Fridlund}, M. and {Georgieva}, I.~Y. and {Bonfanti}, A. and {Alibert}, Y. and {Persson}, C.~M. and {Gandolfi}, D. and {Beck}, M. and {Deline}, A. and {Hoyer}, S. and {Olofsson}, G. and {Wilson}, T.~G. and {Barrag{\'a}n}, O. and {Fossati}, L. and {Mustill}, A.~J. and {Brandeker}, A. and {Hatzes}, A. and {Flor{\'e}n}, H. -G. and {Simola}, U. and {Hooton}, M.~J. and {Luque}, R. and {Sousa}, S.~G. and {Egger}, J.~A. and {Antoniadis-Karnavas}, A. and {Salmon}, S. and {Adibekyan}, V. and {Alonso}, R. and {Anglada}, G. and {B{\'a}rczy}, T. and {Barrado Navascues}, D. and {Barros}, S.~C.~C. and {Baumjohann}, W. and {Beck}, T. and {Benz}, W. and {Bonfils}, X. and {Broeg}, C. and {Cabrera}, J. and {Charnoz}, S. and {Collier Cameron}, A. and {Csizmadia}, Sz. and {Davies}, M.~B. and {Deeg}, H. and {Deleuil}, M. and {Delrez}, L. and {Demangeon}, O.~D.~S. and {Demory}, B. -O. and {Ehrenreich}, D. and {Erikson}, A. and {Esposito}, M. and {Fortier}, A. and {Gillon}, M. and {G{\"u}del}, M. and {Heng}, K. and {Isaak}, K.~G. and {Kiss}, L.~L. and {Korth}, J. and {Laskar}, J. and {Lecavelier des Etangs}, A. and {Lendl}, M. and {Livingston}, J. and {Lovis}, C. and {Magrin}, D. and {Maxted}, P.~F.~L. and {Muresan}, A. and {Nascimbeni}, V. and {Ottensamer}, R. and {Pagano}, I. and {Pall{\'e}}, E. and {Peter}, G. and {Piotto}, G. and {Pollacco}, D. and {Queloz}, D. and {Ragazzoni}, R. and {Rando}, N. and {Rauer}, H. and {Redfield}, S. and {Ribas}, I. and {Santos}, N.~C. and {Scandariato}, G. and {S{\'e}gransan}, D. and {Serrano}, L.~M. and {Simon}, A.~E. and {Smith}, A.~M.~S. and {Steller}, M. and {Szab{\'o}}, Gy. M. and {Thomas}, N. and {Udry}, S. and {Van Eylen}, V. and {Van Grootel}, V. and {Walton}, N.~A.},
        title = "{Planets observed with CHEOPS. Two super-Earths orbiting the red dwarf star TOI-776}",
      journal = {\aap},
         year = 2024,
        month = apr,
       volume = {684},
          eid = {A12},
        pages = {A12},
          doi = {10.1051/0004-6361/202243838},
       adsurl = {https://ui.adsabs.harvard.edu/abs/2024A&A...684A..12F}
}

@ARTICLE{Rauer2024,
       author = {{Rauer}, Heike and {Aerts}, Conny and {Cabrera}, Juan and {Deleuil}, Magali and {Erikson}, Anders and {Gizon}, Laurent and {Goupil}, Mariejo and {Heras}, Ana and {Lorenzo-Alvarez}, Jose and {Marliani}, Filippo and {Martin-Garcia}, C{\'e}sar and {Mas-Hesse}, J. Miguel and {O'Rourke}, Laurence and {Osborn}, Hugh and {Pagano}, Isabella and {Piotto}, Giampaolo and {Pollacco}, Don and {Ragazzoni}, Roberto and {Ramsay}, Gavin and {Udry}, St{\'e}phane and {Appourchaux}, Thierry and {Benz}, Willy and {Brandeker}, Alexis and {G{\"u}del}, Manuel and {Janot-Pacheco}, Eduardo and {Kabath}, Petr and {Kjeldsen}, Hans and {Min}, Michiel and {Santos}, Nuno and {Smith}, Alan and {Suarez}, Juan-Carlos and {Werner}, Stephanie C. and {Aboudan}, Alessio and {Abreu}, Manuel and {Acu a}, Lorena and {Adams}, Moritz and {Adibekyan}, Vardan and {Affer}, Laura and {Agneray}, Fran{\c{c}}ois and {Agnor}, Craig and {Aguirre B{\o}rsen-Koch}, Victor and {Ahmed}, Saad and {Aigrain}, Suzanne and {Al-Bahlawan}, Ashraf and {Alcacera Gil}, M de los Angeles and {Alei}, Eleonora and {Alencar}, Silvia and {Alexander}, Richard and {Alfonso-Garz{\'o}n}, Julia and {Alibert}, Yann and {Allende Prieto}, Carlos and {Almeida}, Leonardo and {Alonso Sobrino}, Roi and {Altavilla}, Giuseppe and {Althaus}, Christian and {Alonso Alvarez Trujillo}, Luis and {Amarsi}, Anish and {Ammler-von Eiff}, Matthias and {Am{\^o}res}, Eduardo and {Andrade}, Laerte and {Antoniadis-Karnavas}, Alexandros and {Ant{\'o}nio}, Carlos and {Aparicio del Moral}, Beatriz and {Appolloni}, Matteo and {Arena}, Claudio and {Armstrong}, David and {Aroca Aliaga}, Jose and {Asplund}, Martin and {Audenaert}, Jeroen and {Auricchio}, Natalia and {Avelino}, Pedro and {Baeke}, Ann and {Bailli{\'e}}, Kevin and {Balado}, Ana and {Ballber Balaguer{\'o}}, Pau and {Balestra}, Andrea and {Ball}, Warrick and {Ballans}, Herve and {Ballot}, Jerome and {Barban}, Caroline and {Barbary}, Ga{\"e}le and {Barbieri}, Mauro and {Barcel{\'o} Forteza}, Sebasti and {Barker}, Adrian and {Barklem}, Paul and {Barnes}, Sydney and {Barrado Navascues}, David and {Barragan}, Oscar and {Baruteau}, Cl{\'e}ment and {Basu}, Sarbani and {Baudin}, Frederic and {Baumeister}, Philipp and {Bayliss}, Daniel and {Bazot}, Michael and {Beck}, Paul G. and {Bedding}, Tim and {Belkacem}, Kevin and {Bellinger}, Earl and {Benatti}, Serena and {Benomar}, Othman and {B{\'e}rard}, Diane and {Bergemann}, Maria and {Bergomi}, Maria and {Bernardo}, Pierre and {Biazzo}, Katia and {Bignamini}, Andrea and {Bigot}, Lionel and {Billot}, Nicolas and {Binet}, Martin and {Biondi}, David and {Biondi}, Federico and {Birch}, Aaron C. and {Bitsch}, Bertram and {Bluhm Ceballos}, Paz Victoria and {B{\'o}di}, Attila and {Bogn{\'a}r}, Zs{\'o}fia and {Boisse}, Isabelle and {Bolmont}, Emeline and {Bonanno}, Alfio and {Bonavita}, Mariangela and {Bonfanti}, Andrea and {Bonfils}, Xavier and {Bonito}, Rosaria and {Bonomo}, Aldo Stefano and {B{\"o}rner}, Anko and {Boro Saikia}, Sudeshna and {Borreguero Mart{\'\i}n}, Elisa and {Borsa}, Francesco and {Borsato}, Luca and {Bossini}, Diego and {Bouchy}, Francois and {Bou{\'e}}, Gwena{\"e}l and {Boufleur}, Rodrigo and {Boumier}, Patrick and {Bourrier}, Vincent and {Bowman}, Dominic M. and {Bozzo}, Enrico and {Bradley}, Louisa and {Bray}, John and {Bressan}, Alessandro and {Breton}, Sylvain and {Brienza}, Daniele and {Brito}, Ana and {Brogi}, Matteo and {Brown}, Beverly and {Brown}, David J.~A. and {Brun}, Allan Sacha and {Bruno}, Giovanni and {Bruns}, Michael and {Buchhave}, Lars A. and {Bugnet}, Lisa and {Buldgen}, Ga{\"e}l and {Burgess}, Patrick and {Busatta}, Andrea and {Busso}, Giorgia and {Buzasi}, Derek and {Caballero}, Jos{\'e} A. and {Cabral}, Alexandre and {Cabrero Gomez}, Juan-Francisco and {Calderone}, Flavia and {Cameron}, Robert and {Cameron}, Andrew and {Campante}, Tiago and {Campos Gestal}, N{\'e}stor and {Canto Martins}, Bruno Leonardo and {Cara}, Christophe and {Carone}, Ludmila and {Carrasco}, Josep Manel and {Casagrande}, Luca and {Casewell}, Sarah L. and {Cassisi}, Santi and {Castellani}, Marco and {Castro}, Matthieu and {Catala}, Claude and {Catal{\'a}n Fern{\'a}ndez}, Irene and {Catelan}, M{\'a}rcio and {Cegla}, Heather and {Cerruti}, Chiara and {Cessa}, Virginie and {Chadid}, Merieme and {Chaplin}, William and {Charpinet}, Stephane and {Chiappini}, Cristina and {Chiarucci}, Simone and {Chiavassa}, Andrea and {Chinellato}, Simonetta and {Chirulli}, Giovanni and {Christensen-Dalsgaard}, J{\o}rgen and {Church}, Ross and {Claret}, Antonio and {Clarke}, Cathie and {Claudi}, Riccardo and {Clermont}, Lionel and {Coelho}, Hugo and {Coelho}, Joao and {Cogato}, Fabrizio and {Colom{\'e}}, Josep and {Condamin}, Mathieu and {Conde Garc{\'\i}a}, Fernando and {Conseil}, Simon},
        title = "{The PLATO Mission}",
      journal = {arXiv e-prints},
         year = 2024,
        month = jun,
          eid = {arXiv:2406.05447},
        pages = {arXiv:2406.05447},
          doi = {10.48550/arXiv.2406.05447},
archivePrefix = {arXiv},
       eprint = {2406.05447},
 primaryClass = {astro-ph.IM},
       adsurl = {https://ui.adsabs.harvard.edu/abs/2024arXiv240605447R}
}

@ARTICLE{Kokori2023,
       author = {{Kokori}, A. and {Tsiaras}, A. and {Edwards}, B. and {Jones}, A. and {Pantelidou}, G. and {Tinetti}, G. and {Bewersdorff}, L. and {Iliadou}, A. and {Jongen}, Y. and {Lekkas}, G. and {Nastasi}, A. and {Poultourtzidis}, E. and {Sidiropoulos}, C. and {Walter}, F. and {W{\"u}nsche}, A. and {Abraham}, R. and {Agnihotri}, V.~K. and {Albanesi}, R. and {Arce-Mansego}, E. and {Arnot}, D. and {Audejean}, M. and {Aumasson}, C. and {Bachschmidt}, M. and {Baj}, G. and {Barroy}, P.~R. and {Belinski}, A.~A. and {Bennett}, D. and {Benni}, P. and {Bernacki}, K. and {Betti}, L. and {Biagini}, A. and {Bosch}, P. and {Brandebourg}, P. and {Br{\'a}t}, L. and {Bretton}, M. and {Brincat}, S.~M. and {Brouillard}, S. and {Bruzas}, A. and {Bruzzone}, A. and {Buckland}, R.~A. and {Cal{\'o}}, M. and {Campos}, F. and {Carre{\~n}o}, A. and {Carrion Rodrigo}, J.~A. and {Casali}, R. and {Casalnuovo}, G. and {Cataneo}, M. and {Chang}, C. -M. and {Changeat}, L. and {Chowdhury}, V. and {Ciantini}, R. and {Cilluffo}, M. and {Coliac}, J. -F. and {Conzo}, G. and {Correa}, M. and {Coulon}, G. and {Crouzet}, N. and {Crow}, M.~V. and {Curtis}, I.~A. and {Daniel}, D. and {Dauchet}, B. and {Dawes}, S. and {Deldem}, M. and {Deligeorgopoulos}, D. and {Dransfield}, G. and {Dymock}, R. and {Eenm{\"a}e}, T. and {Esseiva}, N. and {Evans}, P. and {Falco}, C. and {Farf{\'a}n}, R.~G. and {Fern{\'a}ndez-Laj{\'u}s}, E. and {Ferratfiat}, S. and {Ferreira}, S.~L. and {Ferretti}, A. and {Fio{\l}ka}, J. and {Fowler}, M. and {Futcher}, S.~R. and {Gabellini}, D. and {Gainey}, T. and {Gaitan}, J. and {Gajdo{\v{s}}}, P. and {Garc{\'\i}a-S{\'a}nchez}, A. and {Garlitz}, J. and {Gillier}, C. and {Gison}, C. and {Gonzales}, J. and {Gorshanov}, D. and {Grau Horta}, F. and {Grivas}, G. and {Guerra}, P. and {Guillot}, T. and {Haswell}, C.~A. and {Haymes}, T. and {Hentunen}, V. -P. and {Hills}, K. and {Hose}, K. and {Humbert}, T. and {Hurter}, F. and {Hynek}, T. and {Irzyk}, M. and {Jacobsen}, J. and {Jannetta}, A.~L. and {Johnson}, K. and {J{\'o}{\'z}wik-Wabik}, P. and {Kaeouach}, A.~E. and {Kang}, W. and {Kiiskinen}, H. and {Kim}, T. and {Kivila}, {\"U}. and {Koch}, B. and {Kolb}, U. and {Ku{\v{c}}{\'a}kov{\'a}}, H. and {Lai}, S. -P. and {Laloum}, D. and {Lasota}, S. and {Lewis}, L.~A. and {Liakos}, G. -I. and {Libotte}, F. and {Lomoz}, F. and {Lopresti}, C. and {Majewski}, R. and {Malcher}, A. and {Mallonn}, M. and {Mannucci}, M. and {Marchini}, A. and {Mari}, J. -M. and {Marino}, A. and {Marino}, G. and {Mario}, J. -C. and {Marquette}, J. -B. and {Mart{\'\i}nez-Bravo}, F.~A. and {Ma{\v{s}}ek}, M. and {Matassa}, P. and {Michel}, P. and {Michelet}, J. and {Miller}, M. and {Miny}, E. and {Molina}, D. and {Mollier}, T. and {Monteleone}, B. and {Montigiani}, N. and {Morales-Aimar}, M. and {Mortari}, F. and {Morvan}, M. and {Mugnai}, L.~V. and {Murawski}, G. and {Naponiello}, L. and {Naudin}, J. -L. and {Naves}, R. and {N{\'e}el}, D. and {Neito}, R. and {Neveu}, S. and {Noschese}, A. and {{\"O}{\u{g}}men}, Y. and {Ohshima}, O. and {Orbanic}, Z. and {Pace}, E.~P. and {Pantacchini}, C. and {Paschalis}, N.~I. and {Pereira}, C. and {Peretto}, I. and {Perroud}, V. and {Phillips}, M. and {Pintr}, P. and {Pioppa}, J. -B. and {Plazas}, J. and {Poelarends}, A.~J. and {Popowicz}, A. and {Purcell}, J. and {Quinn}, N. and {Raetz}, M. and {Rees}, D. and {Regembal}, F. and {Rocchetto}, M. and {Rocci}, P. -F. and {Rockenbauer}, M. and {Roth}, R. and {Rousselot}, L. and {Rubia}, X. and {Ruocco}, N. and {Russo}, E. and {Salisbury}, M. and {Salvaggio}, F. and {Santos}, A. and {Savage}, J. and {Scaggiante}, F. and {Sedita}, D. and {Shadick}, S. and {Silva}, A.~F. and {Sioulas}, N. and {{\v{S}}koln{\'\i}k}, V. and {Smith}, M. and {Smolka}, M. and {Solmaz}, A. and {Stanbury}, N. and {Stouraitis}, D. and {Tan}, T. -G. and {Theusner}, M. and {Thurston}, G. and {Tifner}, F.~P. and {Tomacelli}, A. and {Tomatis}, A. and {Trnka}, J. and {Tyl{\v{s}}ar}, M. and {Valeau}, P. and {Vignes}, J. -P. and {Villa}, A. and {Vives Sureda}, A. and {Vora}, K. and {Vra{\v{s}}t'{\'a}k}, M. and {Walliang}, D. and {Wenzel}, B. and {Wright}, D.~E. and {Zambelli}, R. and {Zhang}, M. and {Z{\'\i}bar}, M.},
        title = "{ExoClock Project. III. 450 New Exoplanet Ephemerides from Ground and Space Observations}",
      journal = {\apjs},
         year = 2023,
        month = mar,
       volume = {265},
       number = {1},
          eid = {4},
        pages = {4},
          doi = {10.3847/1538-4365/ac9da4},
archivePrefix = {arXiv},
       eprint = {2209.09673},
 primaryClass = {astro-ph.EP},
       adsurl = {https://ui.adsabs.harvard.edu/abs/2023ApJS..265....4K}
}

@ARTICLE{Cutri2003,
       author = {{Cutri}, R.~M. and {Skrutskie}, M.~F. and {van Dyk}, S. and {Beichman}, C.~A. and {Carpenter}, J.~M. and {Chester}, T. and {Cambresy}, L. and {Evans}, T. and {Fowler}, J. and {Gizis}, J. and {Howard}, E. and {Huchra}, J. and {Jarrett}, T. and {Kopan}, E.~L. and {Kirkpatrick}, J.~D. and {Light}, R.~M. and {Marsh}, K.~A. and {McCallon}, H. and {Schneider}, S. and {Stiening}, R. and {Sykes}, M. and {Weinberg}, M. and {Wheaton}, W.~A. and {Wheelock}, S. and {Zacarias}, N.},
        title = "{VizieR Online Data Catalog: 2MASS All-Sky Catalog of Point Sources (Cutri+ 2003)}",
      journal = {VizieR Online Data Catalog},
         year = 2003,
        month = jun,
          eid = {II/246},
        pages = {II/246},
       adsurl = {https://ui.adsabs.harvard.edu/abs/2003yCat.2246....0C}
}

@ARTICLE{pyaneti2,
       author = {{Barrag{\'a}n}, Oscar and {Aigrain}, Suzanne and {Rajpaul}, Vinesh M. and {Zicher}, Norbert},
        title = "{PYANETI - II. A multidimensional Gaussian process approach to analysing spectroscopic time-series}",
      journal = {\mnras},
         year = 2022,
        month = jan,
       volume = {509},
       number = {1},
        pages = {866-883},
          doi = {10.1093/mnras/stab2889},
archivePrefix = {arXiv},
       eprint = {2109.14086},
       adsurl = {https://ui.adsabs.harvard.edu/abs/2022MNRAS.509..866B}
}

@ARTICLE{Stassun2019,
       author = {{Stassun}, Keivan G. and {Oelkers}, Ryan J. and {Paegert}, Martin and {Torres}, Guillermo and {Pepper}, Joshua and {De Lee}, Nathan and {Collins}, Kevin and {Latham}, David W. and {Muirhead}, Philip S. and {Chittidi}, Jay and {Rojas-Ayala}, B{\'a}rbara and {Fleming}, Scott W. and {Rose}, Mark E. and {Tenenbaum}, Peter and {Ting}, Eric B. and {Kane}, Stephen R. and {Barclay}, Thomas and {Bean}, Jacob L. and {Brassuer}, C.~E. and {Charbonneau}, David and {Ge}, Jian and {Lissauer}, Jack J. and {Mann}, Andrew W. and {McLean}, Brian and {Mullally}, Susan and {Narita}, Norio and {Plavchan}, Peter and {Ricker}, George R. and {Sasselov}, Dimitar and {Seager}, S. and {Sharma}, Sanjib and {Shiao}, Bernie and {Sozzetti}, Alessandro and {Stello}, Dennis and {Vanderspek}, Roland and {Wallace}, Geoff and {Winn}, Joshua N.},
        title = "{The Revised TESS Input Catalog and Candidate Target List}",
      journal = {\aj},
         year = 2019,
        month = oct,
       volume = {158},
       number = {4},
          eid = {138},
        pages = {138},
          doi = {10.3847/1538-3881/ab3467},
archivePrefix = {arXiv},
       eprint = {1905.10694},
 primaryClass = {astro-ph.SR},
       adsurl = {https://ui.adsabs.harvard.edu/abs/2019AJ....158..138S}
}

@ARTICLE{2013A&A...553A...6H,

       author = {{Husser}, T. -O. and {Wende-von Berg}, S. and {Dreizler}, S. and {Homeier}, D. and {Reiners}, A. and {Barman}, T. and {Hauschildt}, P.~H.},

        title = "{A new extensive library of PHOENIX stellar atmospheres and synthetic spectra}",

      journal = {\aap},

         year = 2013,

        month = may,

       volume = {553},

          eid = {A6},

        pages = {A6},

          doi = {10.1051/0004-6361/201219058},

archivePrefix = {arXiv},

       eprint = {1303.5632},

 primaryClass = {astro-ph.SR},

       adsurl = {https://ui.adsabs.harvard.edu/abs/2013A&A...553A...6H}

}

@ARTICLE{2012RSPTA.370.2765A,

       author = {{Allard}, F. and {Homeier}, D. and {Freytag}, B.},

        title = "{Models of very-low-mass stars, brown dwarfs and exoplanets}",

      journal = {Philosophical Transactions of the Royal Society of London Series A},

         year = 2012,

        month = jun,

       volume = {370},

       number = {1968},

        pages = {2765-2777},

          doi = {10.1098/rsta.2011.0269},

archivePrefix = {arXiv},

       eprint = {1112.3591},

 primaryClass = {astro-ph.SR},

       adsurl = {https://ui.adsabs.harvard.edu/abs/2012RSPTA.370.2765A}

}

@ARTICLE{Castelli2004,

   author = {{Castelli}, F. and {Kurucz}, R.~L.},

    title = "{New Grids of ATLAS9 Model Atmospheres}",

  journal = {astro-ph/0405087},

   eprint = {astro-ph/0405087},

     year = 2004,

    month = may,

   adsurl = {http://adsabs.harvard.edu/abs/2004astro.ph..5087C}

}

@ARTICLE{kurucz2005,
       author = {{Kurucz}, Robert L.},
        title = "{ATLAS12, SYNTHE, ATLAS9, WIDTH9, et cetera}",
      journal = {Memorie della Societa Astronomica Italiana Supplementi},
         year = 2005,
        month = jan,
       volume = {8},
        pages = {14},
       adsurl = {https://ui.adsabs.harvard.edu/abs/2005MSAIS...8...14K}
}

@BOOK{Kurucz-93,
   author = {{Kurucz}, R.~L.},
    title = "{SYNTHE spectrum synthesis programs and line data}",
booktitle = {Kurucz CD-ROM, Cambridge, MA: Smithsonian Astrophysical Observatory, |c1993, December 4, 1993},
     year = 1993,
Publisher = "Kurucz CD-ROM", 
   adsurl = {http://adsabs.harvard.edu/abs/1993sssp.book.....K}
}

@ARTICLE{SERVAL,
       author = {{Zechmeister}, M. and {Reiners}, A. and {Amado}, P.~J. and {Azzaro}, M. and {Bauer}, F.~F. and {B{\'e}jar}, V.~J.~S. and {Caballero}, J.~A. and {Guenther}, E.~W. and {Hagen}, H. -J. and {Jeffers}, S.~V. and {Kaminski}, A. and {K{\"u}rster}, M. and {Launhardt}, R. and {Montes}, D. and {Morales}, J.~C. and {Quirrenbach}, A. and {Reffert}, S. and {Ribas}, I. and {Seifert}, W. and {Tal-Or}, L. and {Wolthoff}, V.},
        title = "{Spectrum radial velocity analyser (SERVAL). High-precision radial velocities and two alternative spectral indicators}",
      journal = {\aap},
         year = 2018,
        month = jan,
       volume = {609},
          eid = {A12},
        pages = {A12},
          doi = {10.1051/0004-6361/201731483},
archivePrefix = {arXiv},
       eprint = {1710.10114},
 primaryClass = {astro-ph.IM},
       adsurl = {https://ui.adsabs.harvard.edu/abs/2018A&A...609A..12Z}
}

@ARTICLE{Barragan2022,
       author = {{Barrag{\'a}n}, O. and {Armstrong}, D.~J. and {Gandolfi}, D. and {Carleo}, I. and {Vidotto}, A.~A. and {Villarreal D'Angelo}, C. and {Oklop{\v{c}}i{\'c}}, A. and {Isaacson}, H. and {Oddo}, D. and {Collins}, K. and {Fridlund}, M. and {Sousa}, S.~G. and {Persson}, C.~M. and {Hellier}, C. and {Howell}, S. and {Howard}, A. and {Redfield}, S. and {Eisner}, N. and {Georgieva}, I.~Y. and {Dragomir}, D. and {Bayliss}, D. and {Nielsen}, L.~D. and {Klein}, B. and {Aigrain}, S. and {Zhang}, M. and {Teske}, J. and {Twicken}, J.~D. and {Jenkins}, J. and {Esposito}, M. and {Van Eylen}, V. and {Rodler}, F. and {Adibekyan}, V. and {Alarcon}, J. and {Anderson}, D.~R. and {Akana Murphy}, J.~M. and {Barrado}, D. and {Barros}, S.~C.~C. and {Benneke}, B. and {Bouchy}, F. and {Bryant}, E.~M. and {Butler}, R.~P. and {Burt}, J. and {Cabrera}, J. and {Casewell}, S. and {Chaturvedi}, P. and {Cloutier}, R. and {Cochran}, W.~D. and {Crane}, J. and {Crossfield}, I. and {Crouzet}, N. and {Collins}, K.~I. and {Dai}, F. and {Deeg}, H.~J. and {Deline}, A. and {Demangeon}, O.~D.~S. and {Dumusque}, X. and {Figueira}, P. and {Furlan}, E. and {Gnilka}, C. and {Goad}, M.~R. and {Goffo}, E. and {Guti{\'e}rrez-Canales}, F. and {Hadjigeorghiou}, A. and {Hartman}, Z. and {Hatzes}, A.~P. and {Harris}, M. and {Henderson}, B. and {Hirano}, T. and {Hojjatpanah}, S. and {Hoyer}, S. and {Kab{\'a}th}, P. and {Korth}, J. and {Lillo-Box}, J. and {Luque}, R. and {Marmier}, M. and {Mo{\v{c}}nik}, T. and {Muresan}, A. and {Murgas}, F. and {Nagel}, E. and {Osborne}, H.~L.~M. and {Osborn}, A. and {Osborn}, H.~P. and {Palle}, E. and {Raimbault}, M. and {Ricker}, G.~R. and {Rubenzahl}, R.~A. and {Stockdale}, C. and {Santos}, N.~C. and {Scott}, N. and {Schwarz}, R.~P. and {Shectman}, S. and {Raimbault}, M. and {Seager}, S. and {S{\'e}gransan}, D. and {Serrano}, L.~M. and {Skarka}, M. and {Smith}, A.~M.~S. and {{\v{S}}ubjak}, J. and {Tan}, T.~G. and {Udry}, S. and {Watson}, C. and {Wheatley}, P.~J. and {West}, R. and {Winn}, J.~N. and {Wang}, S.~X. and {Wolfgang}, A. and {Ziegler}, C.},
        title = "{The young HD 73583 (TOI-560) planetary system: two 10-M$_{{\ensuremath{\oplus}}}$ mini-Neptunes transiting a 500-Myr-old, bright, and active K dwarf}",
      journal = {\mnras},
         year = 2022,
        month = aug,
       volume = {514},
       number = {2},
        pages = {1606-1627},
          doi = {10.1093/mnras/stac638},
archivePrefix = {arXiv},
       eprint = {2110.13069},
 primaryClass = {astro-ph.EP},
       adsurl = {https://ui.adsabs.harvard.edu/abs/2022MNRAS.514.1606B}
}

@ARTICLE{Batalha2019,
       author = {{Batalha}, Natasha E. and {Lewis}, Taylor and {Fortney}, Jonathan J. and {Batalha}, Natalie M. and {Kempton}, Eliza and {Lewis}, Nikole K. and {Line}, Michael R.},
        title = "{The Precision of Mass Measurements Required for Robust Atmospheric Characterization of Transiting Exoplanets}",
      journal = {\apjl},
         year = 2019,
        month = nov,
       volume = {885},
       number = {1},
          eid = {L25},
        pages = {L25},
          doi = {10.3847/2041-8213/ab4909},
archivePrefix = {arXiv},
       eprint = {1910.00076},
 primaryClass = {astro-ph.EP},
       adsurl = {https://ui.adsabs.harvard.edu/abs/2019ApJ...885L..25B}
}

@ARTICLE{Gaia2020,
       author = {{Gaia Collaboration}},
        title = "{VizieR Online Data Catalog: Gaia EDR3 (Gaia Collaboration, 2020)}",
      journal = {VizieR Online Data Catalog},
         year = 2020,
        month = nov,
          eid = {I/350},
        pages = {I/350},
       adsurl = {https://ui.adsabs.harvard.edu/abs/2020yCat.1350....0G}
}

@ARTICLE{Nicholson2022,
       author = {{Nicholson}, B.~A. and {Aigrain}, S.},
        title = "{Quasi-periodic Gaussian processes for stellar activity: From physical to kernel parameters}",
      journal = {\mnras},
         year = 2022,
        month = oct,
       volume = {515},
       number = {4},
        pages = {5251-5266},
          doi = {10.1093/mnras/stac2097},
archivePrefix = {arXiv},
       eprint = {2207.12164},
 primaryClass = {astro-ph.SR},
       adsurl = {https://ui.adsabs.harvard.edu/abs/2022MNRAS.515.5251N}
}

@ARTICLE{NASAexoplanet,
       author = {{Christiansen}, Jessie L. and {McElroy}, Douglas L. and {Harbut}, Marcy and {Ciardi}, David R. and {Crane}, Megan and {Good}, John and {Hardegree-Ullman}, Kevin K. and {Kesseli}, Aurora Y. and {Lund}, Michael B. and {Lynn}, Meca and {Muthiar}, Ananda and {Nilsson}, Ricky and {Oluyide}, Toba and {Papin}, Michael and {Rivera}, Amalia and {Swain}, Melanie and {Susemiehl}, Nicholas D. and {Tam}, Raymond and {van Eyken}, Julian and {Beichman}, Charles},
        title = "{The NASA Exoplanet Archive and Exoplanet Follow-up Observing Program: Data, Tools, and Usage}",
      journal = {PSJ},
         year = 2025,
        month = aug,
       volume = {6},
       number = {8},
          eid = {186},
        pages = {186},
          doi = {10.3847/PSJ/ade3c2},
archivePrefix = {arXiv},
       eprint = {2506.03299},
 primaryClass = {astro-ph.EP},
       adsurl = {https://ui.adsabs.harvard.edu/abs/2025PSJ.....6..186C}
}

@ARTICLE{matplotlib,

  author={Hunter, John D.},

  journal={Computing in Science & Engineering}, 

  title={Matplotlib: A 2D Graphics Environment}, 

  year={2007},

  volume={9},

  number={3},

  pages={90-95},

  doi={10.1109/MCSE.2007.55}}

@misc{pandas,
    author       = {The pandas development team},
    title        = {pandas-dev/pandas: Pandas},
    month        = feb,
    year         = 2020,
    publisher    = {Zenodo},
    version      = {latest},
    doi          = {10.5281/zenodo.3509134},
    url          = {https://doi.org/10.5281/zenodo.3509134}
}

@ARTICLE{Hog2000,
       author = {{H{\o}g}, E. and {Fabricius}, C. and {Makarov}, V.~V. and {Urban}, S. and {Corbin}, T. and {Wycoff}, G. and {Bastian}, U. and {Schwekendiek}, P. and {Wicenec}, A.},
        title = "{The Tycho-2 catalogue of the 2.5 million brightest stars}",
      journal = {\aap},
         year = 2000,
        month = mar,
       volume = {355},
        pages = {L27-L30},
       adsurl = {https://ui.adsabs.harvard.edu/abs/2000A&A...355L..27H}
}

@ARTICLE{Nardiello2022,
       author = {{Nardiello}, D. and {Malavolta}, L. and {Desidera}, S. and {Baratella}, M. and {D'Orazi}, V. and {Messina}, S. and {Biazzo}, K. and {Benatti}, S. and {Damasso}, M. and {Rajpaul}, V.~M. and {Bonomo}, A.~S. and {Dolcetta}, R. Capuzzo and {Mallonn}, M. and {Cale}, B. and {Plavchan}, P. and {El Mufti}, M. and {Bignamini}, A. and {Borsa}, F. and {Carleo}, I. and {Claudi}, R. and {Covino}, E. and {Lanza}, A.~F. and {Maldonado}, J. and {Mancini}, L. and {Micela}, G. and {Molinari}, E. and {Pinamonti}, M. and {Piotto}, G. and {Poretti}, E. and {Scandariato}, G. and {Sozzetti}, A. and {Andreuzzi}, G. and {Boschin}, W. and {Cosentino}, R. and {Fiorenzano}, A.~F.~M. and {Harutyunyan}, A. and {Knapic}, C. and {Pedani}, M. and {Affer}, L. and {Maggio}, A. and {Rainer}, M.},
        title = "{The GAPS Programme at TNG. XXXVII. A precise density measurement of the young ultra-short period planet TOI-1807 b}",
      journal = {\aap},
         year = 2022,
        month = aug,
       volume = {664},
          eid = {A163},
        pages = {A163},
          doi = {10.1051/0004-6361/202243743},
archivePrefix = {arXiv},
       eprint = {2206.03496},
 primaryClass = {astro-ph.EP},
       adsurl = {https://ui.adsabs.harvard.edu/abs/2022A&A...664A.163N}
}

@ARTICLE{owenwu2017,
       author = {{Owen}, James E. and {Wu}, Yanqin},
        title = "{The Evaporation Valley in the Kepler Planets}",
      journal = {\apj},
         year = 2017,
        month = sep,
       volume = {847},
       number = {1},
          eid = {29},
        pages = {29},
          doi = {10.3847/1538-4357/aa890a},
archivePrefix = {arXiv},
       eprint = {1705.10810},
 primaryClass = {astro-ph.EP},
       adsurl = {https://ui.adsabs.harvard.edu/abs/2017ApJ...847...29O}
}
%

\bsp	
\label{lastpage}
\end{document}